 \documentclass[final,5p,times,twocolumn]{elsarticle}

\usepackage{amssymb}
\usepackage{amsmath}
\usepackage{mathrsfs}
\usepackage{color}
\usepackage{booktabs}
\usepackage{graphicx}
\usepackage{mdwlist}

\usepackage{epstopdf}

\usepackage{hyperref}
\usepackage{csquotes}
\usepackage{subscript}

\newcommand{\refEq}[1] {(\ref{#1})}
\newcommand{\superscript}[1]{\ensuremath{^{\textrm{#1}}}}
\newcommand{\subscript}[1]{\ensuremath{_{\textrm{#1}}}}

\journal{Fusion Engineering and Design}

\begin{document}

\begin{frontmatter}



\title{ARC: A compact, high-field, fusion nuclear science facility and demonstration power plant with demountable magnets}


\author{B.N. Sorbom, J. Ball, T.R. Palmer, F.J. Mangiarotti, J.M. Sierchio, P. Bonoli, C. Kasten, D.A. Sutherland, H.S. Barnard, C.B. Haakonsen, J. Goh, C. Sung, and D.G. Whyte}

\address{Plasma Science and Fusion Center, Massachusetts Institute of Technology, Cambridge, MA 02139, USA}

\begin{abstract}

The affordable, robust, compact (ARC) reactor is the product of a conceptual design study aimed at reducing the size, cost, and complexity of a combined fusion nuclear science facility (FNSF) and demonstration fusion Pilot power plant. ARC is a $\sim 200-250$ MWe tokamak reactor with a major radius of $3.3$ m, a minor radius of $1.1$ m, and an on-axis magnetic field of $9.2$ T. ARC has rare earth barium copper oxide (REBCO) superconducting toroidal field coils, which have joints to enable disassembly. This allows the vacuum vessel to be replaced quickly, mitigating first wall survivability concerns, and permits a single device to test many vacuum vessel designs and divertor materials. The design point has a plasma fusion gain of $Q_p$ $\approx$ 13.6, yet is fully non-inductive, with a modest bootstrap fraction of only $\sim$63$\%$. Thus ARC offers a high power gain with relatively large external control of the current profile. This highly attractive combination is enabled by the $\sim$23 T peak field on coil achievable with newly available REBCO superconductor technology. External current drive is provided by two innovative inboard RF launchers using 25 MW of lower hybrid and 13.6 MW of ion cyclotron fast wave power. The resulting efficient current drive provides a robust, steady state core plasma far from disruptive limits. ARC uses an all-liquid blanket, consisting of low pressure, slowly flowing fluorine lithium beryllium (FLiBe) molten salt. The liquid blanket is low-risk technology and provides effective neutron moderation and shielding, excellent heat removal, and a tritium breeding ratio $\geq 1.1$. The large temperature range over which FLiBe is liquid permits an output blanket temperature of $900$ K, single phase fluid cooling, and a high efficiency helium Brayton cycle, which allows for net electricity generation when operating ARC as a Pilot power plant. 

\end{abstract}


\end{frontmatter}

% \linenumbers

\section{Introduction}
\label{sec:Introduction}

Most fusion reactor designs, such as the ARIES studies \cite{NajmabadiARIES_I1991,NajmabadiARIES_AT2006,NajmabadiARIES_RS1997,NajmabadiARIES_ST2003}, assume a large, fixed 1000 MWe output for a power plant.  However, large-scale designs make fusion engineering research and development difficult because of the high cost and long construction time of experiments. This paper presents a smaller, less costly, timelier, and lower risk alternative, the $200$ MWe ARC reactor. ARC is a conceptual point design of a fusion nuclear science facility/Pilot power plant that demonstrates the advantages of a compact, high-field design utilizing REBCO superconducting magnets and inboard launched lower hybrid current drive (LHCD). The design was carried out as a follow-on to the Vulcan conceptual design; a tokamak for studying plasma-material interaction (PMI) physics that also utilized the demountable REBCO tape and high-field side LHCD \cite{olynyk2012vulcan}. A goal of the ARC design is to minimize the reactor size in order to reduce the plant capital cost. Like Vulcan and several other proposed tokamaks \cite{NajmabadiARIES_AT2006, coppi1999critical, ando2005design, kingham2012efficient}, ARC makes use of high-temperature superconductors (HTS), which enables large on-axis magnetic fields and ultimately reduces the size of the reactor. It is important to emphasize that ARC represents one of many possible compact, high-field design configurations.  As discussed later in this paper, the modular nature of ARC allows it to change experimental direction and pursue the nuclear materials and vacuum vessel configurations that are determined to be most promising. This enables more innovative and speculative designs because the cost and operational implications of failure are reduced. Indeed a starting design philosophy of ARC is that failure should and will occur as various fusion materials and power exhaust technologies are tried and tested. However, because they can be readily fixed, these failures should not compromise the overall capacity of the device to produce fusing plasmas.

This paper is organized in the following way. Section \ref{sec:DesignOverview} presents an overview of the ARC design. Section \ref{sec:PhysicsBasis} describes the plasma physics basis for the reactor and discusses the current drive system. Section \ref{sec:Magnets} details the design of the magnet system. Section \ref{sec:PowerCore} presents the design of the fusion power core, consisting of the tritium breeding/heat exchange blanket and the neutron shield. Section \ref{sec:Economics} presents a simple costing estimate. Section \ref{sec:RandDReq} briefly lists the most vital research and development necessary to enable a design similar to ARC. Lastly, Section \ref{sec:Conclusions} provides some concluding remarks.

\section{Design motivation and overview}
\label{sec:DesignOverview}

\begin{figure*}[htbf]
 \centering
 \includegraphics[width=0.7\textwidth]{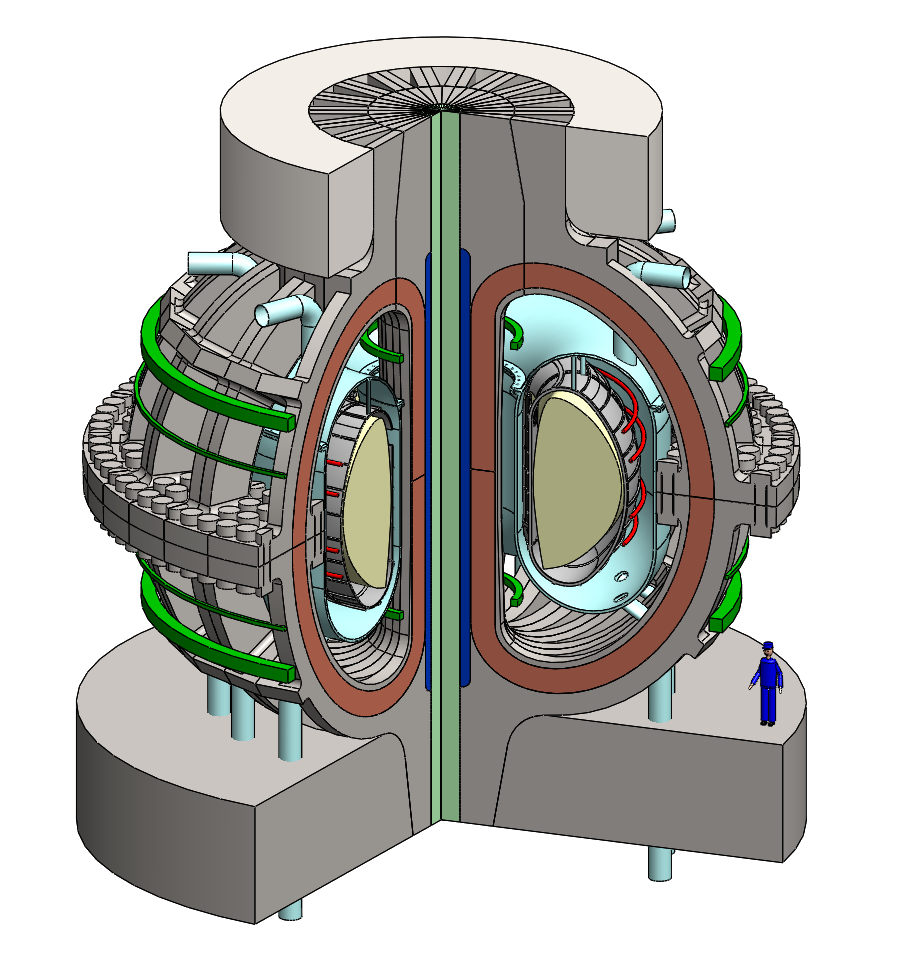}
 \caption{The ARC reactor, shown with the plasma in yellow and the TF superconducting tape in brown. Note the neutron shield is omitted for viewing clarity. Also note that although the ARC design is based on a diverted plasma, the physical divertor design was left for later study and a simplified representation of the vacuum vessel is shown here.}
 \label{fig:FullDesign}
\end{figure*}

The ARC reactor is a conceptual tokamak design that can function as both a demonstration fusion power plant for energy generation and a fusion nuclear science facility (FNSF) for integrated materials and component irradiation testing in a D-T neutron field. The starting objective of the ARC study was to determine if a reduced size D-T fusion device (fusion power $\leq$ 500 MW) could benefit from the high magnetic field technology offered by recently developed high temperature superconductors.  The reasoning was that a high magnitude magnetic field in a compact, superconducting device might offer not only access to high plasma gain $Q_p$, but also enable net electric gain $Q_e > 1$. This specific option has not been explored previously in design studies, although the recent advanced tokamak (AT) Pilot ($Q_e$=1) study of Menard et al. \cite{menard2011prospects} had similar design goals, but used conventional superconductor technology. A recent FNSF study is the FDF design \cite{stambaugh2010candidates}, which is a similar size to ARC, but consumes $> 500$ MW of electricity because it does not use superconducting magnets.

\begin{figure*}[htbf]
 \centering
 \includegraphics[width=0.55\textwidth]{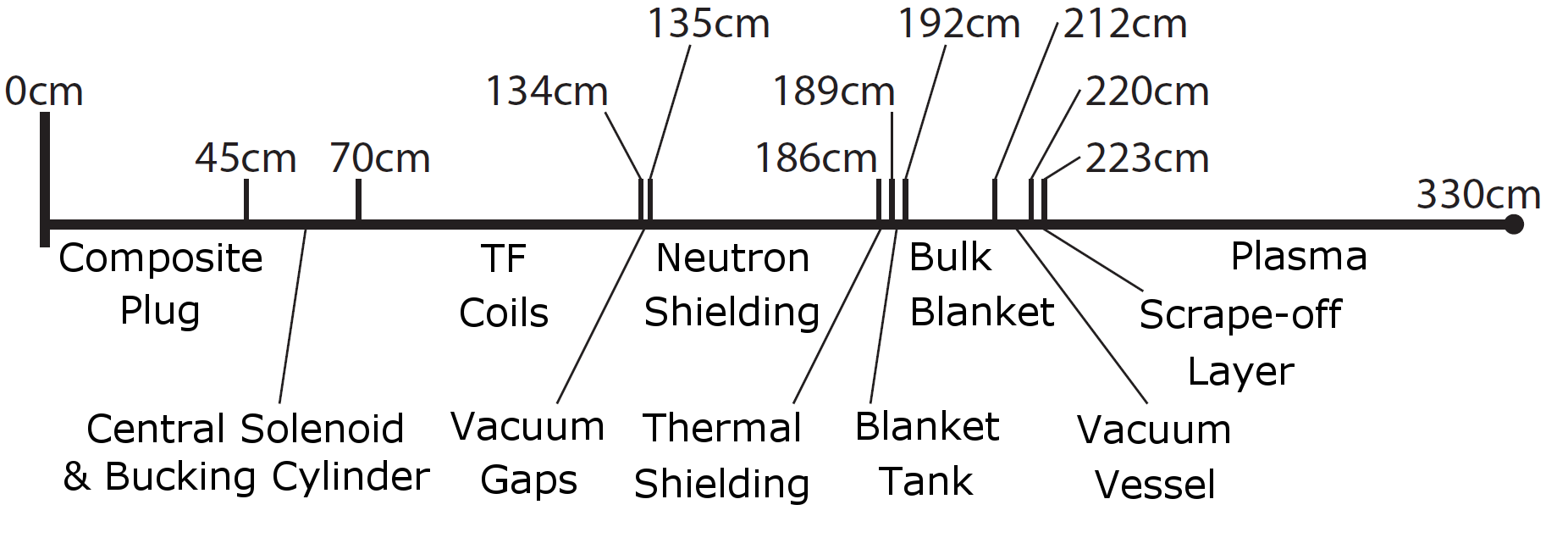}
 \caption{The ARC reactor inboard radial build.}
 \label{fig:inboardRadialBuild}
\end{figure*}

The reactor design is shown in Fig. \ref{fig:FullDesign}, the inboard radial build in Fig. \ref{fig:inboardRadialBuild}, and the most significant design parameters are given in Table \ref{tab:DesignParameters}. Another unique feature of the ARC design is that significant margin to disruptive operational limits was enforced from the start, i.e. strict limits on the edge safety factor (kink limit), Greenwald fraction (density limit) \cite{GreenwaldDensityLimit1988}, and normalized beta below the no-wall limit (pressure limit) \cite{TroyonMHDLimit1984} were imposed. This followed from the logic that high field designs should provide scenarios less prone to disruptions, which are nearly intolerable in burning plasmas because of internal material damage. Thus they should be strongly avoided in any tokamak FNSF/Pilot plant. 

ARC explores an innovative approach to current drive in burning plasma. Lower hybrid waves, launched from the high field side (HFS) of the tokamak, are used to noninductively drive plasma current. High field side launch is shown in modeling to increase the current drive efficiency, which is crucial to maximizing the power plant gain and providing better external control of the radial current profile.  Also, launching from the more quiescent HFS of the plasma is expected to reduce damage to the launcher \cite{LHCD_Vulcan} from plasma-material interactions.

\begin{table*}
  \centering
  \begin{tabular}{| l | c | c |}
    \hline
    Design Parameter & Symbol & Value \\
    \hline
    Fusion power & $P_{f}$ & $525$ MW \\
    Total thermal power & $P_{tot}$ & $708$ MW \\
    Plant thermal efficiency & $\eta_{elec}$ & $0.40$ \\
    Total electric power & $P_{e}$ & $283$ MW \\
    Net electric power & $P_{net}$ & $190$ MW \\
    LHCD coupled power & $P_{LH}$ & $25$ MW \\
    ICRF coupled power & $P_{IC}$ & $13.6$ MW \\
    Power multiplication factor & $Q_e$ & $3.0$ \\
    Major radius & $R_{0}$ & $3.3$ m \\
    Plasma semi-minor radius & $a$ & $1.13$ m \\
    Plasma elongation & $\kappa$ & $1.84$ \\
    Plasma volume & $V_p$ & $141$ $m^3$\\
    Toroidal magnetic field & $B_{0}$ & $9.2$ T \\
    Peak on-coil magnetic field & $B_{max}$ & $23$ T \\
    Plasma current & $I_{p}$ & $7.8$ MA \\
    Bootstrap fraction & $f_{BS}$ & $0.63$ \\
    Tritium Breeding Ratio & TBR & $1.1$ \\
    Avg. temperature & $\left\langle T \right\rangle$ & $14$ keV \\
    Avg. density & $\left\langle n \right\rangle$ & $1.3 \times 10^{20}$ m\superscript{-3}\\
    On-axis temperature & $T_0$ & $27$ keV \\
    On-axis density & $n_0$ & $1.8 \times 10^{20}$ m\superscript{-3}\\
    Greenwald fraction & $f_{Gr}$ & 0.67\\
    Toroidal beta & $\beta_{T}$ & $1.9\%$ \\
    Internal inductance & $l_{i}$ & $0.67$ \\
    Normalized beta & $\beta_{N}$ & $2.59$ \\
    Safety factor at $r/a= 0.95$ & $q_{95}$ & $7.2$ \\
    Edge safety factor & $q_{a}$ & $4.7$ \\
    Minimum safety factor & $q_{min}$ & $3.5$ \\
    Fusion power wall loading & $P_f/S_{b}$ & $2.5$ MW/m\superscript{2} \\
    Energy confinement time & $\tau_{E}$ & $0.64$ sec \\
    H89 confinement factor & $H_{89}$ & $2.8$\\
    H98(y,2) confinement factor & $H_{98,y2}$ & $1.8$\\
    G89 gain factor & $G_{89}$ & $0.14$\\
    \hline
  \end{tabular}
  \caption{List of significant ARC design parameters.}
  \label{tab:DesignParameters}
\end{table*}

\begin{figure}[htbf]
 \centering
 \includegraphics[width=0.3\textwidth]{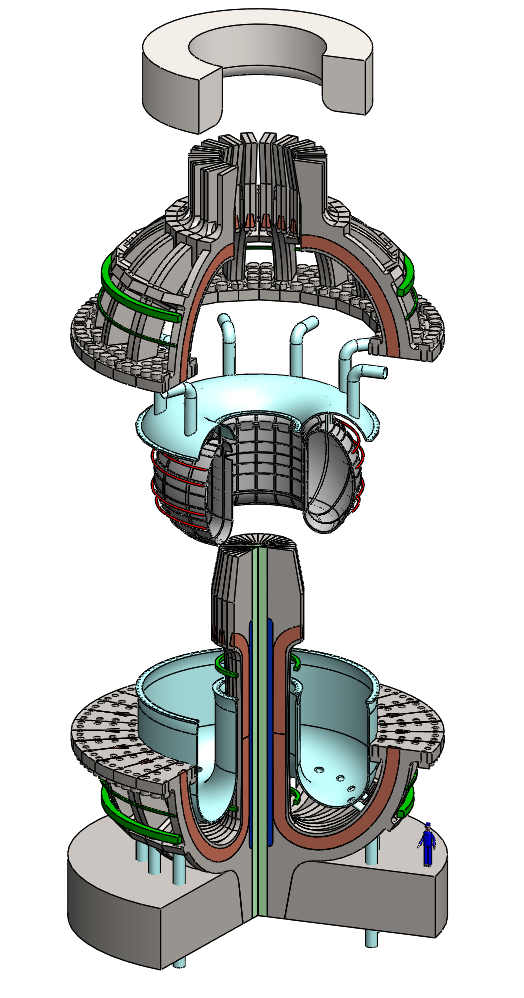}
 \caption{The upper half of ARC's superconducting coils can be removed, allowing the vacuum vessel to be removed from the blanket tank as a single piece.}
 \label{fig:FullDesignDisassembled}
\end{figure}

The use of REBCO superconducting technology in the toroidal field (TF) coils permits significantly higher on-axis magnetic fields than standard Nb\subscript{3}Sn superconductors. High magnetic field strength is essential in small reactor designs in order to achieve the necessary poloidal field/plasma current needed for sufficient confinement and stability against beta (pressure) limits. In addition, when holding beta constant the volumetric fusion power scales as $\sim B_{0}^{4}$ and, at constant safety factor, the plasma confinement strongly improves with magnetic field strength \cite{ITERoldTransport2007}. Since REBCO tapes allow the use of resistive joints in the superconducting coils \cite{vulcan_magnets}, the TF coils can be made demountable, meaning the coils can be split into two pieces (see Fig. \ref{fig:FullDesignDisassembled}). As discussed below, demountability can provide a dramatically different and likely more attractive, modular maintenance scheme for magnetic fusion devices.  The tradeoffs between modular component replacement \cite{barnard2012assessing} and power dissipation in the TF joints \cite{vulcan_magnets} have only been explored at small size in the Vulcan D-D device ($R_{0} \approx 1.2 \text{m}$), which motivates our exploration of demountability in a D-T reactor.

In sector maintenance, it is necessary to split all components located inside the TF coils into toroidal sections that can fit between the gaps in the TF coils, which is complex and time-intensive. It necessitates significantly larger TF coils to allow for space to remove sections of the vacuum vessel (e.g. ARIES-AT \cite{NajmabadiARIES_AT2006}). With joints, the TF coils, which are the most expensive component of a reactor \cite{AymarITERSummary2001}, can be made smaller. Furthermore, the entire vacuum vessel (including all internal components) can be externally constructed and tested as one modular part. This module can then, in principle, be relatively quickly lowered into place with the TF coils demounted, minimizing or eliminating the maintenance that must be performed inside the TF volume itself. The relative ease of installation and external testing of all internal components should greatly increase the simplicity and reliability of component replacement. This is a particularly attractive feature for an FNSF and was the motivation for using demountable copper coils in the FDF design \cite{stambaugh2010candidates}. Additionally, demountable copper TF coils have already been used in experimental devices such as COMPASS \cite{CrosslandcompassTFcoils1990} and Alcator C-Mod \cite{beck1991alcator}. A major motivation for this study was to explore the benefits of modular replacement versus the issues associated with TF joints in REBCO superconductors. The preliminary conclusion is that it represents an attractive alternative to sector maintenance in an FNSF and may also be the optimal choice for future commercial reactors.

The replaceable vacuum vessel is made of corrosion-resistant Inconel 718. Although the high nickel content of Inconel 718 makes this alloy much more prone to nuclear activation, it was chosen as a ``first-round'' material due to its ability to maintain high strength and corrosion resistance at elevated temperatures. Ideally, further materials research will identify a more suitable material for future iterations of the vacuum vessel design. The vessel is approximately shaped like an elliptical torus. It is double-walled and contains a channel through which FLiBe flows for cooling and tritium breeding. The vacuum vessel is attached to the blanket tank from above by 18 support columns, which are evenly spaced between the 18 TF coils. All connections needed for in-vessel components (such as waveguides, vacuum ports, etc.) run though these columns, which are also curved to reduce the flux of neutrons streaming through.  Thus, the vessel is isolated from the permanent tokamak components, so it can be designed to fail without damaging lifetime reactor components in the worst case of a full, unmitigated plasma disruption.

Making the TF coils demountable has a direct impact on the design of the breeding blanket. In order to permit modular maintenance, the blanket is composed entirely of liquid FLiBe that acts as a neutron moderator, shield, and breeder. The FLiBe is contained in a large low-pressure tank, referred to as the blanket tank, and flows slowly past the vacuum vessel. The blanket tank is a robust lifetime component and serves as the primary nuclear containment boundary, as opposed to the vacuum vessel. Neutrons created by the deuterium-tritium fusion reaction are captured in the FLiBe, transferring their energy and breeding tritium to fuel the reactor. Tritium can then be extracted from the liquid FLiBe after it flows out of the blanket tank.

A neutron shield made of titanium dihydride (TiH\subscript{2}) surrounds the blanket tank. This is to protect the inboard leg of the superconducting TF coil, which is particularly space constrained and susceptible to neutron radiation damage. Effective neutron shielding and survivable TF superconducting material is crucial to enable small reactor designs. A detailed MCNP neutronics analysis of the reactor  (see Section \ref{sec:TF_Shielding}) shows that the blanket/neutron shield combination reduces the neutron flux to the TF coil by a factor of $9 \times 10^{-5}$. This ensures at least 9 full-power years (FPY) of operation based on the TF fluence limits currently available.

We estimated the thermal conversion efficiency of the fusion power core (FPC) with a simple, non-ideal Brayton cycle and used this to approximate $Q_e$. It should be noted that the $Q_e$ of the entire power plant will be lower than this estimate, but requires a full site design, which is beyond the scope of this paper. The analysis assumed component efficiencies for the compressor and turbine of 95$\%$ (the expected state of the art of next-generation large-scale power turbomachinery components \cite{mattingly2005elements}) to obtain the cycle thermal efficiencies. Three cases were considered: an FNSF phase, a conservative Pilot phase, and an ``aggressive'' Pilot phase.  In the FNSF phase, the blanket outlet temperature is set at ~900 K, based on a maximum FLiBe flow rate of 0.2 m/s (see  Section \ref{subsec:Blanket_Thermal_Analysis}). This temperature is considered conservative with respect to material limits, but reduces the Brayton cycle efficiency to $\sim 40\%$, resulting in $P_{net}$ = 190 MW and $Q_{e}$ = 3. The next two phases are more speculative, and would require an evolution to higher temperature materials informed by the FNSF stage. The purpose of these phases is to illustrate that first wall/vacuum vessel research during the FNSF stage is crucial to allow a higher blanket temperature, which would greatly increase the total plant efficiency. In the conservative Pilot phase, the blanket outlet temperature is set at 1100 K, for a Brayton cycle efficiency of $\sim 46\%$, resulting in $P_{net}$ = 233 MW and $Q_e$ = 3.5. Finally, in the aggressive Pilot phase, the blanket outlet temperature is set at 1200 K, for a Brayton cycle efficiency of $\sim 50\%$, resulting in $P_{net}$ = 261 MW and $Q_e$ = 3.8. It is noted that decreasing the FLiBe flow rate allows access to higher exit temperatures and may reduce corrosion. Recent molten salt fission studies \cite{clarno2007trade} indicate that many material candidates in molten salts up to temperatures of $\sim$ 1120 K are possible but require further testing in reactor environments, particularly at higher temperatures. Because the FNSF phase of ARC is the obvious first iteration of the design, all further material analysis in this paper (see Section \ref{sec:PowerCore}) is done assuming a blanket outlet temperature of 900 K.

\section{Core plasma physics}
\label{sec:PhysicsBasis}

\subsection{0-D point design optimization}
\label{sec:0D_Design}

In order to determine a starting point for the ARC parameters, a 0-D design exercise was performed. After the initial parameters in this section were determined, the design was iterated several times using codes such as ACCOME, MCNP, and COMSOL.  Note that in many cases the final design parameters (e.g. in Table \ref{tab:DesignParameters} and in the sections following this one) differ from the initial parameters calculated in this section.
 
A fundamental equation for any magnetic fusion reactor design is the scaling \cite{WessonTokamaks2004pg11}
\begin{align}
   \frac{P_{f}}{V_P} \propto 8 \left\langle p \right\rangle^{2} \propto \beta_{T}^2 B_{0}^4 , \label{eq:p_fusion_den}
\end{align}
\noindent
for volumetric fusion power density $P_{f}/V_P$, where $\left\langle p \right\rangle$ is the volume-averaged plasma pressure in MPa. This equation provides two strategies to achieve the high fusion power density desirable for an FNSF and required for economical fusion power: high $\beta_{T} \equiv \frac{2 \mu_{0} \langle p \rangle}{B_{0}^2}$ or large $B_0$. However these two strategies are dramatically different. Increasing $\beta_T$ up to or past its intrinsic limit comes at the risk of exciting MHD modes \cite{TroyonMHDLimit1984} and increasing the frequency of disruptions in devices that have almost no tolerance to disruption damage (see Section \ref{sec:PowerCore_disruption}). Instead, the ARC reactor exploits the quartic dependence on the magnetic field in Eq. \refEq{eq:p_fusion_den} through the use of REBCO superconducting tapes, which provide access to approximately double the magnetic field magnitude of conventional niobium-based superconductors (see Section \ref{sec:Mag_SuperconductorChoice}). 

As with any tokamak, four principal stability considerations restrict the ARC reactor design parameter space: the external kink, Greenwald density, Troyon beta, and elongation (vertical stability) limits. The simplified rules given here were used to provide a 0-D scoping of the operating space available at high B and small size. Operating at the Troyon beta limit \cite{TroyonMHDLimit1984}, given by
\begin{align}
   \beta_{N} \equiv \frac{a  B_{0} }{I_{p}} \beta_{T} = 3 , \label{eq:troyon_limit}
\end{align}
\noindent
where $I_p$ is in MA and $\beta_{T}$ is in percent, allows for a safety margin to pressure-driven instabilities and disruptions. Second, the edge safety factor was constrained to be well above the disruptive kink limit \cite{KinkLimit2004}, which is approximated by
\begin{align}
   q_{a} \equiv \frac{5 \epsilon a B_{0}}{I_p} S \gtrsim 2.2 , \label{eq:kink_limit}
\end{align}
\noindent
where
\begin{align}
	S = \frac{1 + \kappa^2}{2} \label{eq:eliptical_shaping_factor}
\end{align}
\noindent
is the leading order shaping term \cite{jardin2000physics}. The minimum value of 2.2 includes a safety margin of $10\%$ above the hard disruption limit of $q_{a} = 2$ as violating this limit would result in large, damaging disruptions. Third, operating at a $10\%$ safety margin below the disruptive Greenwald density limit \cite{GreenwaldDensityLimit1988},
\begin{align}
   n_{20} = 0.9 \frac{I_{p}}{\pi a^{2}} , \label{eq:density_limit}
\end{align}
\noindent
is enforced to allow for unexpected excursions of the plasma density during steady state operation. Lastly, operation at the empirical elongation limit for vertical stability of
\begin{align}
   \kappa \approx 5.4 \epsilon . \label{eq:elongation_limit}
\end{align}
\noindent
The elongation is chosen in order to gain the benefits of shaping (e.g. higher current at lower safety factor) without the use of complex position stabilization schemes, which fits the ARC design philosophy of minimizing operational limits. This empirical elongation limit is based on the standard elongations of existing devices with divertors and is valid over the range of $ 0.2 \leq \epsilon \leq 0.55$. A more rigorous treatment, given in Ref. \cite{StambaughMaxElongation1992}, provides justification that the chosen elongation is conservative.

The desire to increase elongation as much as possible is well understood by combining Eqs. \refEq{eq:p_fusion_den} through \refEq{eq:eliptical_shaping_factor} to yield
\begin{align}
	\frac{P_{f}}{V_P} \propto \left ( \frac{\beta_N \epsilon \left( 1 + \kappa^{2} \right)}{4 q_{a}} \right )^{2} B_{0}^{4} . \label{eq:pf}
\end{align}
\noindent
We see the fusion power density can also be optimized through the choice of geometry, in particular the aspect ratio. Although, we will see that the on-coil magnetic field is constrained rather than the on-axis field, which introduces a factor of $\left( 1 - \epsilon \right)^{4}$ into Eq. \refEq{eq:pf}. 

In addition to the four plasma physics constraints stated above, it is necessary to approximate nuclear engineering limitations, structural limitations, and current drive accessibility conditions to further constrain the design parameter space. Firstly, the limit on the minimum possible inboard blanket thickness is estimated to be
\begin{align}
   \Delta_{b} \approx 0.5 \text{m} , \label{eq:inboard_blanket_limit}
\end{align}
\noindent
in order to sufficiently moderate and absorb fusion neutrons. This distance becomes critical on the inboard side in compact tokamaks because it constrains the achievable on-axis magnetic field. It should be noted that Eq. \refEq{eq:inboard_blanket_limit} is an intentionally simple estimate used only to constrain the 0-D design. This limit is investigated in detail in Section \ref{sec:TF_Shielding}. Secondly, increasing the on-coil magnetic field to obtain high fusion power densities increases the mechanical stresses on the TF coils. As a result of space constraints and the $1/R$ dependence of the toroidal field, the mechanical constraints are most severe on the inboard leg. Therefore, the on-axis toroidal magnetic field was limited to be
\begin{align}
B_{0} = B_{coil,max}(1 - \epsilon - \frac{\Delta_b}{R_{0}}) , \label{eq:max_mag_stress}
\end{align}
\noindent
where $B_{coil,max} \approx 18$ T is the estimated maximum allowable on-coil toroidal magnetic field estimated from simplistic mechanical stress limits \cite{thome1982mhd}. It should again be noted that Eq. \refEq{eq:max_mag_stress} is an intentionally simplified estimate used for the purpose of assessing the 0-D design. A full stress analysis, dependent on the particular support structure is performed in Section \ref{sec:coilStress}. For REBCO superconductors operating far below their critical temperatures, the toroidal field is generally limited by mechanical stress rather than the critical current density, which typically limits standard Nb\subscript{3}Sn. 

Finally a requirement for achieving a non-inductive scenario was added to the 0-D scoping by considering current drive (CD) and bootstrap current. For reactor-relevant RF current drive schemes in the lower hybrid and electron cyclotron range of frequencies it is highly desirable, or required, to keep the plasma under-dense such that
\begin{align}
\frac{f_{pe}}{f_{ce}} = \frac{89.9 \sqrt{n_{20}} }{28 B_{0}} < 1
\label{eq:under_dense}
\end{align}
\noindent
is satisfied \cite{LHCD_Vulcan}, where $f_{pe}$ is the electron plasma frequency and $f_{ce}$ is the electron cyclotron frequency. It should be noted that Eq. \refEq{eq:under_dense} is evaluated on axis for simplicity. In addition, since the ECRH wave access condition is more limiting than that of LH, it is was chosen to constrain the 0-D analysis. This allowed for flexibility in the choice of the current drive method, even though LH was ultimately chosen.

If the plasma is under-dense then the accessibility condition for LHCD determines the minimum allowable parallel index of refraction, $n_\parallel$ (see Section \ref{sec:CD_Physics} and Eq. \refEq{eq:waveAccessLimit} evaluated at $R = R_{0}$). This determines the maximum CD efficiency, using the analytic estimate \cite{fisch1980creating}
\begin{align}
   \eta_{20,CD} = \frac{31}{ln\Lambda}\frac{4}{(5+Z_{eff})}\frac{1}{n_\parallel^2} ,
\label{eq:CD_efficiency}
\end{align}
\noindent
where $\ln \Lambda \sim 17$ and $Z_{eff} \sim 1.2$ are assumed and $\eta_{20,CD}$ is in units of $10^{20} A/W/m^2$. A standard empirical characterization of current drive efficiency is given as 
\begin{align}
   \eta_{20,CD} = \frac{I_{CD} n_{20} R_{0}}{P_{ext}} , \label{eq:etaDef}
\end{align}
where $I_{CD}$ is the externally driven current (in MA), $n_{20}$ is the density (in units of $10^{20}$ m\superscript{-3}), and $P_{ext}$ is the external heating power (in MW). We can use this to calculate the current drive fraction, $f_{CD} \equiv I_{CD} / I_{p}$, from $P_{ext}$ (which is assumed to be entirely LHCD). The bootstrap fraction, $f_{bs}$, can be estimated from a global formula \cite{NajmabadiARIES_AT2006, PomphreyBootstrapCurEst1992, kessel1994bootstrap} as
\begin{align}
f_{bs}=0.04 \frac{\beta_N q_a}{\sqrt{\epsilon}} . \label{eq:Kessel_fbs}
\end{align}
\noindent
Since ARC must be steady state, we know that there is no inductive current, so
\begin{align}
f_{CD} + f_{bs} = 1 . \label{eq:noninductSols}
\end{align}
This gives the external heating power as a function of the plasma current. Plugging it into
\begin{align}
   Q_{p} & \equiv \frac{P_{f}}{P_{ext}} \geq 25 ,
\end{align}
we arrive our final condition which sets a lower limit on the desired plasma gain, $Q_p$.

\begin{figure}[htbf]
   \begin{center}
   \includegraphics[width=0.5\textwidth]{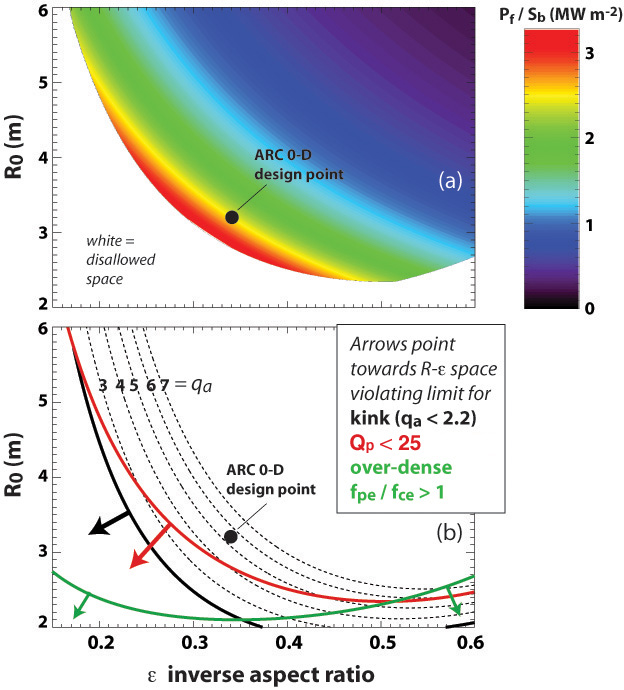}
   \caption{The (a) reactor design $R_{0}$-$\epsilon$ parameter space and (b) the most limiting constraints for $P_{f} = 500$ MW with contours of blanket power loading (in MW/m\superscript{2}). Colored regions in (a) indicate allowable $R_{0}$-$\epsilon$ combinations and the point indicates the chosen initial design point assuming a maximum on-coil magnetic field of 18 T.}
   \label{fig:arc_solution_space_bpd}
\end{center}
\end{figure}

These 0-D constraints were imposed to scope $R_{0} - \epsilon$ parameter space, as shown in Fig. \ref{fig:arc_solution_space_bpd}. As can be seen from the governing limit equations (see Eqs. \refEq{eq:p_fusion_den} through \refEq{eq:noninductSols}), $R_{0}$ and $\epsilon$ play critical roles in setting the physics limits and directly determine the size of the device. The use of demountable coil magnets (detailed in Section \ref{sec:Magnets}) strongly motivated this 0-D study and optimization. With demountable magnets, sector maintenance requirements (e.g. ensuring that components can fit between the TF coils) no longer constrain the aspect ratio. This allowed the design to forgo the standard AT aspect ratio of 4 (e.g. Refs. \cite{NajmabadiARIES_AT2006} and \cite{menard2011prospects}). Also, a very low aspect ratio ($\epsilon \leq 1.5$) was prohibited since it does not allow room for inboard shielding of the superconducting tapes for a Pilot power plant. The purpose of the 0-D study was then two-fold: identify a minimum size for ARC to meet its FNSF/Pilot fusion mission and determine a reasonable choice for the aspect ratio. 
 
The scoping study used the following fixed parameters: $B_{coil,max}$ = 18 T, $\Delta_b$ = 0.5 m, $f_{RF}$ = 5 GHz, $P_f$ = 500 MW, and $\beta_N$ = 3. The $R_0$-$\epsilon$ space allowed by the above constraints is plotted in Fig. \ref{fig:arc_solution_space_bpd}(a). Fig. \ref{fig:arc_solution_space_bpd}(b) indicates that the boundary of allowed space is mostly set by the $Q_{p}$ limit (although the over-dense limit becomes limiting at $\epsilon \geq 0.5$).  Because total fusion power is fixed, the contours of constant areal power density in Fig. \ref{fig:arc_solution_space_bpd}(a) are also contours of constant blanket area, $S_{b}$, where red denotes the smallest blanket and blue the largest. Blanket area is a good measure of device ``size'', since with a fixed blanket thickness it sets the volume of the blanket. Fig. \ref{fig:arc_solution_space_bpd}(a) shows the design goal of ARC to produce 500 MW at the smallest size is thus met at $\epsilon \sim$ 0.3 and $R_{0} \sim 3$ m (red contour). 

The ARC 0-D design point of $R_0$ = 3.2 m, $\epsilon$ = 0.34 (the point in Fig. \ref{fig:arc_solution_space_bpd}) was chosen because, at $P_{f}/S_{b} = 2.5$ MW/m\superscript{2}, it meets the FNSF/Pilot mission requirement for power density. It should be noted that a wide range of aspect ratios, 0.2 $\leq \epsilon \leq$ 0.5 could satisfy the power density requirement (yellow contour of Fig. \ref{fig:arc_solution_space_bpd}(a)) at fixed blanket size.  The choice of our operating point at $\epsilon$ = 0.34 was determined by locating the $R_0-\epsilon$ point on the $P_{f}/S_{b} = 2.5$ MW/m\superscript{2} contour that was furthest from the operating boundary.  Of course this optimization will change with different assumptions, particularly blanket thickness. Thus there are exciting opportunities to explore the use REBCO demountable magnet technology at various aspect ratios.

The 0-D design point has the following parameters:  $R_0$ = 3.2 m, a = 1.1 m, $B_0$ = 9 T, $P_{f}/S_{b} = 2.42$ MW/m\superscript{2}, $n_{20}$ = 1.6, T = 13 keV, and estimated CD efficiency of $\eta_{CD} = 0.5 \times 10^{20} A W^{-1} m^{-2}$. The plasma current at the design point shown in Fig. \ref{fig:arc_solution_space_bpd}(b) is $I_p \sim$ 6.75 MA and the safety factor is $q_a \sim$ 5.6 or $q_{95} \sim$ 7.6. However these values are problematic for the starting design point because they require the current to be over-driven (meaning the externally driven current must be used to partially cancel the bootstrap current). This occurs because the scoping algorithm only assessed operational limits at the maximum $\beta_N$. The design point was determined by increasing the current until a self-consistent non-inductive fraction of unity was obtained. To solve this problem the plasma current was increased to $I_p \sim$ 8.4 MA, which decreased the safety factor to $q_a \sim$ 4.5 (still well away from the kink limit) and the normalized beta to $\beta_N \sim 2.4$ (further from the Troyon limit since the pressure is fixed because the fusion power is fixed). These values in turn set the bootstrap fraction $f_{bs} \sim$ 0.76 and current drive fraction $f_{CD} \sim$ 0.24, without any need for cancellation. Due to the limited accuracy of achieving this balance the 0-D design point was estimated to have $I_{p} \sim$ 8-8.5 MA and $q_{95} \sim$ 6. The 0-D design point is the starting point for the detailed 1-D plasma profile design and current drive/equilibrium simulations below.

Due to its simplicity and transparency, it is worthwhile to discuss the 0-D results in the context of ``wins'' gained by using the high-field approach of ARC. A natural comparison is ITER which also produces 500 MW of fusion power with a similar shaping ($\epsilon \sim$ 0.33), but with $B_0 \sim$ 5.3 T. As expected from the $B_{0}^{4}$ dependence in fusion power density, the peak on-coil field of $B_{coil,max} \sim 20$ T enabled by REBCO technology allows ARC to achieve a FNSF/Pilot-relevant areal fusion power density ($\sim$3 MW/m\superscript{2}) in a device with roughly a tenth of ITER's volume. Additionally, as a consequence of the high toroidal field, the ARC design point has double the safety factor of ITER, making it more robust against disruptions. Looking at Eq. \refEq{eq:Kessel_fbs}, we see the high safety factor permits a reasonable bootstrap fraction of $\sim$75$\%$, while staying below the no-wall beta limit. Thus, the high toroidal field increases the bootstrap fraction, as well as improves LHCD accessibility and efficiency (see Section \ref{sec:CD_Physics}). This simultaneously provides the non-inductive solutions critical for an FNSF and an attractive $Q_p \geq$ 25, critical to a Pilot power plant. It is worth noting that the above advantages of the ARC over ITER come from a peak field ratio of only $\sim$ 1.5 between the two designs (this ratio becomes closer to $\sim$ 2 with the ARC field from the more detailed analysis in Section \ref{sec:Magnets}). Energy confinement has not been considered within this scoping because it is not a disruptive or operating limit. The effect of confinement is discussed in Section \ref{sec:PlasmaProfiles}.

\subsection{Plasma profiles and characteristics}
\label{sec:PlasmaProfiles}

Due to the high field, compact nature of ARC we have chosen to explore the I-mode \cite{Whyte_2010_NF, Hubbard_2011_POP} regime, which has produced excellent absolute and scaled performance in the high-field, compact tokamak Alcator C-Mod. I-mode is characterized by L-mode-like particle confinement and H-mode-like energy confinement \cite{wagner1982regime}, making it an attractive regime for reactor operations because it may allow for easier control of density and impurities, critical control features for burning plasmas. Another intriguing feature of I-mode is that it features weak degradation of energy confinement time with heating power \cite{Whyte_2010_NF}, a highly desirable feature in a self-heated plasma. A recent study \cite{WalkThesis}  has confirmed a $\tau_e \propto P_{heat}^{-0.27}$ scaling by examining a large database of I-mode plasmas, in comparison to $\tau_{e} \propto P_{heat}^{-0.69}$ for standard H-mode. Critically, because I-mode has L-mode-like particle confinement properties, Edge Localized Modes (ELMs) are not required to control impurity content. ELMs are a relatively violent mechanism that regulates impurities in H-mode discharges \cite{Becoulet_2003_PPCF}, but will be unacceptable in burning plasma devices because they would likely damage plasma-facing components. I-mode has its own high frequency instability, the weakly coherent mode, which is suspected to regulate edge impurity transport \cite{Dominguez_2012_PHD} while the plasma regime is stationary, making it attractive for non-inductive operation. Simultaneously the lack of a density gradient results in stationary regimes that are far from the ELM stability limit \cite{WalkThesis}. Therefore, I-mode has a much lower risk of large transient plasma-material interactions, which improves the wall and divertor lifetimes.

While I-mode has some attractive features for a fusion reactor regime, it must also be realized that there is significantly less information regarding I-mode energy transport scaling, particularly with device size (although efforts are underway). Therefore, the ARC design will simply explore the use of scaled density, temperature, and pressure radial profiles from I-mode on C-Mod, rather than directly relying on global confinement scaling laws for predicting performance. This approach also allows us to evaluate how appropriate I-mode profiles are for non-inductive reactor scenarios, since its weak density gradient will have a strong effect on the bootstrap fraction compared to standard H-mode. The resulting profiles needed to achieve the design point will then be evaluated after the fact with respect to required global scaling laws such as the H89 and H98(y,2) scalings as well as the total plasma gain ($\beta_N H / q^2$). We note that this is a standard procedure for assessing fusion reactor performance (e.g. Ref. \cite{NajmabadiARIES_AT2006}).

\subsubsection{Density and temperature profile scalings}
\label{subsubsec:densityAndTempProfiles}

\begin{figure}[htbf]
   \begin{center}
   \includegraphics[width=0.50\textwidth]{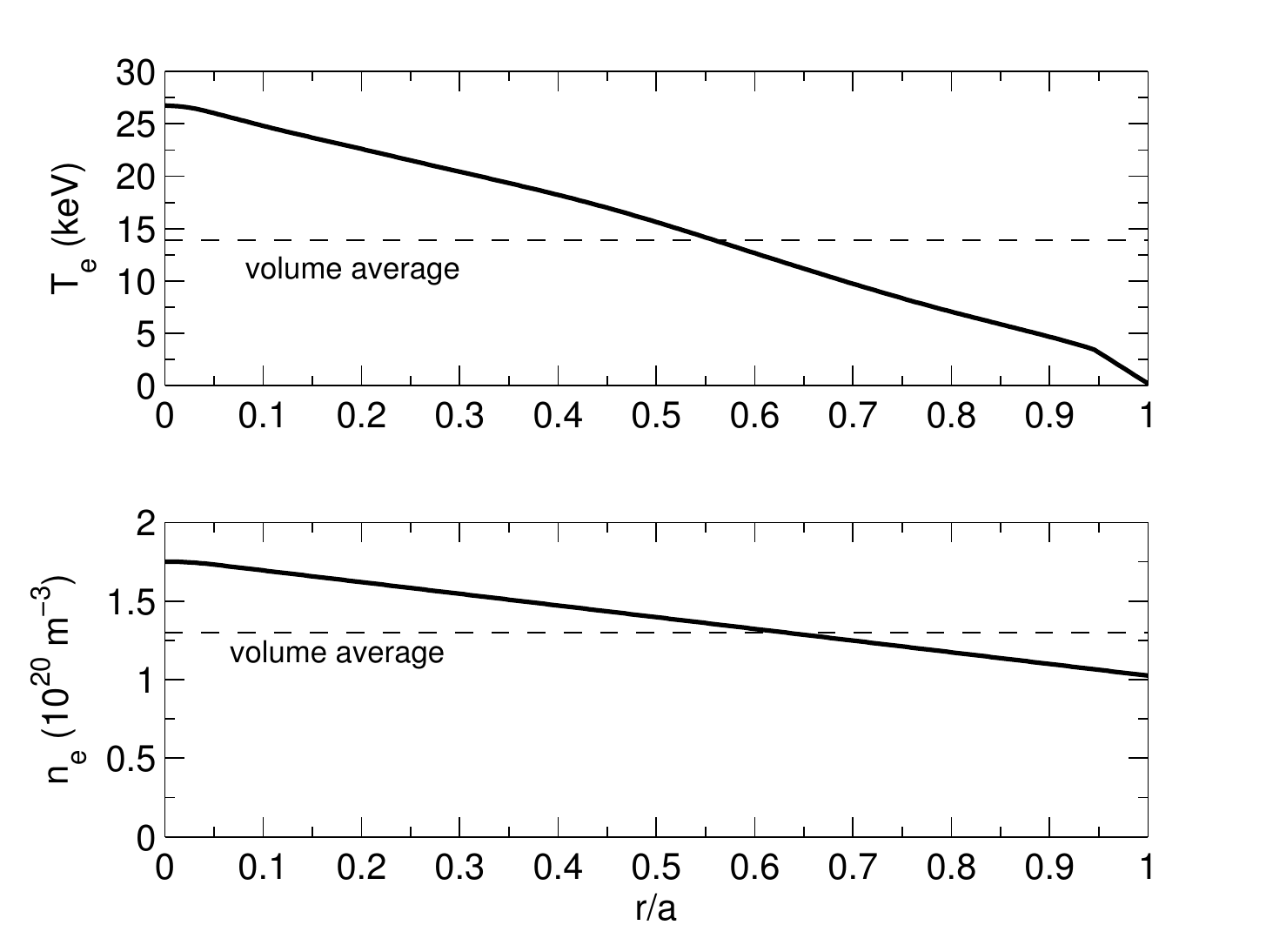}
   \caption{Radial profiles of electron temperature and electron density in ARC.}
   \label{fig:profiles}
   \end{center}
\end{figure}

The density and temperature profiles (see Fig. \ref{fig:profiles}) are generated using experimental scalings from Alcator C-Mod I-mode profiles, with the assumption that $T_{i} = T_{e}$. The density profile is calculated by setting an almost triangular profile achieving $n_{0} / n_{e,average} \sim 1.3$ equal to the average of the C-Mod data \cite{Whyte_2010_NF, Whyte_ITER_IMode_2011} from $0< \rho <1$, where $\rho \equiv r/a$. This omits the density flattening effects of core sawteeth, which is appropriate because ARC has $q>1$ everywhere. As in the 0-D design the line-averaged density is not allowed above $90\%$ of the Greenwald density limit. The gradient is rolled off to zero inside of $\rho = 0.05$. Note that the extension of a constant slope $dn/dr$ to $\rho=1$ is simply consistent with the lack of a particle transport barrier in the edge, a feature that distinguishes I-mode from H-mode.

The electron temperature profiles are constructed inwards, starting at $\rho = 1.0$, where the temperature is fixed to be $200$ eV based on simple parallel heat conduction limits of the two-point model \cite {stangeby2000plasma}. From $0.95< \rho <1.0$, the radial temperature gradient is set according to an experimentally observed C-Mod pedestal scaling at B = 5.4 T and $q_{95} \sim 3$ \cite{Whyte_2010_NF, Whyte_ITER_IMode_2011},
\begin{align}
	\nabla T_{ped} \approx 70 \frac{B_0}{q_{95}}\frac{P_{heat}/S_{p}}{n_{20,ped}},
	\label{eq:gradte_ped}
\end{align}
\noindent
where $n_{20,ped}$ is the pedestal density, $P_{heat} = P_{ext} + P_{\alpha}$ is the total heating power, $S_{p}$ is the plasma surface area, and $\nabla T_{ped}$ is in units of keV/m. The factor $B_0/q_{95}$ accounts for the experimentally measured linear increase of the pedestal gradient with plasma current \cite{WalkThesis}. Note that $B_0/q_{95}$ scales as $I_p$ since ARC and C-Mod I-mode shots have very similar aspect ratio and shaping \cite{Whyte_2010_NF}.

From $0.05 < \rho < 0.95$, a different core temperature gradient scaling is used. This scaling is also based on experimentally measured gradients in C-Mod \cite{Whyte_2010_NF, Whyte_ITER_IMode_2011},
\begin{align}
	\nabla T_{core,ARC}=\nabla{T_{core,CMod}}\frac{\left(\frac{B_0}{q_{95}}\sqrt{P_{heat}/S_{p}}\right)_{ARC}}{\left(\frac{B_0}{q_{95}}\sqrt{P_{heat}/S_{p}}\right)_{CMod}}.
	\label{eq:gradte_core}
\end{align}
As with the density, the temperature gradient is rolled off to zero inside of $\rho = 0.05$. The C-Mod core gradient is $\sim$ 22 keV/m for B = 5.4 T and $q_{95} \sim$ 3. The scaling in Eq. \refEq{eq:gradte_core} reflects the expectation that stored energy scales as $I_p$ (at fixed density), but that the temperature profile scales weaker than linear with heating power due to critical-gradient physics. Combining Eqs. \refEq{eq:gradte_ped} and \refEq{eq:gradte_core} near the ARC design point leads to $W_{th} \propto P_{heat}^{0.7}$, which is again consistent with C-Mod data \cite{Whyte_2010_NF, WalkThesis}.

The temperature gradient scalings depend on $P_{heat}=P_{ext}+P_{\alpha}$ and $P_{\alpha}$ depends on the pressure profile. To determine $T_{e} \left( r \right)$ (and $T_{i}$ since we have assumed that $T_i = T_e$), the profile is self-consistently iterated in the following way. Initially, the total heating power is set to be the externally applied power, $P_{heat}=P_{ext}=25$ MW based on the 0-D scaling. The temperature profile is then built as described above, and the fusion power, $P_{f}$, is computed. The alpha heating power, $P_{\alpha} = P_{f} / 5$, is then added to the external heating power to compute a new $P_{heat}$, which is used to build a new electron temperature profile at a fixed density. The process is repeated until the fusion power converges to within a few percent, indicating that the temperature profile and the heating power are consistent. The value of $q_{95}$ is chosen in these scalings such that the resulting heating power is $\sim 500$ MW. 

The target final density and temperature profiles are shown in Fig. \ref{fig:profiles} and the principle core parameters are listed in Table \ref{tab:DesignParameters}. Slight alterations to the 0-D point (see Section \ref{sec:0D_Design}) were made to accommodate evolving design choices. The inside blanket/shield width was increased to $\Delta_b = 0.85$ m for magnet shielding (see Section \ref{sec:TF_Shielding}). The major radius was increased from $R_0$ = 3.2 m to 3.3 m to help accommodate the larger $\Delta_b$. Simultaneously, the peak field on coil was increased to $B_{max} \sim$ 23 T based on a more detailed examination of the REBCO magnet limits (Section \ref{sec:Mag_SuperconductorChoice}). This resulted in an on-axis field $B_0$ = 9.2 T which was then fixed in the design. The core density was decreased slightly to $n_{20}$ = 1.3 for better CD efficiency.

The temperature and density profiles were built using the rules stated above based on $B_0$ = 9.2 T, the 0-D estimate $q_{95} \sim$ 6, and total heating power $P_{heat}$ = 500 MW/5 + 25 MW = 125 MW. These values lead to a heating power density of $P_{heat}/S_{p} \sim$ 0.63 MW/m\superscript{2}. Equations \refEq{eq:gradte_ped} and \refEq{eq:gradte_core} result in a pedestal temperature of $\sim$ 4 keV and a central temperature of $T_0 \sim$ 26 keV (see Fig. \ref{fig:profiles}). Coincidentally, the values of $B_0/q_{95} \sim 1.6$, $P_{heat}/S_{p}$, $n_{20}$, and plasma shape are very close to the C-mod I-mode shots used for the scaling (which was not by design), and thus the assumed temperature gradients are also very close. Therefore, at fixed gradient, the scaled temperature profiles in ARC are $\sim$ 5 times larger than C-Mod simply due to the 5-fold increase in linear size between ARC and C-Mod (C-Mod I-mode has pedestal $\sim$ 800 keV and $T_0 \sim$ 6 keV \cite{Whyte_2010_NF}).

The temperature and density profiles were required for input into the ACCOME current drive and equilibrium code. Due to the profile and geometry effects from the ACCOME equilibrium solution the design values were slightly increased: fusion power $P_f$ = 500 to 525 MW, external power $P_{ext}$ = 25 to 38 MW (for sufficient current drive, see Section \ref{sec:CD_Physics}) and safety factor $q_{95} \sim$ 6 to 7.2. These equilibrium results increase $P_{heat}/S_{p}$ by $15\%$ and decrease $B_0/q_{95}$ by $15\%$ as compared to the starting assumptions for developing the profiles. Since these effects nearly cancel out and result in $<10\%$ changes in the temperature and density profiles (which is within the uncertainty of the scaling accuracy), no further iterations were performed. The sensitivity of ARC performance to these uncertainties is addressed in Section \ref{subsec:ConfinementSensitivity}. 

\begin{figure}[htbf]
   \begin{center}
   \includegraphics[width=0.50\textwidth]{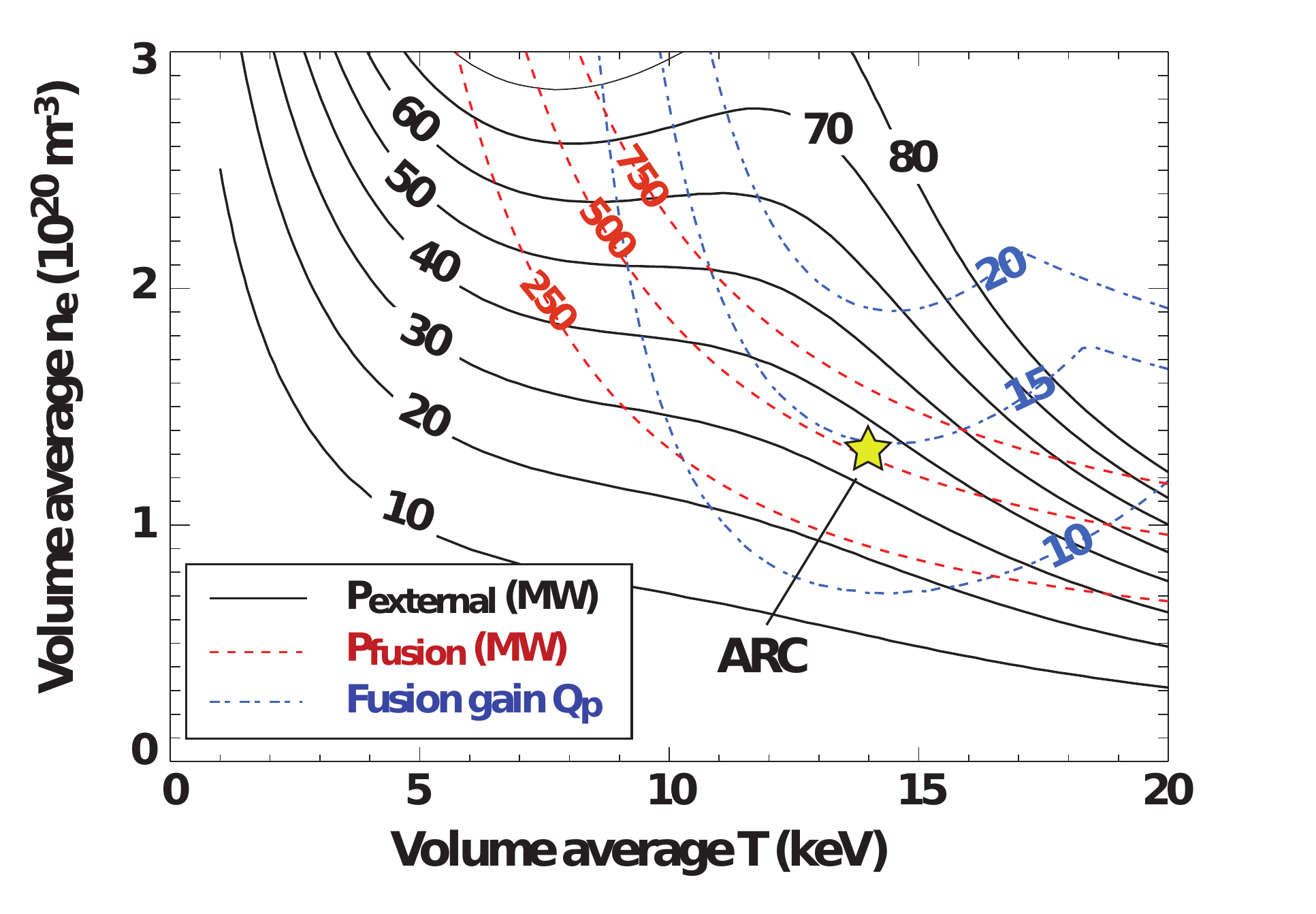}
   \caption{Plasma operating contour plot, where the operating point, indicated by the star, requires an $H_{89}$ factor of 2.78 and is accessible and stable.}
   \label{fig:popcon}
   \end{center}
\end{figure}

The ARC operating point has a volume-averaged temperature $\left\langle T \right\rangle \sim 13.9$ keV and volume-averaged density $\left\langle n_{20} \right\rangle \sim 1.3$. The on-axis temperature is $T_0 \sim 27$ keV and density $n_{20} \sim 1.75$. ARC has a $\beta_{N} = 2.59$, which respects the Troyon limit and a 1-D variant \cite{TaylorOneDBetaLimit1991},
\begin{align}
   \beta_N \equiv \frac{a B_{0}}{I_p} \beta_{T} \leq 4 l_{i} , \label{troyon_limit1D}
\end{align}
\noindent
where $l_{i} \equiv \left\langle B_{p}^{2} \right\rangle/B_{p}^{2} \left( a \right) = 0.67$ is the normalized inductance. The assumed I-mode pressure peaking $p_0/\left\langle p \right\rangle \sim 2.6$ is modest and also aids stability. Additionally, it should be noted that, unlike other aggressive reactor designs, this volume-average density is only $64\%$ of the Greenwald density limit. This indicates ARC can readily explore various densities and associated CD efficiency and divertor heat exhaust solutions around its design point without fear of a density limit disruption.  Additionally, the operating point is accessible with the installed external power and thermally stable, as shown by the plasma operating contour plot given in Fig. \ref{fig:popcon}. Based on the heating power density, $P_{heat}/S_{p}$, and volume-average plasma density at the operating point as well as I-mode experiments performed on C-Mod \cite{Hubbard_IModeTransition} indicate that I-mode should be accessible in ARC. At the minimum threshold of the experiments presented, I-mode may be accessed by initially lowering the plasma density to $n_{20} \sim 1$ and applying the installed heating power of $\sim 40$ MW. The operating point is then reached by increasing the density through fueling (due to L-mode particle transport) and the aid of the alpha heating. In fact, it may even be possible to access I-mode directly at the operating volume-averaged plasma density given the expected installed heating power with conditioned waveguides (see Section \ref{subsec:LHCD_Engineering}). Given its recent discovery, research into I-mode is still required, as discussed in Section \ref{sec:RandDReq}. It is important to note that the use of I-mode in this study is not primarily motivated by core fusion performance, but rather by the absence of ELMs in a stationary regime. Stability analysis of C-Mod I-mode pedestals \cite{walk2014edge} indicates they have considerable margin to the peeling ($\sim$ factor of 2) and ballooning ($\sim$ factor of 3) limits. While a dedicated pedestal stability analysis has not yet been performed for ARC, simple scalings indicate it will also be stable to ELMs. The ARC pedestal features $\beta_{ped} \approx 0.4\%$, an increase of only $60\%$ from the C-Mod $q_{95} \sim$ 3.2 I-mode cases. The most universal metric for stability is the Troyon-normalized pedestal pressure
\begin{align}
\beta_{N,ped} \equiv \beta_{ped}\frac{a B}{I_p} ,
\label{eq:Troyon_normalized_ped_pres}
\end{align} 
\noindent
in units of \%-m-T/MA. For an assumed pedestal width of r/a $\sim 5\%$ (typical of I-mode \cite{walk2014edge}), ITER and FIRE \cite{snyder2004elms} reach stability limits at $\beta_{N,ped}$ = 1.09 and 1.16 respectively, while ARC is only at $\beta_{N,ped} \sim$ 0.5, again indicating stability. While this treatment is overly simple and does not consider the global stability of the pedestal based on the pressure and current profiles, these trends suggest the pedestal in ARC is away from ELM stability limits. A critical open question is the expected pedestal width. 

Despite these uncertainties, because of the absence of ELMs, it is interesting to assess the compatibility of ``I-mode-like'' temperature and density profiles with current drive and bootstrap current. The following sections investigate this compatibility as part of designing a non-inductive scenario at modest $\beta_N$.

\subsection{Current drive physics}
\label{sec:CD_Physics}

The ARC reactor design utilizes a combination of RF power in the ``fast-wave'' ion cyclotron range of frequencies (ICRF) and the lower hybrid range of frequencies (LHRF) to heat the plasma and shape the $q$ profile. ICRF is required to drive current efficiently in the core while lower hybrid current drive (LHCD) provides increased efficiency for driving current near mid-radius and beyond. The goal of this combination of current drive methods is to create an \enquote{advanced tokamak} (AT) q-profile, characterized by weak reverse magnetic shear.  This provides self-consistency to higher confinement and also avoids dangerous instabilities.

LHCD is better than neutral beams or ICRF at driving current at mid-radius because of its high efficiency. The strategy for driving current at mid-radius is guided by the Vulcan study \cite{LHCD_Vulcan}, which found a higher current drive efficiency from launching in regions of high magnetic field and better radial penetration from launching in a region of low poloidal field. This motivates HFS launch in regions of high flux expansion, such as the upper vertex of ARC's triangular plasma cross section. The physical basis for this, as previously described in the Vulcan study, is briefly reviewed here.

A standard empirical characterization of current drive efficiency is given by Eq. \refEq{eq:etaDef}. For the case of LH, the efficiency is determined in part by the phase velocity of the waves parallel to $\vec{B}$ as they damp on electrons \cite{FreidbergPlasmaPhysics}, and follows 
\begin{align}
   \eta_{LHCD} \propto \frac{1}{n_{\parallel}^2} .  
   \label{eq:etaDependency}
\end{align}
Thus, it is advantageous to reduce $n_{\parallel} \equiv c k_{\parallel} / \omega$ as much as possible. The accessibility condition \cite{Porkolab_accessibility} provides the lower bound on $n_{\parallel}$, which limits the maximum achievable efficiency, and is given by
\begin{align}
   n_{\parallel} \geq \frac{\omega_{pe}}{\omega_{ce}} + \sqrt{1+\frac{\omega_{pe}^{2}}{\omega_{ce}^{2}}-\frac{\omega_{pi}^{2}}{\omega_{RF}^{2}}} \propto \frac{\sqrt{n_{e}}}{B} , \label{eq:waveAccessLimit}
\end{align}
where $\omega_{pe}$ is the electron plasma frequency, $\omega_{ce}$ is the electron cyclotron frequency, $\omega_{pi}$ is the ion plasma frequency, and $\omega_{RF}$ is the frequency of the LHRF waves. Thus, from Eq. \refEq{eq:etaDependency}, we find that
\begin{align}
  \eta_{LHCD} \sim \frac{B^2}{n_e} .
\end{align}
This dependence on $B$ as well as the analysis below motivates the HFS launch of lower hybrid waves and the use of LHCD in a high-field tokamak. It should be noted that the choice of density is quite constrained in reactor regimes by the required plasma pressure, so lowering $n_{e}$ to increase efficiency is limited.

The physical motivation for launching near regions of high flux expansion is a direct result of the slow wave branch of the cold, electrostatic lower hybrid dispersion relation \cite{LHCD_Vulcan},
\begin{align}
   \frac{\omega^{2}}{\omega_{LH}^{2}} = 1 + \frac{k_{\parallel}^{2}}{k^{2}} \frac{m_{i}}{m_{e}} , \label{eq:LH_dispersion}
\end{align}
in the limit of $\omega^{2} \ll \Omega_{c e}^{2}$. Differentiating Eq. \refEq{eq:LH_dispersion} with respect to the wavenumber $k$ yields the group velocities in a given direction. Of particular concern is the radial, $v_{g r}$, and poloidal, $v_{g p}$, propagation velocities as these determine how far the wave will penetrate into the plasma before damping. The ratio of these velocities can be shown \cite{LHCD_Vulcan} to be
\begin{align}
   \frac{v_{g r}}{v_{g p}} \approx - \frac{\omega^{2}}{\omega_{pe}^{2}} \frac{n_{r}}{n_{\parallel}} \frac{B}{B_{p}} ,
\end{align}
where $n_{r} \equiv c k_{r}/ \omega$, $k_{r}$ is the radial wavenumber, and $B_{p}$ is the poloidal magnetic field. This equation shows that for more effective radial penetration, lower hybrid systems should tend toward higher launch frequency, lower $n_{\parallel}$, larger $B$, and lower $B_{p}$. Thus near the high-field poloidal null point, where lower $n_{\parallel}$ is accessible and $1/B_{p}$ is maximum, is optimal for the best radial penetration of the LH slow wave rays. However, the resonant Landau damping condition \cite{LandauDampCondition},
\begin{align}
n_{\parallel}^2 \leq \frac{35}{T_{keV}} , \label{eq:LandauDamping}
\end{align}
limits the radial penetration of slow lower hybrid waves. At the magnetic axis of ARC, the maximum $n_{\parallel}$ that is not Landau damped is $1.2$, while the minimum accessible $n_{\parallel}$ on axis is approximately $1.6$. Therefore, slow lower hybrid waves will damp at mid-radius and cannot penetrate to the magnetic axis. Therefore, fast-wave current drive using frequencies near the ion cyclotron resonance have been chosen for on-axis CD. However it should be noted that EC current drive would also be attractive for central current drive if the high frequency sources ($\sim$ 300 GHz) were available to avoid cutoff issues.

The decay wavenumber for ICRF waves increases significantly on axis (implying a significant increase in absorption) because of the dependence on density and temperature given by \cite{Porkolab1994}
\begin{align}
    2 k_{\perp, \text{Im}} = \sqrt{\frac{\pi}{2}} \frac{\omega_{ICRF}}{c} \frac{\omega_{pi}}{\omega_{ci}} \beta_{e} \zeta_{e} \text{exp} \left( - \zeta_{e}^{2} \right), \label{eq:ICRF_Damping}
\end{align}
where $\zeta_e = \omega_{ICRF}/k_{\parallel} v_{te}$ and $\beta_{e} \equiv 2 \mu_{0} n_{e} T_{e} / B^{2}$ is the local electron plasma beta. This shows the absorption at $\zeta_{e} \sim 1$ is proportional to $\omega_{ICRF}$, $n_{e}^{3/2}$, $T_{e}$, and $B^{-3}$. The dependence on the magnetic field also motivates high field side ICRF launch, as more of the wave energy will penetrate to the axis. Not only does this localize the current drive to the core, but it also increases the current drive efficiency because of the dependence of $\eta_{CD}$ on electron temperature \cite{KarneyFisch1985}. In general, the ICRF fast wave will undergo weaker single pass damping in high-field designs such as ARC because of the $B^{-3}$ dependence of the damping.

\subsection{Current drive modeling using ACCOME} 
\label{subsec:ACCOME_results}

Using HFS launch as a starting point, the current drive and plasma performance were modeled using the ACCOME code \cite{ACCOME_LHCD_Details}, a 2-D, self-consistent, free boundary, magnetic equilibrium solver. The code takes coil locations and plasma parameters, including density and temperature profiles as inputs (see Section \ref{subsubsec:densityAndTempProfiles}). It then iterates with current drive modules to find a self-consistent solution to the MHD equilibrium as given by the Grad-Shafranov equation \cite{GradGradShafranovEq1958, ShafranovGradShafranovEq1966}. The code can model various current drive methods, including LHCD and current drive due to bootstrap effects. Currently, there is no module for simulating ICRF current drive. Instead, a fast wave power deposition profile is assumed that has a volumetric power deposition centered on axis and a broad radial distribution based on evaluating Eq. \refEq{eq:ICRF_Damping}. The magnitude of the power deposition on axis is chosen to give an integrated ICRF-driven current totaling $1.1$ MA.

\begin{figure}[ht]
 \centering
 \includegraphics[width=0.45\textwidth]{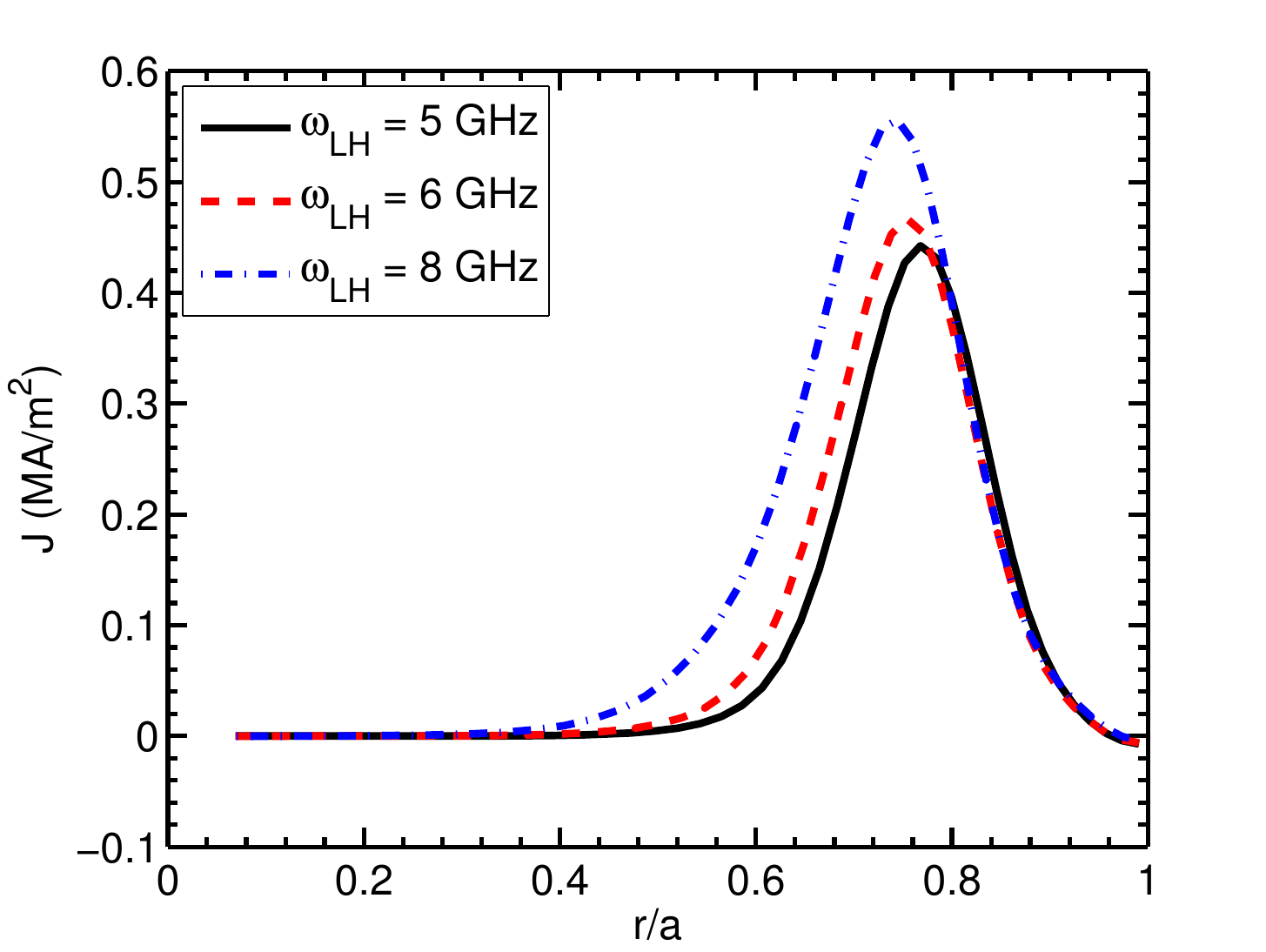}
 \caption{Lower hybrid driven current density for the design source frequency of $8$ GHz and several other frequencies as a function of normalized minor radial location.}
 \label{fig:multi-freq radial LH current profiles}
\end{figure}

For lower hybrid current drive, the source frequency, launcher position, power, and $n_{\parallel}$ are all specified. An ``advanced tokamak'' current profile \cite{NajmabadiARIES_AT2006} is desired, characterized by weak reverse magnetic shear throughout the plasma, so the lower hybrid waves are required to damp primarily at mid-radius to supplement the current drive profile from ICRF and bootstrap current. In the optimization, the lower hybrid source frequency, launched $n_{\parallel}$, and launcher position were all varied. ACCOME results showed that the current drive efficiency is sensitive to the launch frequency and $n_{\parallel}$. A launch frequency of $8$ GHz is chosen to avoid parasitic damping on alpha particles, which occurs when the perpendicular phase velocity of the wave, $v_{\perp}$, matches the alpha birth velocity. For a fixed $n_{\parallel}$, $v_{\perp}$ is proportional to the launch frequency. At $8$ GHz, the entire coupled power of $25$ MW contributes to driving current, while at $5$ GHz as much as $20$\% of the injected power is lost to alpha particles. This shows at higher launch frequencies less power parasitically damps on alpha particles since $v_{\perp}$ is higher than the birth speed of the alphas. Fig. \ref{fig:multi-freq radial LH current profiles} demonstrates that higher frequencies drive more current and penetrate farther radially.

\begin{figure}[ht]
 \centering
 \includegraphics[width=0.45\textwidth]{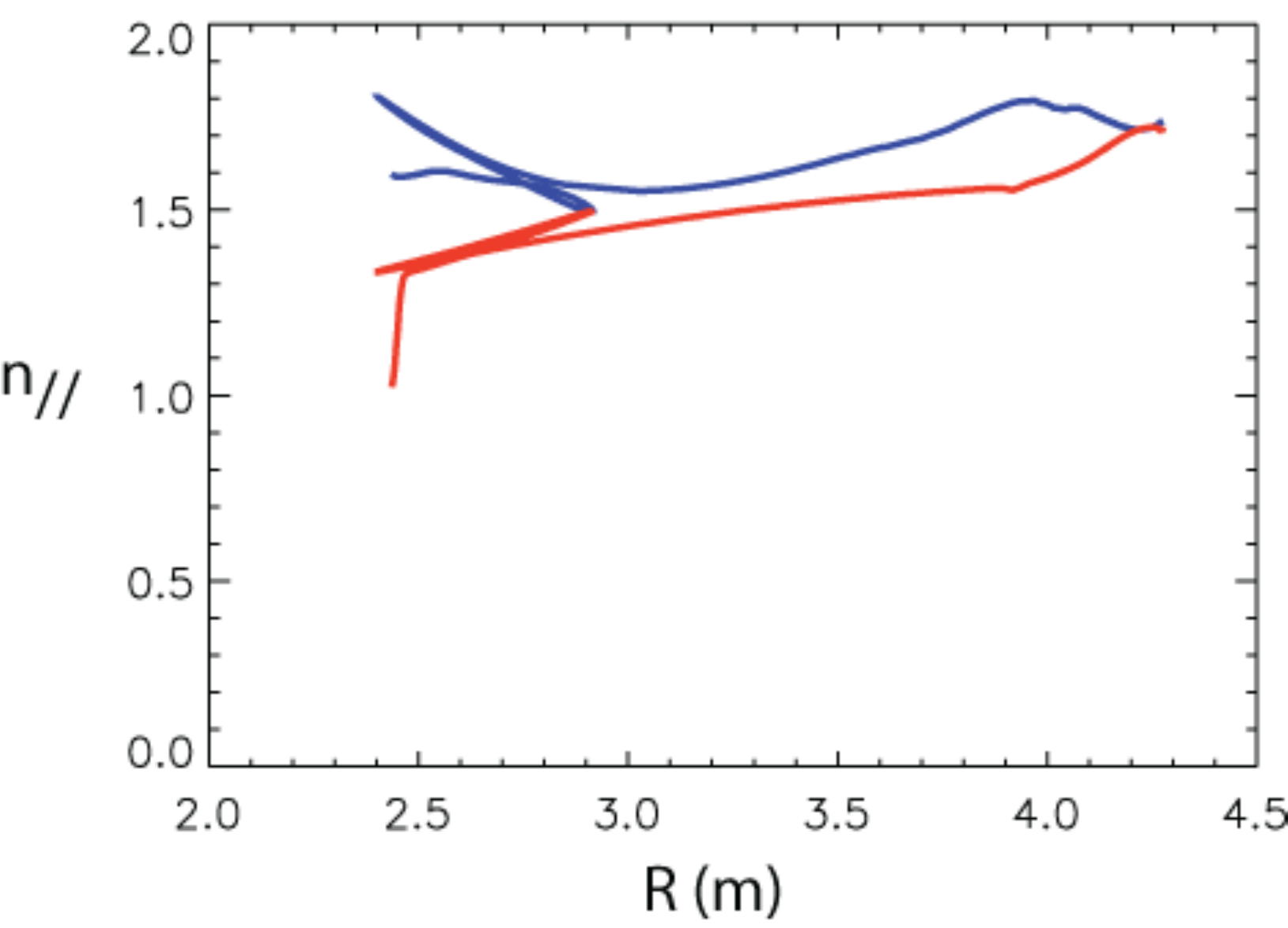}
 \caption{An example of the evolution of the parallel index of refraction when violating the wave accessibility limit (see Eq. \refEq{eq:waveAccessLimit}), where blue represents the wave $n_{\parallel}$ along the trajectory and orange represents the critical value determined by the local accessibility limit.}
 \label{fig:upshift of n||}
\end{figure}

\begin{figure}[ht]
 \centering
 \includegraphics[width=0.45\textwidth]{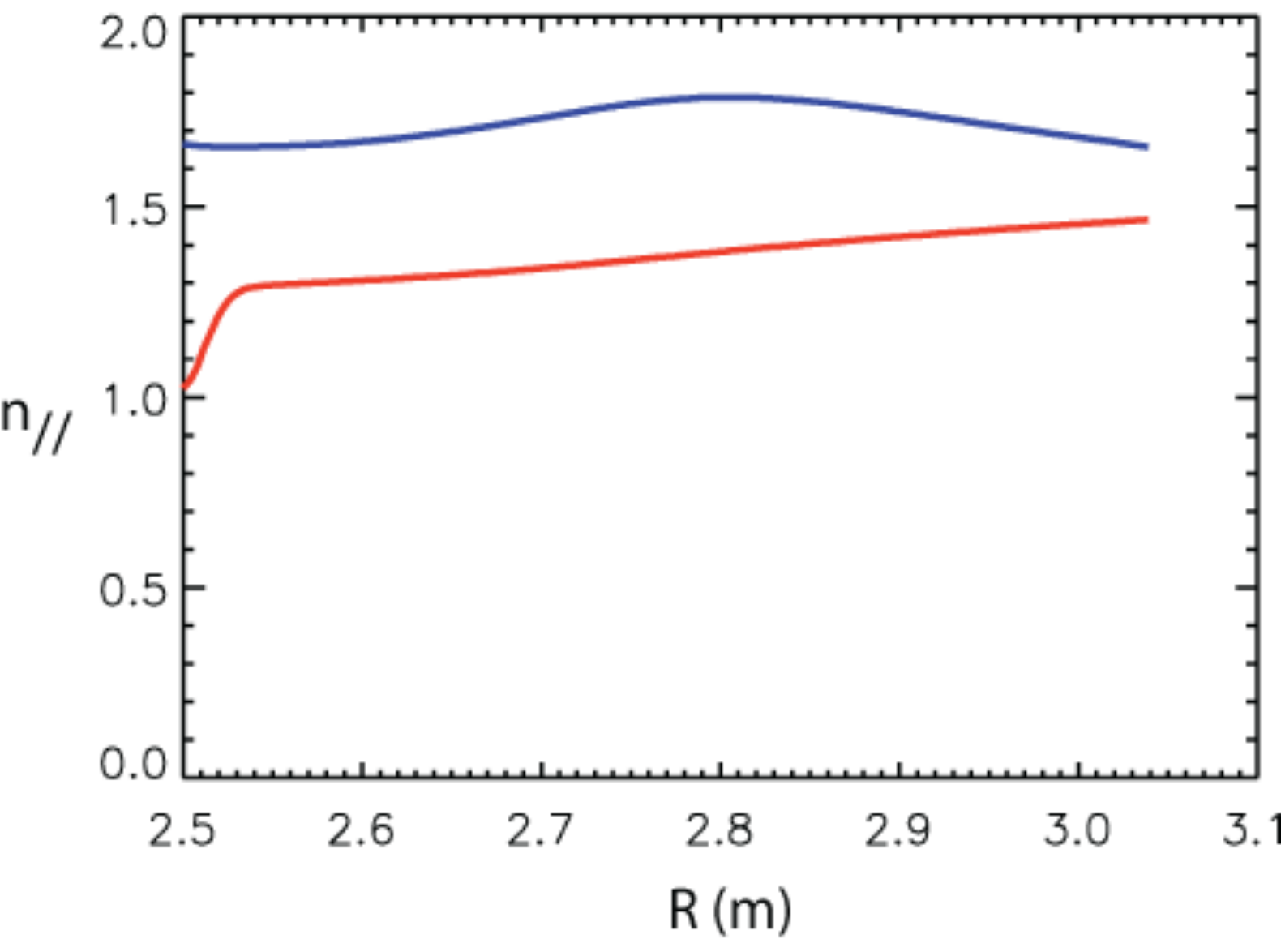}
 \caption{Evolution of the parallel index of refraction with propagation for the launch conditions in ARC, where blue represents the wave value and orange represents the critical value determined by the accessibility limit. Note that this follows the ray until 99$\%$ of its energy is damped.}
 \label{fig:n|| vs R}
\end{figure}

\begin{figure}[ht]
 \centering
 \includegraphics[width=0.45\textwidth]{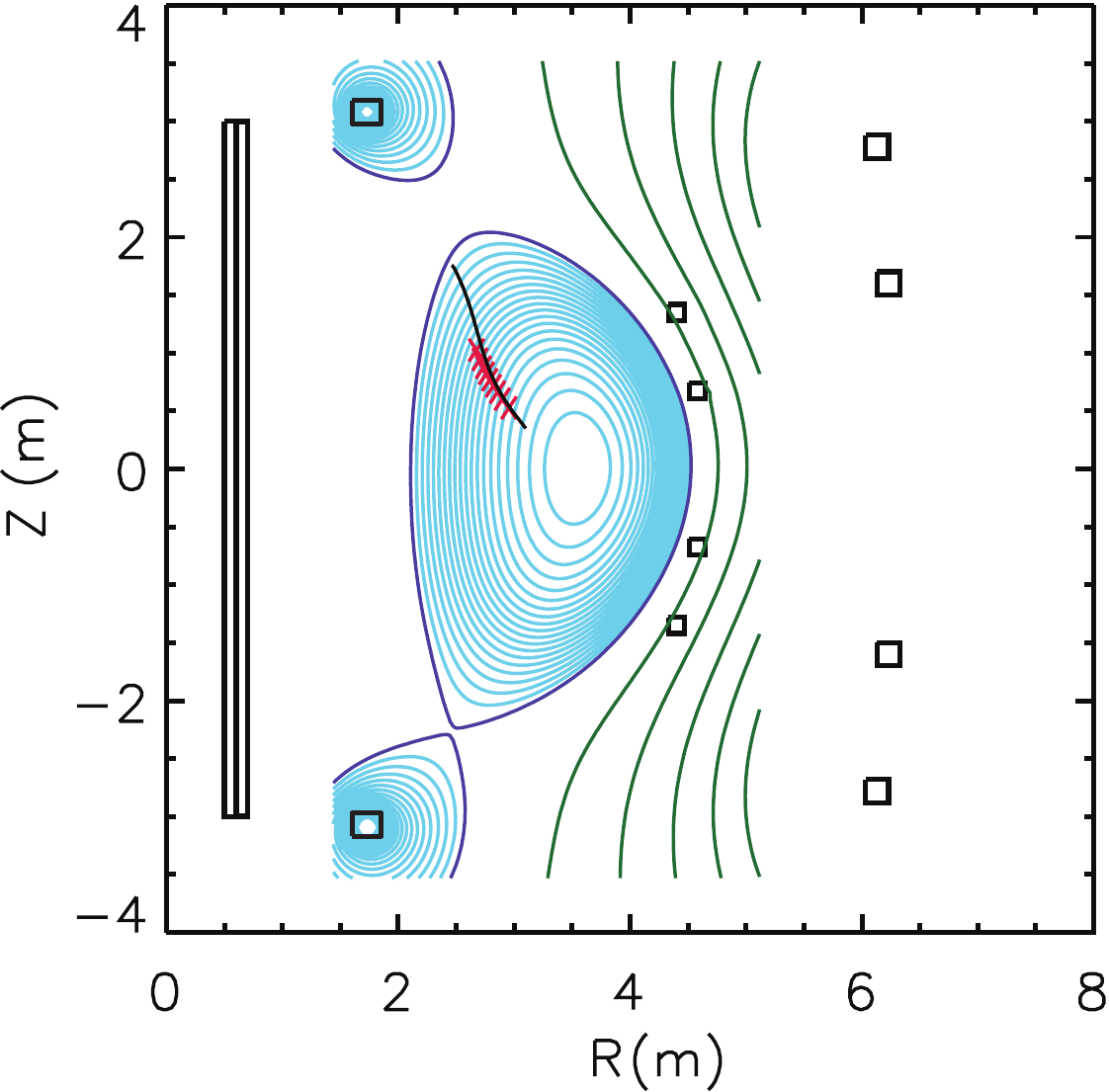}
 \caption{ACCOME plasma equilibrium for ARC with the LHCD wave trajectory indicated in black. Each red tick mark along the ray trajectory indicates a 10$\%$ decrease in the wave power due to electron Landau damping.}
 \label{fig:plasma_equilibrium_with_LH}
\end{figure}

The launched $n_{\parallel}$ was varied between $1.4$ and $1.7$ for the ACCOME calculations. As shown in Eq. \refEq{eq:etaDependency}, LHCD efficiency increases with decreasing $n_{\parallel}$, therefore it is advantageous to minimize $n_{\parallel}$. However, the optimized launched $n_{\parallel}$ is found to be $1.67$ with a small spectral width, $\delta n_{\parallel} = 0.05$. Decreasing the initial $n_{\parallel}$ below this value causes the wave to become inaccessible. In Fig. \ref{fig:upshift of n||} we see the consequences of this. The wave is launched inwards, reflects back to the plasma edge, reflects again, and finally damps. In contrast, Figs. \ref{fig:n|| vs R} and \ref{fig:plasma_equilibrium_with_LH} show the wave trajectory in ARC, which propagates directly towards the magnetic axis and damps at mid-radius. Furthermore, the current drive efficiency depends on the $n_{\parallel}$ where the wave damps, not where it is launched. In ARC, the launched $n_{\parallel}$ of $1.67$ outperforms a lower launched $n_{\parallel}$ as the wave penetrates without dramatic upshifts to a mid-radial location where a combination of poloidal magnetic field and toroidal effects cause a gradual downshift prior to damping (see Figs. \ref{fig:upshift of n||} and \ref{fig:n|| vs R}).

\begin{figure}[ht]
   \centering
   (a) \hspace{0.2\textwidth} (b) \hspace{0.2\textwidth}

   \includegraphics[width=0.2\textwidth]{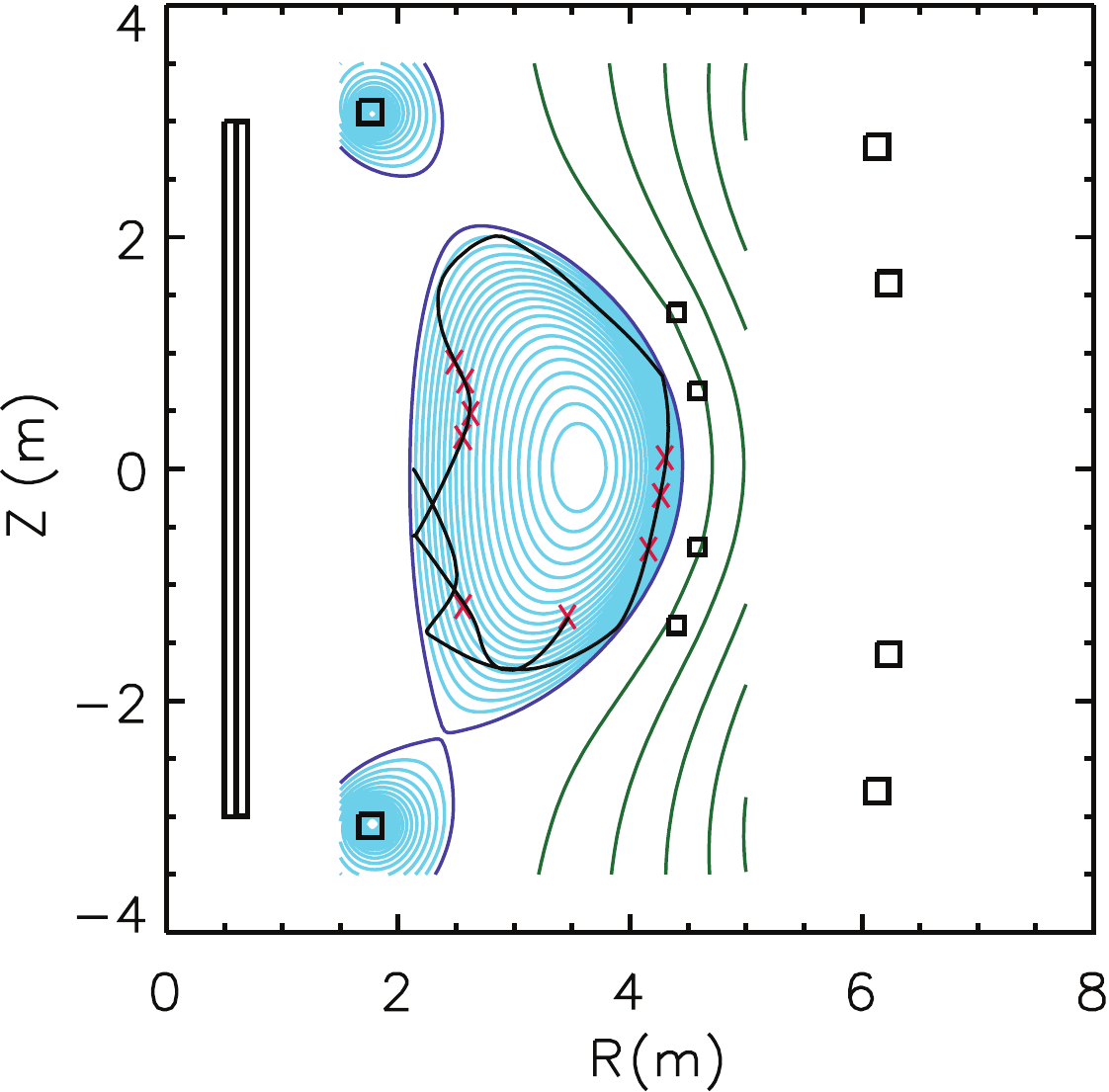}
   \includegraphics[width=0.2\textwidth]{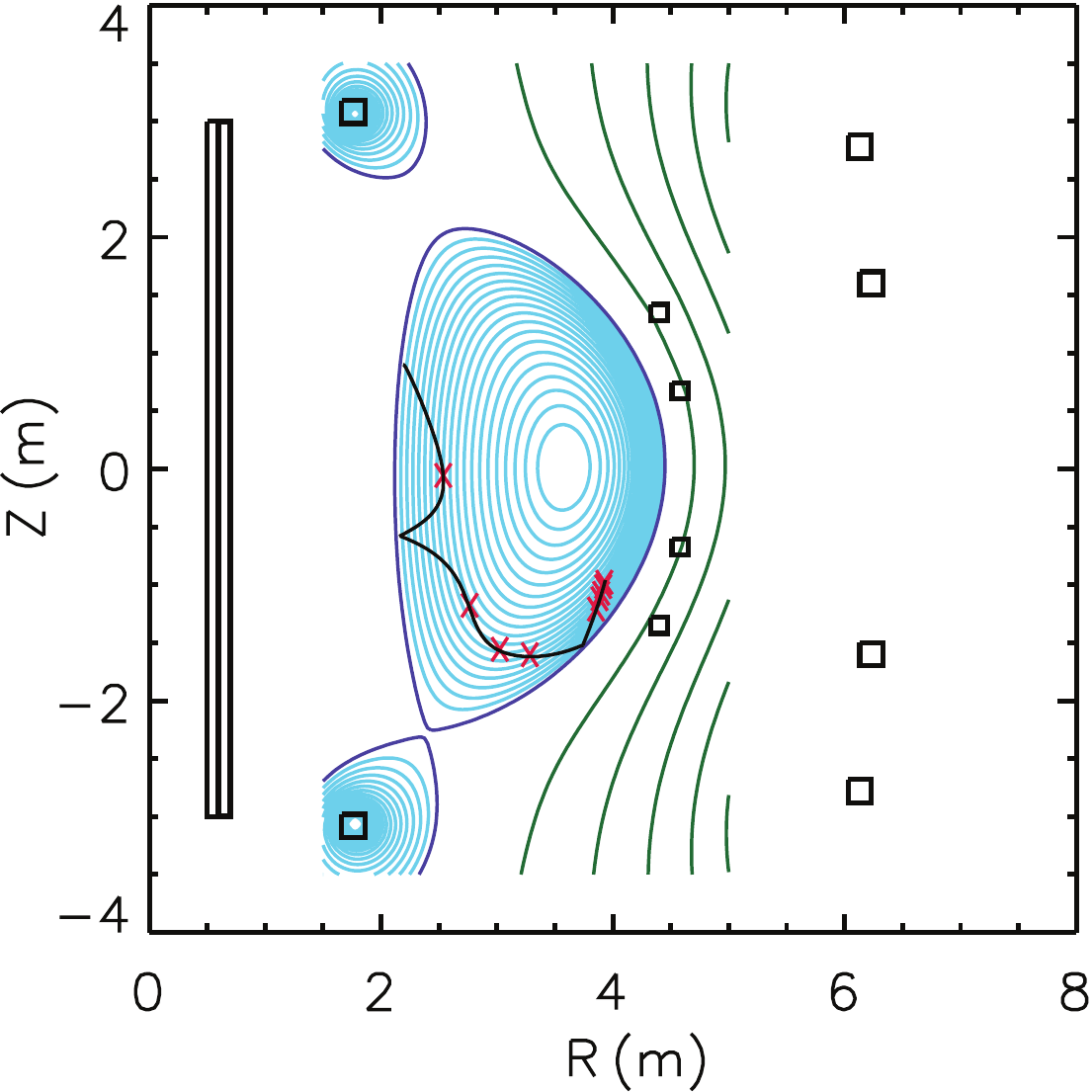}

   (c) \hspace{0.2\textwidth}

   \includegraphics[width=0.2\textwidth]{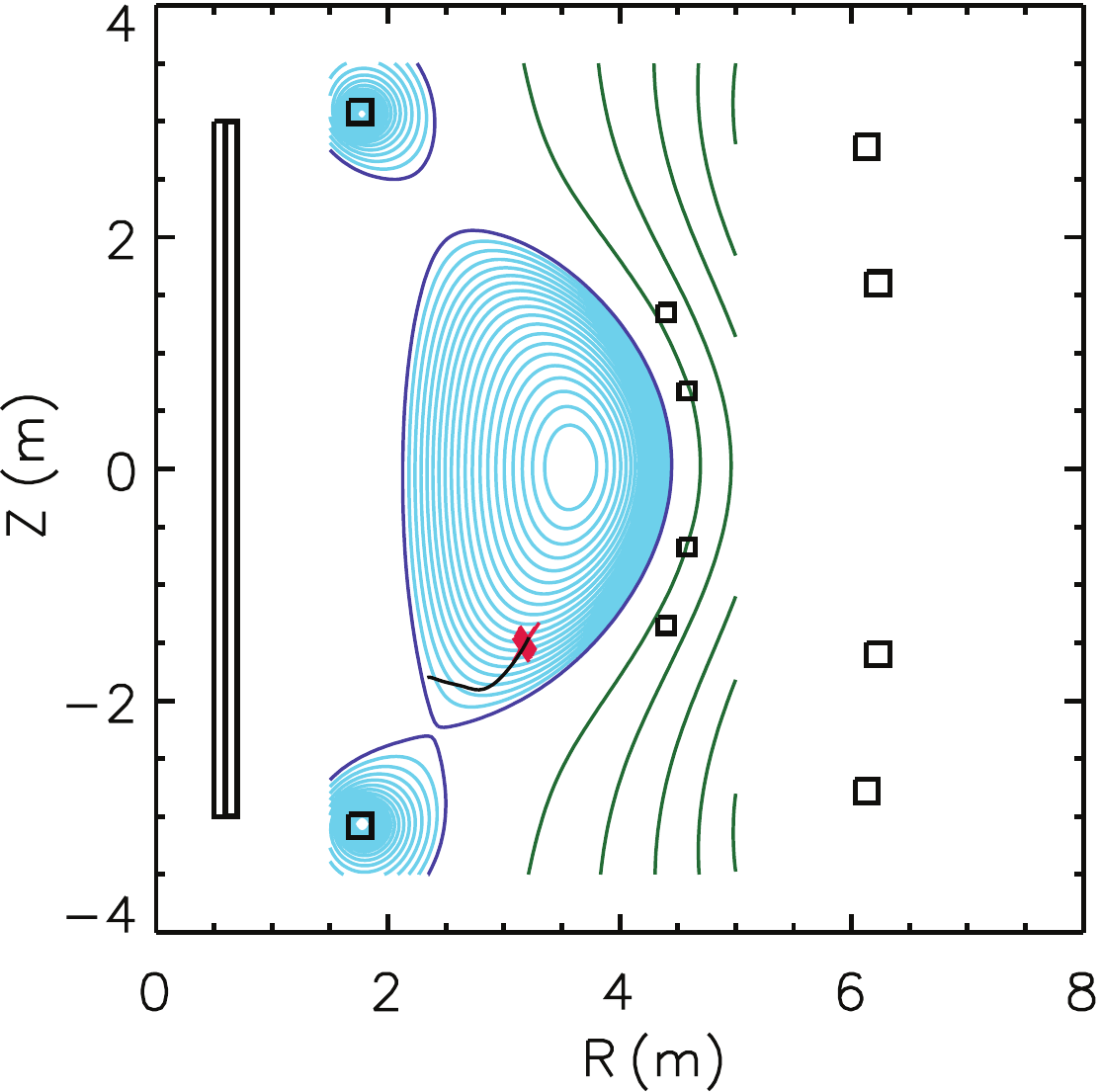}
   \caption{Lower hybrid ray traces for non-optimized launch locations; (a) midplane launch, (b) launch halfway between the midplane and the upper extremity of the plasma, and (c) launch near the lower null. The red X's represent the location of each 10\% reduction in wave power due to damping.}
   \label{fig:multiple ray traces}
\end{figure}

Throughout the ACCOME runs, the launcher position had the most significant effect on the lower hybrid efficiency and ability to drive current at mid-radius. The launcher position determined whether the waves could propagate radially, upshift, convert to fast waves, and/or reflect. Various positions were tested, ranging from the midplane to regions of high flux expansion. Fig. \ref{fig:multiple ray traces} demonstrates the wide variability in ray trajectories resulting from varying only the launcher position. It is noted that ARC overcomes a commonly perceived limitation that LHCD drives most of its current far off-axis (e.g. at r/a $\sim$ 0.9-0.95 in ARIES-AT). Effective current drive at r/a $\sim$ 0.6-0.7 is enabled in ARC by a combination of a) larger magnetic fields improving accessibility, b) employing HFS launch (which further improves accessibility and avoids damping at a low temperature by launching at lower $n_\parallel$), and c) the choice of the poloidal launcher position to optimize the variation in $n_\parallel$ as the wave propagates. 

\begin{figure}[ht]
 \centering
 \includegraphics[width=0.45\textwidth]{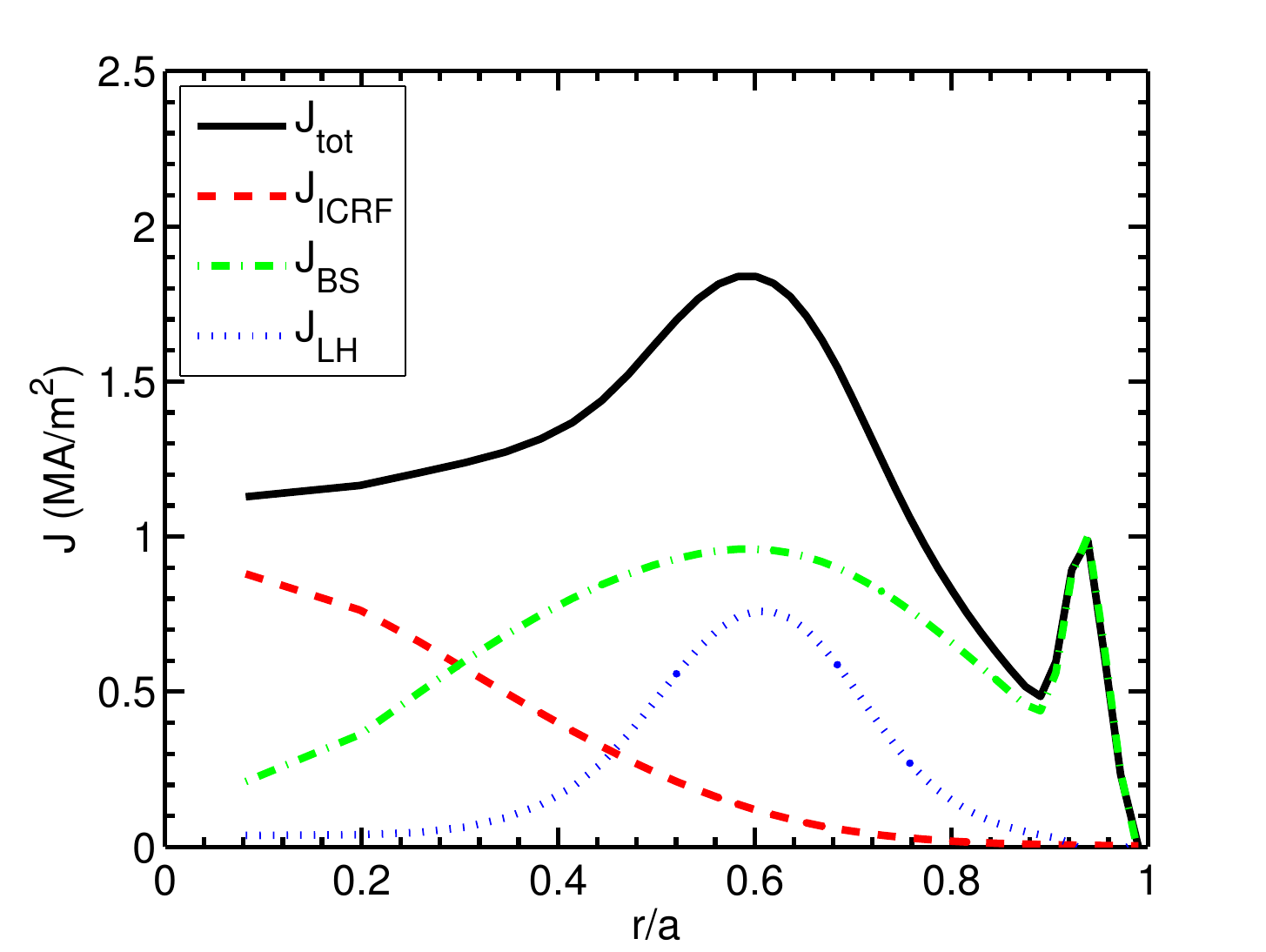}
 \caption{Current profiles in ARC, with the total current (black, solid), the ion cyclotron current drive (red, dashed), the bootstrap current (green, dashed-dotted), and the lower hybrid-driven current (blue, dotted) shown.}
 \label{fig:current_density}
\end{figure}

At an $n_{\parallel}$ of $1.7$, damping will occur on electrons with a temperature of approximately $14$ keV. This can be seen by comparing the peak in the LHCD profile in Fig. \ref{fig:current_density} with the temperature profile in Fig. \ref{fig:profiles}. Note that the minor radius location of the current peak roughly corresponds to the location where $T_{e} = 14$ keV (i.e. $r/a = 0.6$). This indicates that HFS LHCD is well suited to a compact device where, due to confinement concerns, $\left\langle T \right\rangle \sim$ 14 keV is chosen to maximize the Lawson triple product. Since $\left\langle T \right\rangle$ approximately corresponds to the mid-radius T, efficient mid-radius CD naturally follows.

\begin{figure}[ht]
 \centering
 \includegraphics[width=0.45\textwidth]{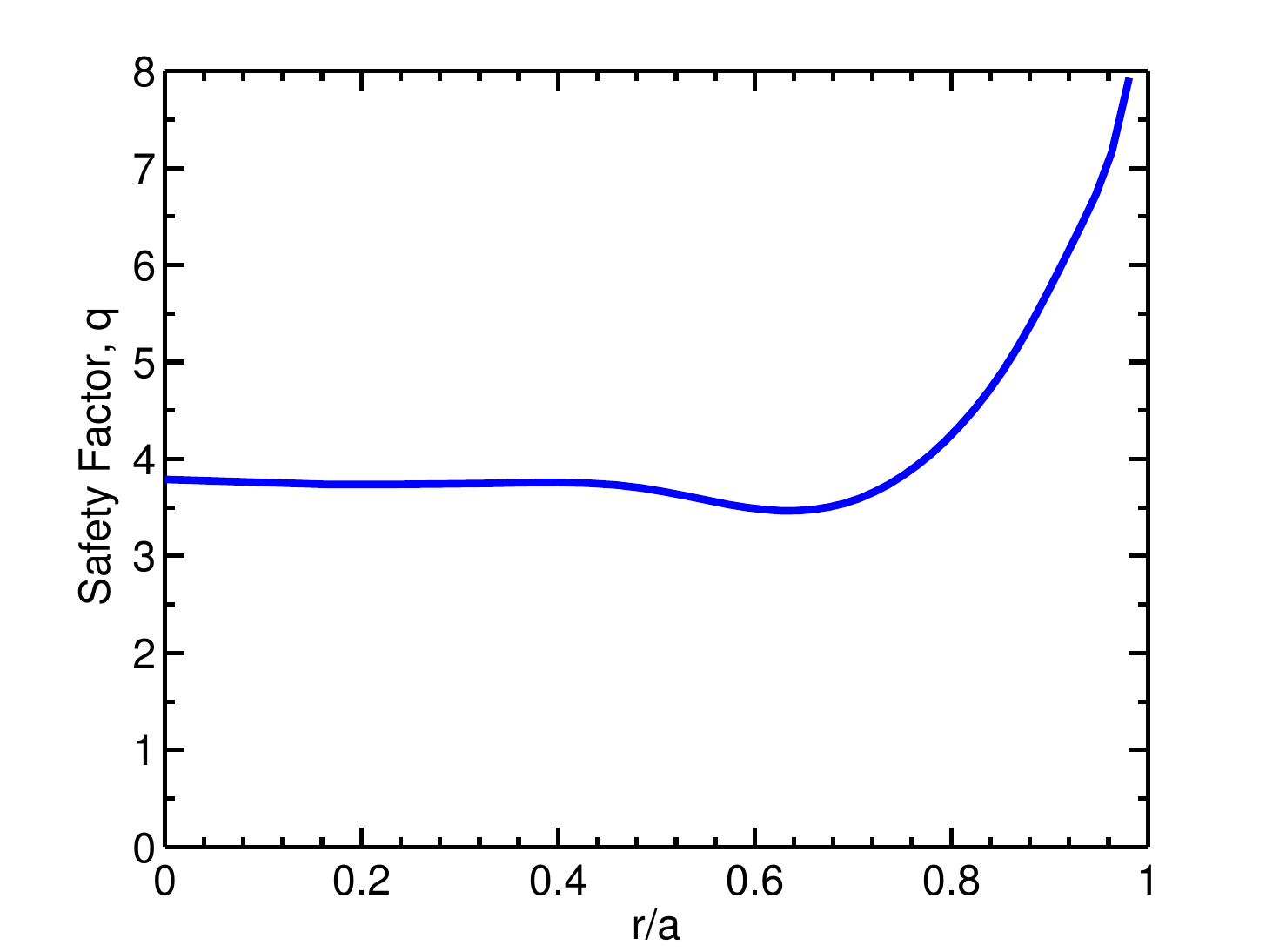}
 \caption{Safety factor profile in ARC.}
 \label{fig:q_profile}
\end{figure}

The safety factor profile calculated by ACCOME is plotted in Fig. \ref{fig:q_profile}, showing an elevated edge safety factor and an on-axis safety factor greater than 3. Thus, ARC should avoid the ballooning kink mode at the edge, the sawtooth instability on axis, and low-order tearing modes (2/1, 3/2) (although an ideal MHD stability analysis would be required to confirm ballooning stability). As previously noted, the profiles operate below the no-wall Troyon limit. At the safety factor profile above, the ratio of the banana orbit width to minor radius is approximately double that of ITER. However, this is not expected to lead to prompt losses of fast-birth alpha particles (with banana orbit width of $\sim$ 0.1m). Furthermore, ARC lacks energetic particles in the edge from beam heating. Therefore it is judged that the high safety factor is justified in order to avoid global disruptions.

In addition to solving the MHD equilibrium equations, ACCOME also calculates the global plasma and current drive performance. The code estimates a fusion power of $525$ MW for the plasma equilibrium obtained. ACCOME calculates that $25$ MW of coupled lower hybrid current drive power will drive $1.77$ MA and the total plasma current will be $7.75$ MA after including 1.1 MA from ICRF. This corresponds to a bootstrap fraction, $f_{BS}$, of approximately 63\% and a lower hybrid efficiency, $\eta_{LHCD}$, of $0.4 \times 10^{20} A W^{-1} m^{-2}$. This efficiency is somewhat below the 0-D estimate of $\sim 0.5 \times 10^{20} A W^{-1} m^{-2}$ because of trapped particle effects (estimated in ACCOME) at r/a $\sim$ 0.6. For a design like ARIES-AT, the loss of efficiency due to trapping would be larger because r/a $\sim$ 0.9-0.95 (although these corrections were not included \cite{JardinARIESATphysicsBasis2006}). Minimizing these deleterious trapping effects is another motivation for high magnetic fields and HFS launch, since edge damping can be avoided.

\begin{figure}[ht]
   \centering
   (a) \hspace{0.2\textwidth} (b) \hspace{0.2\textwidth}

   \includegraphics[width=0.2\textwidth]{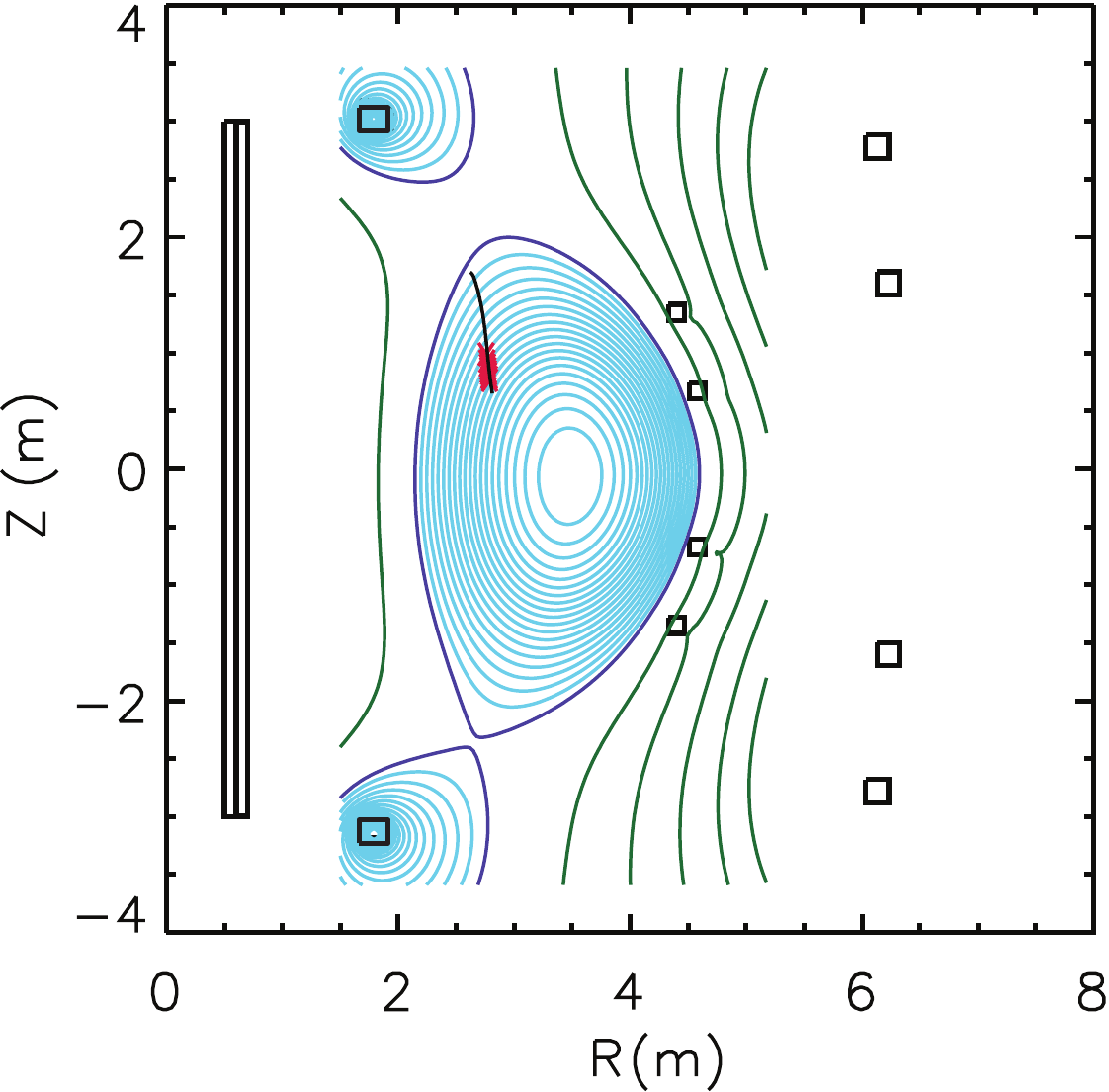}
   \includegraphics[width=0.2\textwidth]{Figures/Core/MHD_Equilibrium_Ray_B_4_3.pdf}
 \caption{The (a) initial and (b) final MHD equilibrium from ACCOME, demonstrating the stability of the wave trajectory to variations in the plasma equilibrium.}
 \label{fig:MHD equilibria}
\end{figure}

The ICRF current drive is assumed to have a similar efficiency to the ideal lower hybrid current drive efficiency of $0.43\times10^{20} A W^{-1} m^{-2}$. This choice is based on the following considerations. The ICRF source frequency was chosen to be 50 MHz (similar to the ITER ICRF system \cite{MessiaenITERicrf2009}) in order to place the wave frequency below any fundamental or second harmonic ion cyclotron resonances. Furthermore, damping of the ICRF wave (given by Eq. \refEq{eq:ICRF_Damping}) maximizes for $\zeta_{e} = \omega/(k_{\parallel}v_{te}) \approx 0.7$. For ARC parameters, this implies that we must have $n_{\parallel} \sim 4.4$ on axis and $n_{\parallel} \sim 3.3$ at the antenna.  Using Fig. 2(a) of Ref. \cite{KarneyFisch1985}, the ICRF current drive efficiency can be estimated to be $0.4 - 0.5\times10^{20} A W^{-1} m^{-2}$ for a narrow spectrum of Landau damped ICRF waves with $T_{e} / ( m_{e} c^{2} ) \approx 0.05$ and $p_{\parallel}/ \left(m_{e} c \right) \sim 1/n_{\parallel} \sim 0.25$. This efficiency leads to a required coupled power, $P_{IC}$, of $13.6$ MW to drive 1.1 MA of ICRF current in ARC. Self-consistency between the MHD equilibrium and the current drive sources was achieved by allowing ACCOME to iterate between the solution of the Grad-Shafranov equation and a re-evaluation of the current drive source terms. The plasma equilibria for several iterations are shown in Fig. \ref{fig:MHD equilibria}, demonstrating that, for the coil configuration and plasma current drive system chosen, the wave trajectory is anticipated to be stable to small changes in the equilibrium. 

At this point, the relatively broad characteristic width assumed for the ICRF current density profile deserves further discussion. It can be seen in Fig. \ref{fig:current_density} that this corresponds to a $\Delta(r/a) \approx 0.4$. The parameters of ARC result in $2 k_{\perp,Im} \Delta R \approx 0.16$ for a single pass. This is calculated using $\Delta R = a / 2 = 0.5$ m and Eq. \refEq{eq:ICRF_Damping}, evaluated on axis with $f_{ICRF} \equiv \omega_{ICRF} / \left( 2 \pi \right) = 50$ MHz, $n_{\parallel} = 4.4$, $\zeta_{e} = 0.7$, and $\beta_{e} = 0.02$. The resulting single pass damping following Ref. \cite{Porkolab1994} is then $1.0 - \text{exp} \left( -2k_{\perp,Im} \Delta R \right) \sim 0.15$. Therefore, the ICRF wave requires several passes through the plasma before the power is absorbed completely. This results in the relatively broad deposition profile with a damping profile that might even be characterized by `eigenmode' features.

\subsection{ARC Sensitivity to Confinement Quality}
\label{subsec:ConfinementSensitivity}

Energy confinement is almost always a constraint on achieving fusion performance, but this is particularly true at small scale where the energy confinement time tends to be smaller due to basic physics considerations. For ARC the $525$ MW design point (see Table \ref{tab:DesignParameters}) minimizes the required energy confinement time,   by operating at the minimum in the triple product Lawson criterion $T \sim 14$ keV. The calculated confinement quality is $H_{89} = 2.8$ for the design point, which is $40\%$ above standard H-mode. However, this is common to reactor designs using AT plasmas \cite{NajmabadiARIES_AT2006,JardinARIESATphysicsBasis2006}. Such high confinement is justified by the theoretical expectation and experimental confirmation that weak shear q profiles and the lack of low-order rational q surfaces ($q_{min} > 2$) lead to enhanced confinement factors. For example, DIII-D achieved $H_{89} \sim 2.5-3$ under such conditions with an internal transport barrier \cite{Garofalo_highBeta_D3D}. Section \ref{subsec:ACCOME_results} demonstrates that the computed q profile in ARC meets this criteria. 

An even better measure of confinement would be the gain factor \cite{LuceGainFactor2004}
\begin{align}
    G \equiv \frac{\beta_{N}}{q_{95}^{2}} H , \label{eqn:gain_factor}
\end{align}
which provides a global assessment of the plasma physics dimensionless parameters required to meet the Lawson criterion according to
\begin{align}
    n T \tau_{E} \sim p_{th}\tau_E \sim \beta_{N} B_{0}^{2} \frac{H B_{0}}{q_{95}} = G B_{0}^{3} . \label{eqn:Lawson_with_Gain}
\end{align}
Substituting the values for ARC (calculated from computational results in Section \ref{subsec:ACCOME_results}) gives an expected gain factor, $G_{89}$, of $0.14$ (or a $G_{98}$ factor of $0.08$) due to the low $\beta_N$ and high $q_{95}$ in ARC. These gain factors have been achieved in non-inductive scenarios in several tokamaks including DIII-D and JT-60 \cite{Garofalo_highBeta_D3D}. In fact, the representative weak shear DIII-D experiment chosen in Ref. \cite{Garofalo_highBeta_D3D} and the stationary states reported in Ref. \cite{ZarnstorffMFE} have nearly identical plasma parameters to the ARC design point. Thus, ARC is unique among recent conceptual tokamak designs (including those cited in Ref. \cite{ZarnstorffMFE}) in operating at previously achieved gain factors. That being said, unlike DIII-D and JT-60, ARC lacks neutral beams to drive plasma rotation, which could have important consequences for confinement.

\begin{figure}[htbfp]
   \begin{center}
   \includegraphics[width=0.50\textwidth]{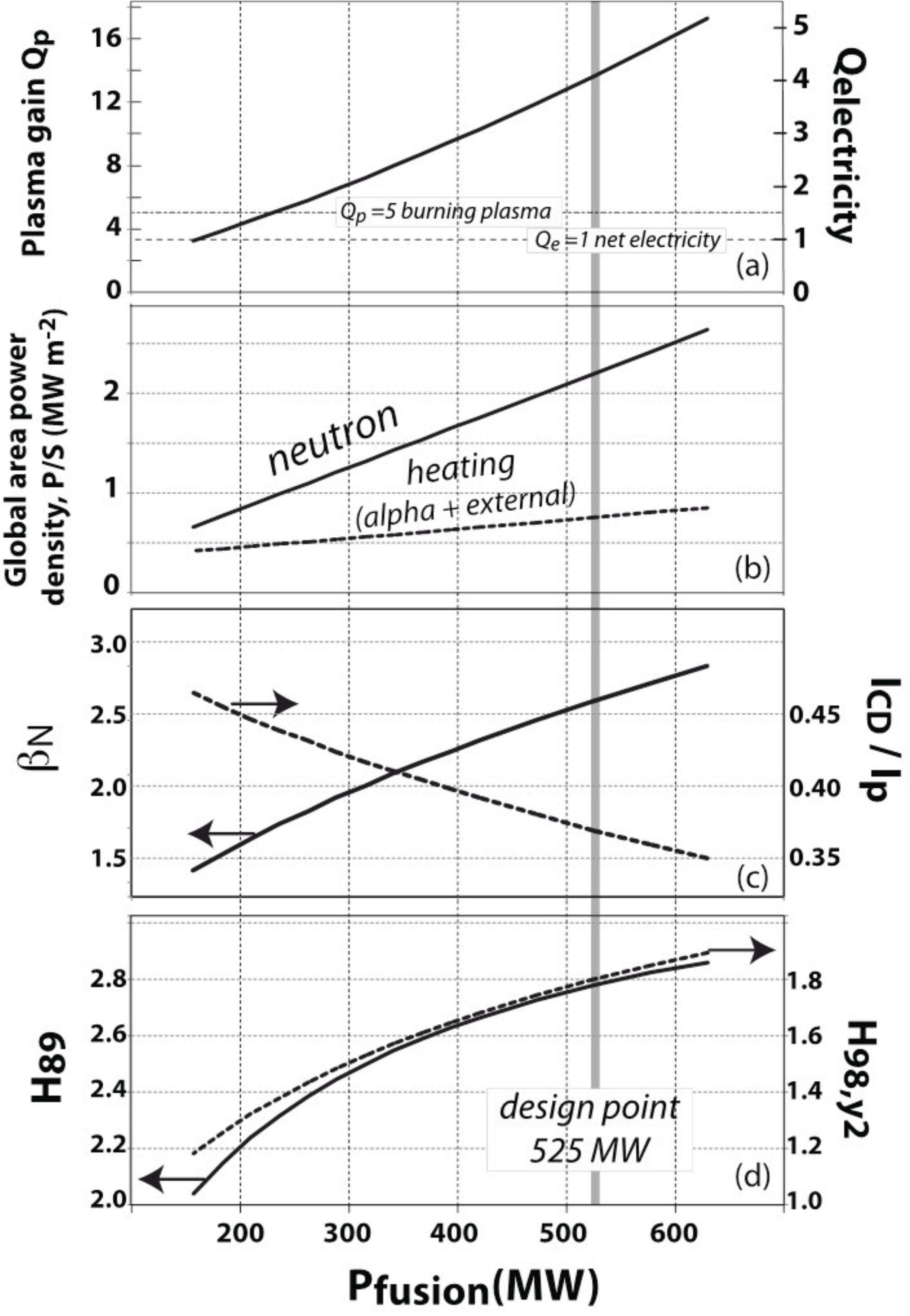}
   \caption{Result of ARC 0-D sensitivity scan organized versus fusion power. The grey vertical line is the design point of $P_{f} =525$ MW.  Table \ref{tab:DesignParameters} shows values of the parameters kept fixed during the scan: $I_{p}$, $B_{0}$, shape, $n_{20}$, and non-inductive fraction=1. The four graphs show (a) the plasma and electricity multiplication factors, (b) the global power density for neutrons and heating, (c) the normalized beta ($\beta_N$) and current drive fraction, and (d) the calculated normalized confinement qualities, H89 and H98(y,2) consistent with fusion performance. Note that $Q_e$ in (a) is calculated using the 50$\%$ thermal efficiency of the ``aggressive'' Pilot phase. The arrows point to the relevant axis for a given curve.}
   \label{fig:ConfinementSensitivityStudy}
   \end{center}
\end{figure}

Nevertheless it is prudent to examine the effect of confinement quality on the ARC FNSF/Pilot plant mission. The results of a 0-D scoping for ARC are shown in Fig. \ref{fig:ConfinementSensitivityStudy} and were carried out in the following manner. The total fusion power is scanned by scaling the volume-averaged pressure ($P_f \propto p_{th}^2$) obtained from the $P_f = 525$ MW baseline case. Simultaneously the plasma current, volume-averaged density and shaping are kept constant. Keeping these parameters fixed has the simplifying benefit of maintaining the same Greenwald density fraction (well away from the limit) and current drive efficiencies. The power/pressure scan is thus equivalent to a $\beta_N$ scan or a $\left\langle T \right\rangle$ scan for the core plasma. This results in a variable bootstrap fraction. Here we have used the approximation from Sauter \cite{Sauter1999} that $2/3$ of the bootstrap current arises from the density gradient, which is fixed during the scan. The external power, which is assumed to be used entirely for current drive, is modified to assure unity non-inductive fraction. The heating power can be set to $P_{\alpha} + P_{CD}$, which are both known, after which the plasma gain and H factors can be recalculated. Since ARC operates far from the kink limit and at the Troyon limit (see Table \ref{tab:DesignParameters}), it is always advantageous to increase the plasma current because it allows the device to confine more plasma pressure. Therefore, any power needed to heat the plasma will always be injected using the current drive system.

As can be seen in Fig. \ref{fig:ConfinementSensitivityStudy}, $H_{89} \sim 2.2$ results in only $\sim200$ MW of fusion, but this still meets general requirements for an FNSF (steady state neutron flux density $\sim$ MW/m\superscript{2}) and a Pilot plant ($Q_e > 1$), albeit with more modest performance.  Therefore a general conclusion is that the ARC mission can be met over a range of confinement quality $H_{89} \sim 2.2 - 2.8$, spanning from roughly standard H-mode confinement to \enquote{AT} confinement. The use of I-mode confinement may be beneficial to achieving a burning plasma because it degrades more weakly with heating power ($\tau_E \sim P_{heat}^{-0.3}$). This partially explains why the C-Mod I-mode, when scaled up to ARC, produces such a high $H_{89}$ factor. Designing ARC at a higher total current would also improve confinement. There is considerable margin to lower safety factor from $q_{95} \sim$ 7, but this could increase the fusion power too far past 500 MW and require reassessing the current drive. This path is left for future studies. Ultimately, global confinement scalings are a crude tool to anticipate performance and what is really required is a better predictive pedestal model coupled to a core gradient model. 

It is noted that improved confinement (over standard H-mode) is also expected from theoretical considerations due to reduced turbulent transport from weak q shear as well as the lack of low order rational q surfaces. In Ref. \cite{Garofalo_highBeta_D3D} the improved confinement factor was observed to decrease back to $H_{89} \sim 2.2$ (from $2.5-3$) when the $q = 5/3$ surface entered the profile. Therefore one would estimate that the most important design tool is sufficient current profile control. This is central to ARC, where optimized CD efficiency (through high-field side launch) and lower bootstrap fraction are paramount to the design. Indeed with reduced fusion power there is a tendency to gain more external control of the q profile (see Fig. \ref{fig:q_profile}), which will permit more control of the current profile. This control can be used to improve confinement, thus providing a positive feedback for fusion power. An additional note is that a variety of fusion powers should be considered because the quantitative limit on plasma heat exhaust, driven by $P_{heat}/S_{p}$ and the radiated power fraction, is unknown.

\subsection{Conceptual engineering of current drive systems}
\label{subsec:LHCD_Engineering}

\begin{figure}[htp]
 \centering
 \includegraphics[width=0.4\textwidth]{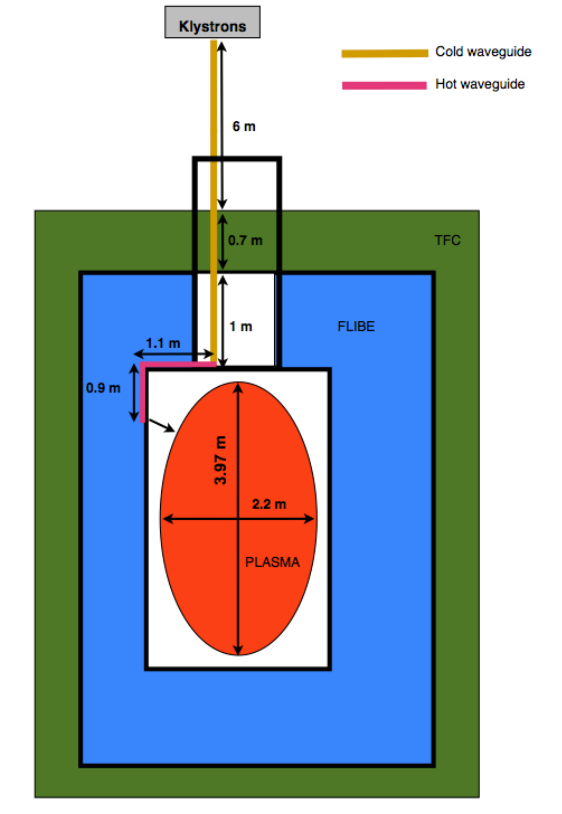}
 \caption{Schematic of the hot and cold portions of the LH waveguides (not to scale and shown as straight for simplicity).}
 \label{fig:LH_Engineering}
\end{figure}

The chosen lower hybrid source frequency of 8 GHz allows the industry standard waveguide, WR-112, to be used for the transmission system. Access to the vacuum vessel will be provided through the hollow posts that support the vacuum vessel in the FLiBe. The launcher horns will be similar to those discussed in the Vulcan study \cite{LHCD_Vulcan}, except that, instead of having discrete launching structures distributed toroidally, the ARC reactor will use two toroidally-continuous strips of alternating active-passive waveguides \cite{ActivePassiveWaveguides, LHCD_Launcher}. The use of toroidal continuous launchers will maximize spectral control, which is desirable for CD control. These two strips provide up to 40 MW of LHCD with conditioned waveguides, nearly double that required for steady state operation. The effective launched power density exceeds 50 MW/m\superscript{2}, a benefit of high frequency RF, and the launchers comprise a small fraction ($\sim 1\%$) of the ARC inner wall. A schematic of LCHD waveguide integration into the ARC reactor design is shown in Fig. \ref{fig:LH_Engineering}, and the power budget is given in Table \ref{tab:lowerHybridPowerInventory}. The klystrons are assumed to have a conversion efficiency of 50\% \cite{MaebaraKlystronEff1994}. The attenuation coefficient, $\alpha_{c}$, in the waveguide depends on resistivity according to \cite{WaveguideTheory}

\begin{align}
   \alpha_c = \frac{1}{\eta b} \sqrt{\frac{\omega\mu}{2\sigma}} \left( \frac{1 + \frac{2b}{a} \frac{\omega_c^2}{\omega^2}}{\sqrt{1 - \frac{\omega_c^2}{\omega^2}}} \right)
   \label{eq:WaveguideAttenuation}
\end{align}
where $\omega_{c}$ is the waveguide cut-off frequency, $\omega$ is the wave launch frequency, $a$ and $b$ are geometric constants of the waveguide, $\mu$ is the conductor permeability, $\sigma$ is the conductivity, $\eta \equiv \sqrt{ \mu / \epsilon}$ is the medium impedance, and $\epsilon$ is the permittivity. Since conductivity decreases (and resistivity increases) with increasing temperature, hot waveguides have significantly higher losses than cold waveguides. This motivates placing the launcher close to the support posts and is the primary reason for optimizing the launch location around the upper region of flux expansion (rather than the lower magnetic X-point). It is important to note that the full effect of high gamma-radiation fluxes to the high frequency, high power density waveguides is unknown and could possibly limit achievable power density in the waveguides. Neutron irradiation will also increase the resistivity of the waveguides, but it may be possible to anneal some of this radiation damage by operating the waveguides at a higher temperature. Both of these effects motivate investigation into finding the optimum temperature to operate the waveguides in a radiation field, and are discussed in further detail in Section \ref{sec:RandDReq}.

\begin{table}
  \centering
  \begin{tabular}{lcc}
    \toprule
    Location & Power Output \\
    \midrule
    \midrule
    Wall plug &  69.6 MW \\
    \midrule
    Klystrons &  34.8 MW \\
    \midrule
    Cold waveguide &  30.0 MW \\
    \midrule
    Hot waveguide &  28.0 MW \\
    \midrule
    LHRF launcher & 25.0 MW \\
    \bottomrule
    \end{tabular}
  \caption{Power inventory throughout the lower hybrid system.}
  \label{tab:lowerHybridPowerInventory}
\end{table}

This LH system has several advantages. First, the extra installed capacity will ensure enough current drive is available in the event of individual waveguide failures. The additional power can also be used to assist in plasma start-up, reducing load on the central solenoid, and to overdrive the plasma to recharge the central solenoid. Other lower hybrid designs incorporate phasing of launchers to better control the plasma current distribution \cite{LHCD_Vulcan, LHCD_Launcher}, but, instead of phasing entire launching structures at discrete toroidal locations, this design allows for continuous phasing of individual waveguides in the toroidal location. This added flexibility could allow current to be driven at a specific poloidal location, in addition to a specific minor radial location. Also, the precise phasing of neighboring waveguides provides the narrow spectral width of the launched power spectrum assumed for the ACCOME simulations. Additionally, the active-passive configuration allows for enhanced cooling of the active waveguides by pumping coolant through the neighboring passive waveguides. The material (often copper and aluminum) and dimensions of the waveguides allow them to behave as fins, leading to high heat transfer rates. As discussed in Vulcan \cite{olynyk2012vulcan}, one of the greatest advantages to HFS launch is that the launcher is in the good curvature region, which minimizes plasma-material interactions.

The ICRF launcher system will follow a similar design to the lower hybrid system. However, since ICRF waves can be transmitted with simple coaxial cables, the transmission line losses will be negligible. Additionally, ICRF sources ($50-80$ MHz) are more efficient than the envisioned sources for the $8$ GHz lower hybrid waves, motivating an assumed source efficiency of approximately $70\%$ \cite{JardinARIESATphysicsBasis2006}. Combined, this results in a required wall-plug power of $19$ MW, only $1.4$ times greater than the plasma-coupled ICRF power. The exact location of the ICRF launchers has not yet been optimized with respect to the device geometry, source frequency, wave trajectory, and wave damping because it requires advanced simulation tools beyond ACCOME. As with the LH launchers, the $\sim$ 13 MW ICRF antennae only occupy a very small fraction of the first wall.

\subsection{ARC as an inductive burning plasma}
\label{subsec:InductiveARC}

While ARC was designed to operate under non-inductive scenarios (see Sec. \ref{sec:0D_Design}), evaluating its performance as an inductive burning plasma experiment is informative. The following 0-D exercise allows for a more straightforward comparison to lower magnetic field, inductive burning plasmas such as ITER ($B_0$ = 5.3 T). Here ARC is assumed to operate with a monotonic sawtoothing ($q_{min} \sim$ 1) current profile and thus have standard confinement factors. The operating current of ARC is scanned while the Greenwald fraction ($f_{Gr} = 0.9$), plasma shape, and auxiliary heating is fixed at the standard ARC values (see Table \ref{tab:DesignParameters}). The required current is determined such that the device produces a fusion power of $P_{f} = 525$ MW (and a plasma gain of $Q_{p} \sim 13.5$), consistent with a burning plasma mission. Following the $H_{89}$ confinement scaling, we obtain: $H_{89} = 2$, $I_{p} \sim 10.8$ MA, $q_{a} \sim 3.6$, $n_{20} \sim 1.9$, and $\left\langle T \right\rangle \sim 10.5$ keV. Following the $H_{98}$ confinement scaling we obtain: $H_{98,y2} = 1$, $I_{p} \sim 12$ MA, $q_{a} \sim 3.3$, $n_{20} \sim 2.1$, and $\left\langle T \right\rangle \sim 9.4$ keV. This compares favorably to ITER: $H_{98,y2} = 1$, $I_{p} \sim 15$ MA, $q_{a} \sim 2.5$, $n_{20} \sim 1$, $f_{Gr} \sim 0.9$, $P_{f}$ = 500 MW, $Q_{p} = 10$. We see that ARC provides larger margins to the disruptive safety factor and density limits, a natural advantage seen in many high-field compact devices (e.g. BPX \cite{goldston1992burning}).

\section{Magnet design}
\label{sec:Magnets}

\begin{figure}[htp]
 \centering
 \includegraphics[width=0.45\textwidth]{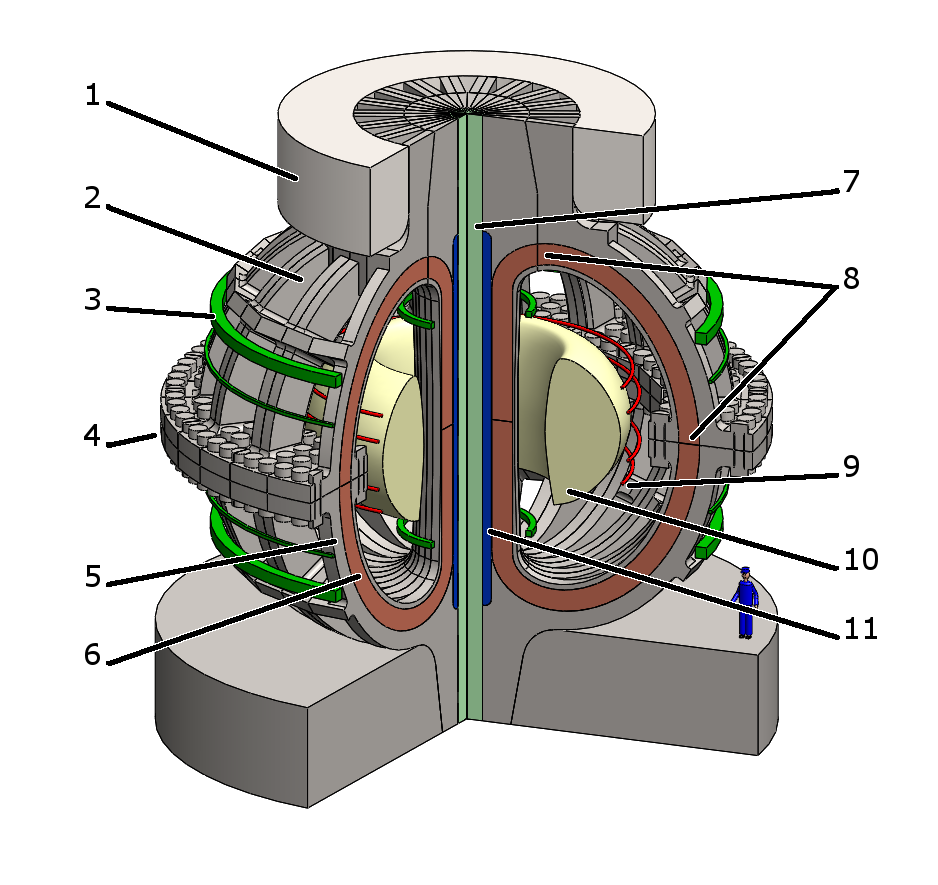}
 \caption{Schematic design of the coil systems, including: 1 -- outward force support ring; 2 -- top demountable leg of the TF coils; 3 -- PF coils (in green); 4 -- outer bolted joint between TF coil legs; 5 -- bottom leg of the TF coils; 6 -- TF coil winding pack; 7 -- glass-filled epoxy reinforcement plug; 8 -- TF electrical joints; 9 -- AUX coils (in red); 10 -- plasma; 11 -- CS and bucking cylinder. The superconducting cables in the TF coils are shown in brown, within the steel support structure.}
 \label{fig:Mag_coils_4groups_1}
\end{figure}

A central aspect of the ARC conceptual design is exploring possible fusion reactor/FNSF scenarios at the much higher field afforded by REBCO superconductors. It is imperative to explore these new magnet designs to understand the tradeoffs and limitations. The magnet system, shown in Fig. \ref{fig:Mag_coils_4groups_1}, is divided into four groups: toroidal field (TF) coils, poloidal field (PF) coils, the central solenoid (CS), and auxiliary (AUX) coils. The first two groups are steady state superconducting magnets that provide the required magnetic fields for stability, shaping, and startup. The large Lorentz forces on the superconducting coils are supported by stainless steel 316LN structure. The demountable TF coils have been designed to provide a magnetic field of 9.2~T on axis, with a peak field of 23~T on coil, and their conceptual design has been introduced in Ref. \cite{YBCO_JOINTS_GOH}. The CS will be used primarily for inductive startup of the plasma current. While ARC is designed for a non-inductive scenario, the CS is very useful for off-normal plasma current control. The auxiliary (AUX) coils are copper magnets that carry relatively small currents for real-time shape adjustments. Located close to the plasma, just on the outside of the vacuum vessel, these coils allow for quick feedback to the plasma shape and constitute the main fast response magnetic control system. 

\subsection{Superconductor choice}
\label{sec:Mag_SuperconductorChoice}

\begin{figure}[htp]
 \centering
 \includegraphics[width=0.45\textwidth]{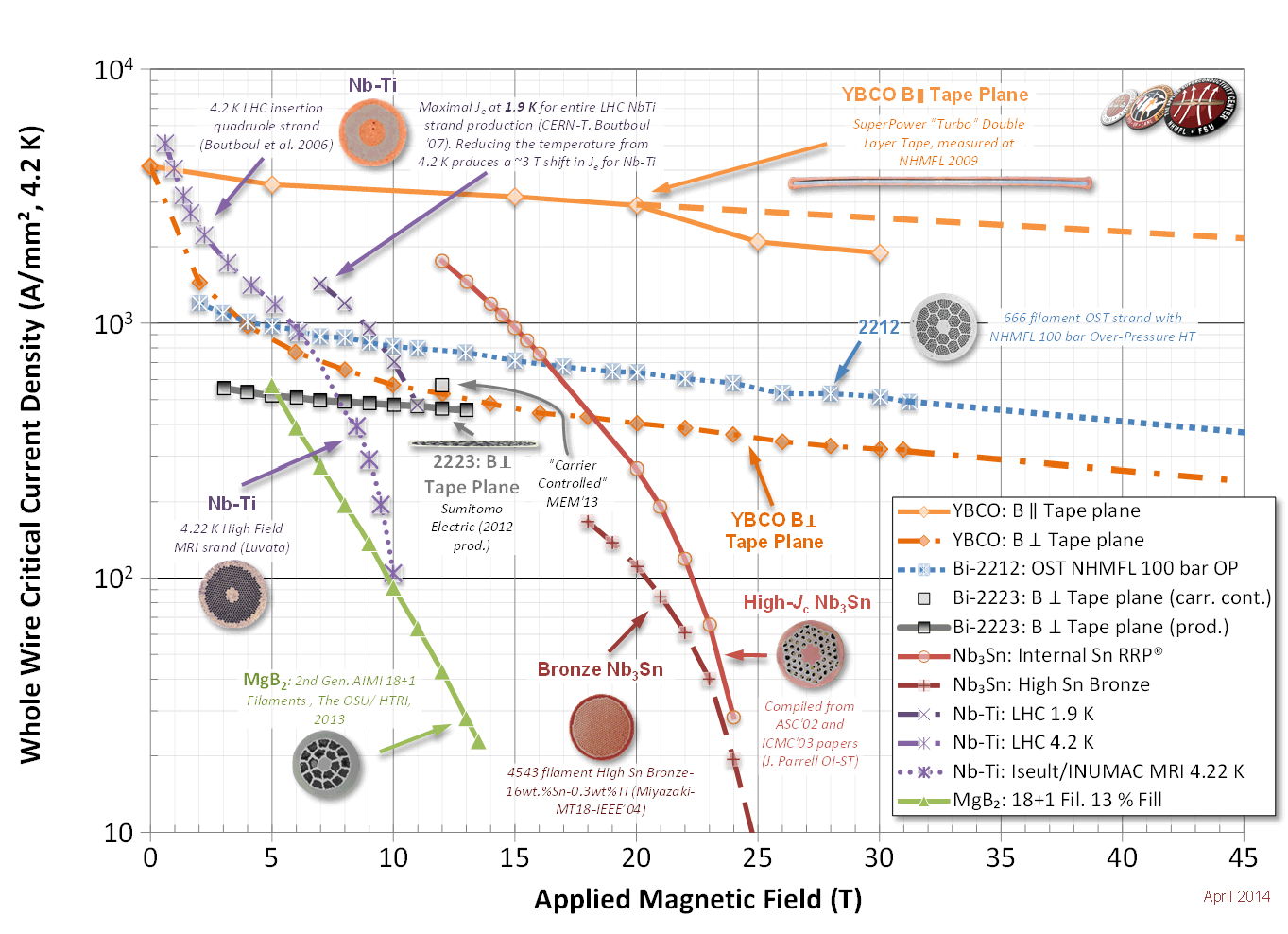}
 \caption{Engineering critical current density as a function of applied magnetic field for commercially available superconductors at 4.2~K. High temperature superconductors (REBCO, shown in the figure as YBCO, and BSSCO) have orders of magnitude higher critical current density than standard Nb\subscript{3}Sn at local fields $>$ 20 T and their critical current decreases weakly with $B$. Note that the orientation of the tape relative to $\vec{B}$ alters the REBCO critical current. Plot compiled by P.J. Lee \cite{PeterLeeASCPlot}.}
 \label{fig:NHMFL_plot}
\end{figure}

To obtain 9.2~T on axis, the maximum magnetic field in the conductors in the TF coils will be 23~T at the inboard midplane. As shown in Fig. \ref{fig:NHMFL_plot} (compiled by P.J. Lee \cite{PeterLeeASCPlot}), at these large magnetic fields, subcooled REBCO outperforms other well-developed superconductors such as NbTi and Nb\subscript{3}Sn. At 23~T and 4.2~K, the critical current density of REBCO tape superconductors produced by SuperPower Inc. is between one and two orders of magnitude higher than Nb\subscript{3}Sn, making REBCO the best superconductor for ARC TF coils.

\begin{figure}[htp]
 \centering
 \includegraphics[width=0.45\textwidth]{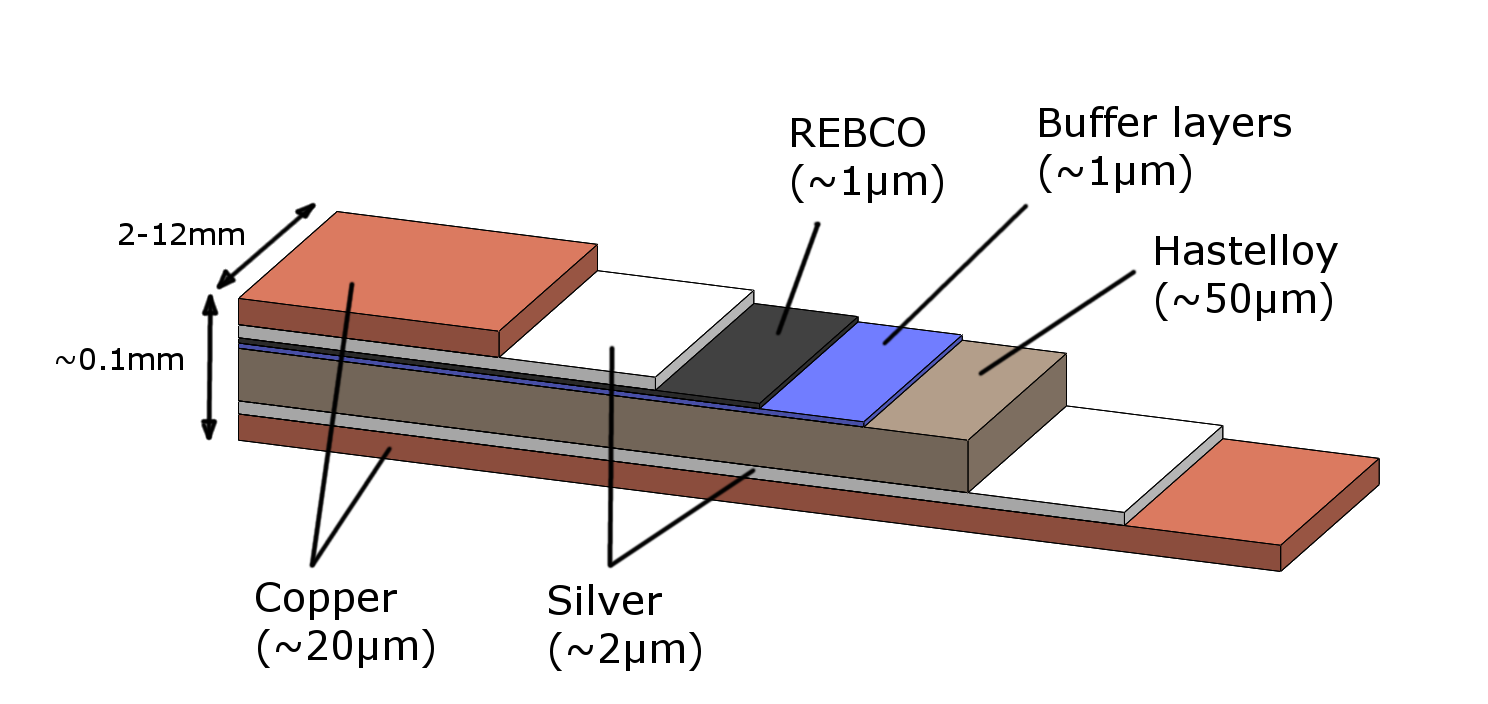}
 \caption{Schematic, not to scale, view of the cross section of REBCO tape superconductor by SuperPower Inc. Each layer is shown with its typical thickness. The tape is typically available in widths between 2 and 12 mm, and a total thickness of about 0.1 mm.}
 \label{fig:REBCO_XS}
\end{figure}

Fig. \ref{fig:REBCO_XS} shows a schematic view of the cross section of commercially available REBCO tape conductor by SuperPower Inc. The conductor is mostly composed of copper and Hastelloy, with a very thin layer (approximately 1\% of the total thickness) of REBCO superconductor. The buffer layers are used during manufacturing to orient the REBCO crystals in a preferred direction, such that the c-axis of the crystals are perpendicular to the face of the tape \cite{YBCO_texture}. The critical current density of REBCO is very sensitive to the orientation of the magnetic field. It is maximized when the magnetic field is parallel to the tape surface ($B \perp c $) and minimum with the magnetic field perpendicular to the tape surface ($B \parallel c$). To achieve the largest possible critical current, ARC uses REBCO tape oriented parallel to the toroidal magnetic field ($B \perp c$) in the inner leg of the TF coils.
% Note: 'B \parallel c' means 'B is parallel to c', not 'the component of B parallel to c' (which would be 'B_{\parallel c}'). Same with 'B \perp c'

REBCO is a high temperature superconductor, meaning it can operate at temperatures up to about 80~K, much higher than the 4.2~K necessary for Nb\subscript{3}Sn. However, we operate the REBCO at 20~K, meaning it is ``sub-cooled'' and far from its critical temperature. Like Vulcan, ARC features finite-resistance joints between REBCO tapes at the locations where the coils demount. Operation at 20~K, rather than 4.2~K, has several operational advantages: a) the overall thermodynamic cost of cooling, including the resistive joints, is reduced, b) the thermal stability of the coil is greatly enhanced because the heat capacity of materials is much higher (nearly eighty times higher in the case of copper \cite{ReedAndClark}, between five and ten times higher for Hastelloy \cite{Hastelloy_properties} and steel \cite{Ekin}), and c) coolants other than liquid helium can be used (e.g. liquid hydrogen, liquid neon, or helium gas). There are a variety of demountable joints, such as butt and edge joints \cite{YBCO_JOINTS_ITO} and bridge lap joints \cite{YBCO_JOINTS_ITO_HTS4Fus}, however ``comb-like'' joints \cite{YBCO_JOINTS_GOH} were ultimately selected for the ARC conceptual design (see Section \ref{sec:Mag_Joint}).

\subsection{Toroidal field coils}
\label{sec:Mag_TF}

The TF coil system is composed of 18 demountable TF coils with stainless steel 316LN structure, which has been well characterized for use in superconducting coils \cite{Nyilas2004_SS316LN_ITER}. The shape of the coils is based on the constant tension Princeton D-shape \cite{Thome1982}. The magnets are cooled to 20~K, an operating point chosen based on the Vulcan findings of the minimum total capital and operating cost for superconductor and cryoplant volume \cite{vulcan_magnets}. The TF coil is divided into two parts: a removable upper leg and a stationary lower leg. The joints are located at the outer midplane and the top of the coils, as shown in Fig. \ref{fig:Mag_coils_4groups_1}. The legs are bolted together in the outer joint and a steel tension ring serves as structural support for the top joint.

The TF coils are composed of a winding pack and a stainless steel coil case. The winding pack consists of a set of 120 jacketed superconducting cables, described in Section \ref{sec:Mag_TF_CICCs}. The inner leg of each TF coil is supported by bucking against the central solenoid and bucking cylinder, and shear keys placed between adjacent TF coils help support the overturning moments in the central column. The vertical axis area ($R$~=~0--0.45~m) is filled with a glass-filled epoxy plug to reduce the maximum stresses in the inboard side of the central solenoid and TF coils. The epoxy was chosen such that it does not influence the magnetic field from the central solenoid, which acts as an air-core solenoid.

\subsubsection{Superconducting cables performance}
\label{sec:Mag_TF_CICCs}

To generate an on-axis toroidal magnetic field of 9.2~T, the net current in each TF coil must be 8.4~MA. This current is carried by 70~kA REBCO cables, each composed of a stack of 12~mm wide, 0.1~mm thick REBCO tapes. The cables are built with the cable-in-conduit conductor (CICC) method: the superconducting tapes lay in grooves in an extruded copper stabilizer, forming a cable that is surrounded by a square steel jacket. The square steel jacket is 40~mm x 40~mm. The coolant flows in cooling channels extruded in the copper stabilizer. 

The coil insulation is similar to the K-STAR design \cite{KSTAR_insulation}. Each CICC cable will be insulated by wrapping Kapton and S2 glass around them. The total stored magnetic energy in the TF coil system is approximately 18~GJ and the maximum expected quench voltage is 2~kV. The required insulation thickness to withstand this voltage is about 2~mm.

\begin{figure}[htp]
 \centering
 \includegraphics[width=0.45\textwidth]{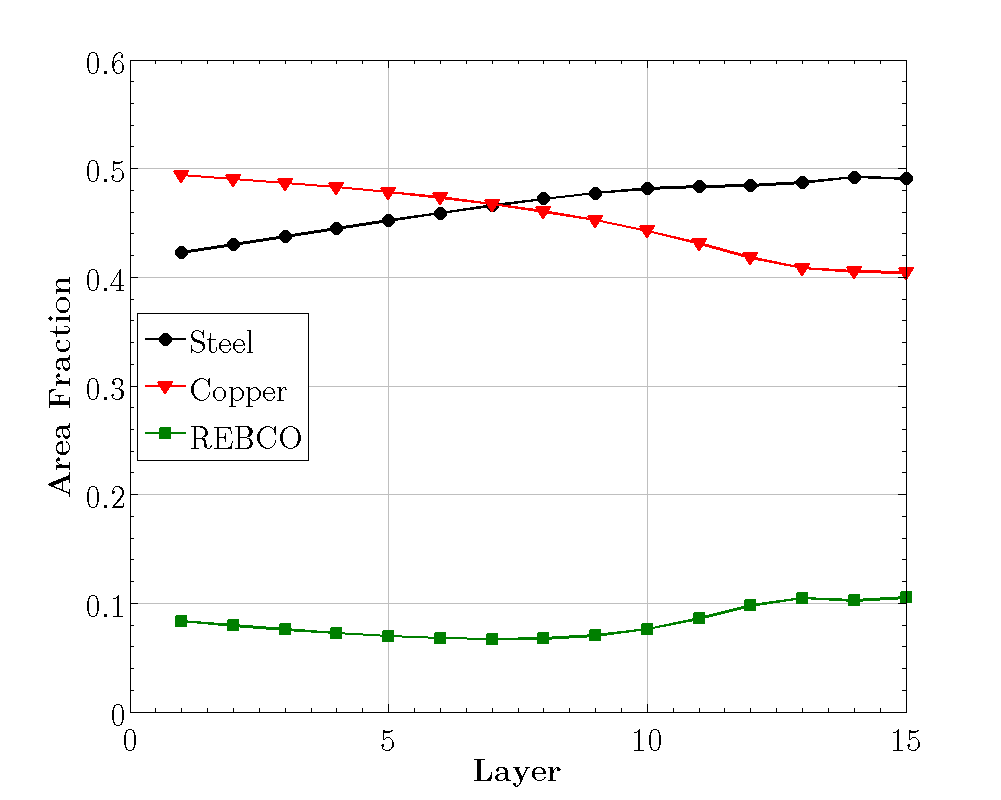}
 \caption{Winding pack composition of the TF coil as a function of layer number. The three components of the winding pack are the steel jacket, copper stabilizer and REBCO tapes. Layer \#1 corresponds to the layer closest to the plasma (subject to the largest magnetic field) and layer \#15 is the outermost. The amount of superconductor required for the outermost layers is slightly larger due to the larger magnetic field perpendicular to the REBCO tape plane near the ends of the central solenoid (see main text for details).}
 \label{fig:CoilComposition}
\end{figure}

The coils are graded, to leverage the increased critical current with lower magnetic fields. The amount of superconductor is chosen such that the current density never exceeds 50\% of critical current density in the layer. The amount of copper in each layer is determined by requiring that the conductor stays below 200~K during a current quench (see Ref. \cite{IWASA}, pp. 471--475). The remainder of the cross section of the cable is made of stainless steel. A plot of the cable composition in each layer, including the copper stabilizer and structural steel, is shown in Fig. \ref{fig:CoilComposition}. The first layer corresponds to the layer closest to the plasma (subject to the largest magnetic field) and the last layer is the outermost. Additional structural steel is required for the coil case and reinforcements, but those components are not taken into account in Fig. \ref{fig:CoilComposition}. The average composition of the winding pack (by area) is: 45.9\% copper, 46.1\% steel, and 8.0\% REBCO tapes (corresponding to a stack of 106 REBCO tapes of 12~mm width). The winding pack current density, including the copper and steel area, is 44~A/mm\superscript{2}. In addition to the factor of two margin already enforced on the critical current, there is substantial margin in the operating temperature. At full current, the temperature can increase by 10~K and the tape will remain superconducting.

The minimum engineering critical current density in the coil occurs in outermost layer (layer \#15 in Fig. \ref{fig:CoilComposition}), at the inner leg of the TF coil, close to the end of the central solenoid. In that area, the toroidal component of the magnetic field (parallel to the REBCO tape plane) is much lower than the radial component caused by the end effects of the CS (perpendicular to the REBCO tape plane). In this situation, at 20~K, with a perpendicular magnetic field of 4.2~T and a parallel magnetic field of 0.8~T, the critical current density of REBCO is estimated to be 815~A/mm\superscript{2}. For comparison, Table \ref{tab:JC_Comparison} shows the largest components of the magnetic field in layers \#1, \#7 and \#15, with the resulting critical current of REBCO oriented parallel to the toroidal field at 20~K.

\begin{table}
  \centering
  \begin{tabular}{lccc}
    \toprule
    Layer & \#1 & \#7 & \#15 \\
    \midrule
    \midrule
    $B_{T}^{max}$ (T) & 23 & 13.5 & 0.8 \\
    $B_{R}^{max}$ (T) & 1.2 & 1.4 & 4.2 \\
    $B^{max}$ (T) & 23 & 13.6 & 4.3 \\
    $J_{C}$ (A/mm\superscript{2}) & 1025 & 1280 & 815 \\
    $J_{op}$ (A/mm\superscript{2}) & 512 & 640 & 407 \\
    $I_{op}$ (kA) & 70 & 70 & 70 \\
    $N^{tapes}$ & 114 & 91 & 144 \\
    \bottomrule
    \end{tabular}
  \caption{The maximum toroidal magnetic field ($B_{T}^{max}$), radial magnetic field ($B_{R}^{max}$), total magnetic field ($B^{max}$), and critical current density ($J_{C}$) in layers \#1, \#7 and \#15. The critical current density was calculated for REBCO tapes at 20~K, oriented parallel to the toroidal magnetic field and perpendicular to the radial magnetic field. The operating current density ($J_{op}$) is calculated to be 50\% of the minimum critical current in the layer. The operating current ($I_{op}$) is necessarily the same for all layers, but the number of 12~mm wide REBCO tapes, $N^{tapes}$, is not.}
  \label{tab:JC_Comparison}
\end{table}

The critical current of REBCO superconductors is expected to increase as the technology continues to evolve. The average critical current density of REBCO tapes produced by SuperPower Inc. increased by over 50\% between 2006 and 2011 \cite{SuperPower2011}. Furthermore, increasing the thickness of the REBCO film in the tape will increase (although not proportionally) the critical current. For example, increasing the REBCO thickness from 1~$\mu$m to 4~$\mu$m increases the critical current by approximately 200\% \cite{SuperPower2012}. 

Each leg of the TF coils will require a total of 120 CICC cables and the total number of superconducting tape turns is 106 tapes/cable x 120 cables/coil x 18 coils $\sim$ 230,000. The total required length of 12~mm wide REBCO tape for the entire TF coil system is about 5730~km. However, the length of the individual CICC cables is 17~m for the bottom leg and 7~m for the top leg. Compared with continuous coil winding such as in ITER (where the length of individual cables is 760~m \cite{Sborchia_2011_ITER_Overview}), ARC requires relatively small lengths of continuous REBCO tape. This allows for more economic quality control of the superconductor, since a defect in a REBCO tape spool can be easily removed with minimal loss of material. 

\subsubsection{Electrical joints}
\label{sec:Mag_Joint}

\begin{figure}[htp]
 \centering
 \includegraphics[width=0.45\textwidth]{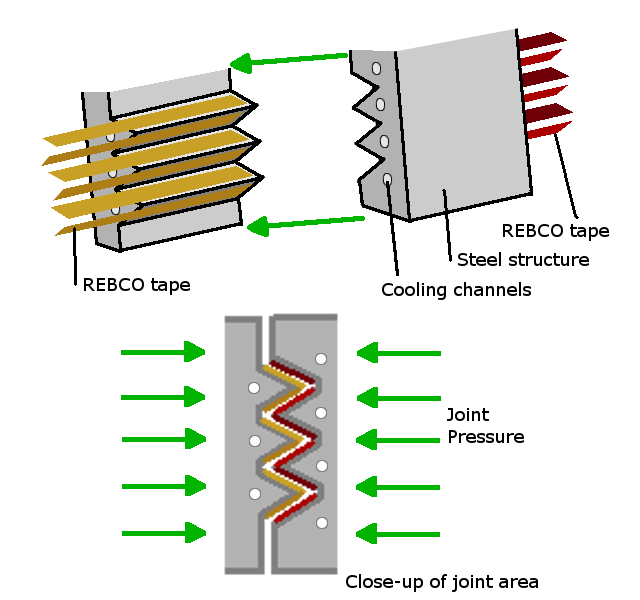}
 \caption{Schematic drawing of comb-style joint before mounting (top) and after mounting (bottom). The REBCO tapes are glued or clamped to the steel structure, oriented such that the Hastelloy layer is between the REBCO layer and the steel structure. This way, the electrical resistance from the REBCO layer to the surface of the tape is minimum. Clamping the two halves together and applying pressure as shown completes the circuit. Cooling channels are located in both structural pieces, at the base of each comb.}
 \label{fig:CombJoint}
\end{figure}

There are a wide variety of electrical and mechanical considerations that go into selecting the type of joint for the superconducting tapes. To assess all of these design choices is beyond the scope of our study. Therefore we choose a rather simple joint topologically in order to provide an estimate of the electrical consumption that arises from the finite resistance of each joint. The electrical joints between the two legs of the TF coils were chosen to be ``comb-style''. A schematic of the joint is shown in Fig. \ref{fig:CombJoint}. Unlike other types of joints, this design is robust to slippage in the direction of the tapes, as slight movement of the tapes does not significantly change the joint resistance. The joint requires pressure normal to the conductors, provided by a mechanical loading structure and the out of plane Lorentz force generated by the PF coils.

\begin{figure}[htp]
 \centering
 \includegraphics[width=0.45\textwidth]{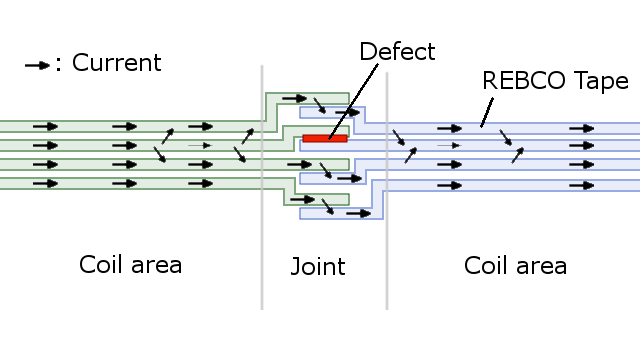}
 \caption{Schematic drawing of expected failure mode of comb-style joint. The current in the coil section of the conductor can redistribute to avoid a defect in one of the tape to tape connections.}
 \label{fig:CurrentRedistribution}
\end{figure}

Each TF coil requires 240 sets of comb joints in series. During steady state operation, the expected failure mode is partial damage, specifically degradation of a small number of tape-to-tape joints in a single comb joint. The comb joints are robust with regards to this scenario, as shown in Fig. \ref{fig:CurrentRedistribution}, because each comb joint is composed of several tape-to-tape joints in parallel. This means damage to a tape will just cause the current to redistribute across the other tapes, bypassing the defect. This will allow the joint and coil to continue operation even with the failure of a tape-to-tape joint.

Each set of comb joints is insulated with Kapton and S2 glass. The insulation of the terminals overlap with each other in each joint, to increase the interface length. The insulation must protect the joint against Paschen and tracking discharges in the event of a magnet quench. Design concepts of joint insulation schemes are under development but not yet demonstrated.

Preliminary experimental measurements of REBCO comb joint resistance have been performed at 77~K without a background magnetic field \cite{YBCO_JOINTS_GOH}. They yield an average contact resistance of 30~$\mu\Omega$-mm\superscript{2}. For the average case of 105~x~12~mm wide tapes per 70~kA cable, each tape pair will have an overlap length of 50~mm, corresponding to a tape-to-tape contact area of 600~mm\superscript{2}. This configuration would require 35 comb teeth per joint, each tooth 150~mm long. 

This geometry yields a tape-to-tape connection resistance of 50~n$\Omega$, which corresponds to a power dissipation of 2.3~W in each comb joint. Coil grading does not have an effect on the total power dissipation in the comb joints, since the length of the tape-to-tape contact area is adjusted according to the number of tapes in each 70~kA cable to keep the power dissipation constant. Thus, each TF coil will dissipate a total of 550~W, for a total of 9.9~kW of Joule power dissipation in all the electrical joints of the TF coil system at 20~K. Taking into consideration the practically achievable Carnot efficiency (calculated with the correlations recommended by Kittel \cite{CryocoolerEfficiency}), it requires a wall-plug power of 0.57~MW to cool the resistive joints. However, this is small compared to the power requirements for the other coil systems and will not affect the overall plant power balance. This strongly contrasts with the copper-based FNSF designs, where the coil electricity consumption is about 500~MW \cite{stambaugh2010candidates}.

\subsubsection{Stress analysis in the coil structure}
\label{sec:coilStress}

\begin{figure}[htp]
 \centering
 \includegraphics[width=0.45\textwidth]{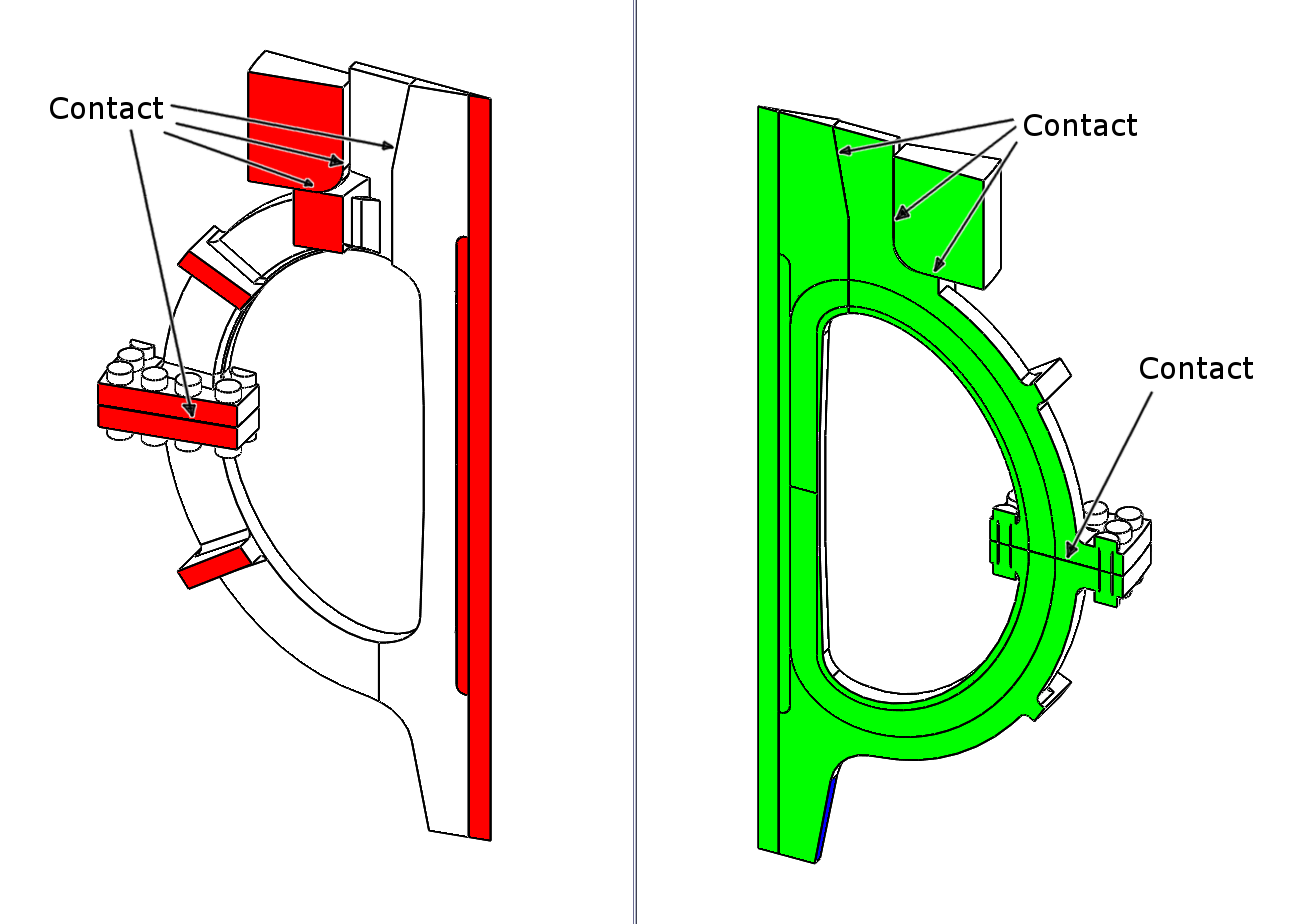}
 \caption{Boundary conditions applied to the stress simulation, from two different angles. Surfaces shown in green have roller boundaries. Surfaces with the same color are located in the same plane. The contact areas between the three bodies (top leg, bottom leg, tension ring) are marked with arrows.}
 \label{fig:FEM_bc}
\end{figure}

A 3-D finite element method (FEM) stress analysis was performed on the TF coils using COMSOL, a multiphysics code \cite{COMSOL}. The simulation domain is a 10 degree section of the TF and CS systems, corresponding to half the toroidal extent of a single TF coil. The out-of-plane forces are not simulated. In Fig. \ref{fig:FEM_bc} the boundary conditions of the simulation are illustrated. Roller boundaries were applied on the side faces of the CS and some of the side faces of the TF, appropriate for the symmetry present in the design. The two legs were analyzed together, with sliding frictionless contact conditions applied between them. The interface between the top leg and the tension ring was also a contact condition. The body load in the superconductors was set to be the Lorentz load from the total toroidal field. The winding pack was simulated as a homogeneous isotropic material with the average composition: 45.9\% copper, 8.0\% REBCO tapes (calculated as 45\% copper, 55\% Hastelloy), and 46.1\% steel.

\begin{figure}[htp]
 \centering
 \includegraphics[width=0.45\textwidth]{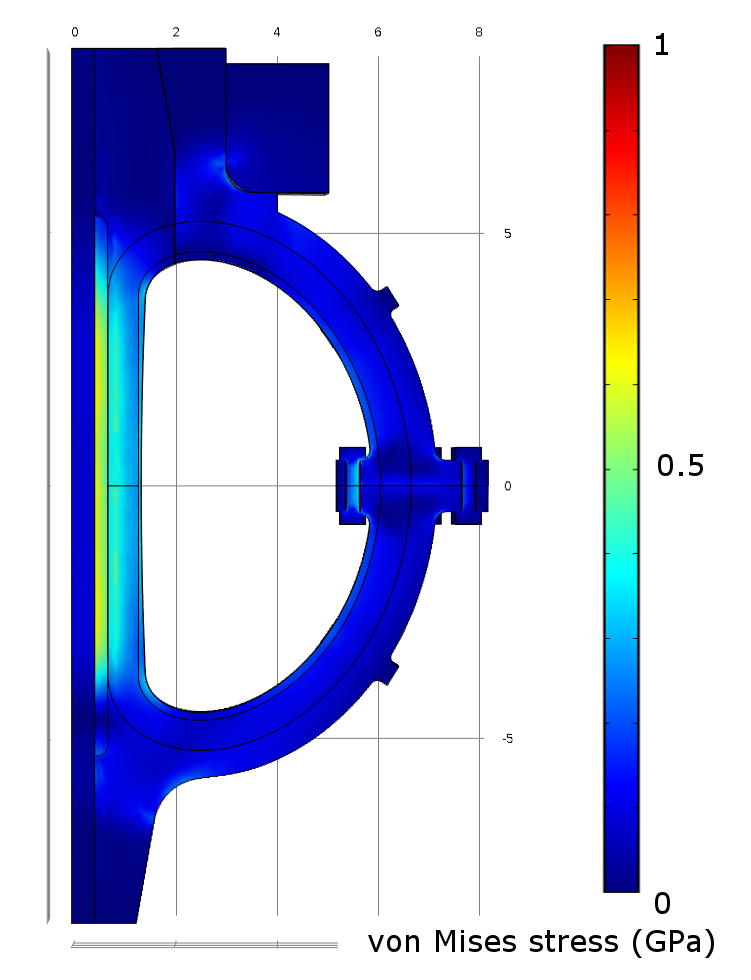}
 \caption{Results of stress simulations in the TF coils. The maximum stress in the stainless steel 316LN structure is 660~MPa, which gives safety margin of approximately 65\%.}
 \label{fig:FEM_results}
\end{figure}

The results of the FEM simulation are shown in Fig. \ref{fig:FEM_results}. The maximum von Mises stress on the structures is roughly 660~MPa in the bucking cylinder (about 65\% of the yield stress of stainless steel 316LN of 1050~MPa at cryogenic temperatures \cite{Nyilas2004_SS316LN_ITER}) and 500~MPa in the TF coils structure. The strain in the superconductor is less than 0.2\%, half the limit of reversible critical current degradation (a 5\% degradation occurs with 0.4\% strain \cite{Shin2007736}). In this analysis the CS was not energized because simulations showed that this yielded slightly lower stresses in the central column. Therefore the scenario without a magnetic field generated by the CS was used as a conservative approximation for the steady state stress. Note that during operation the CS will be energized, thus the actual steady state stresses are lower.

There may be other choices of stainless steel alloys that exhibit higher yield stress at cryogenic temperatures (e.g. AISI 301 with a yield stress of about 1500~MPa \cite{strength301}), but this higher strength comes at the expense of less ductility. The key observation is that structural stress is an important limit to consider for implementing REBCO in high-field applications. While further optimization should be carried out through modeling and experiments (e.g. on the choice of materials and geometry), it is not unreasonable to expect 20--23~T peak field on coil, given the known strength limit of various non-magnetic alloys.

\subsection{PF, CS and AUX coil systems}

\begin{figure}[htp]
 \centering
 \includegraphics[width=0.45\textwidth]{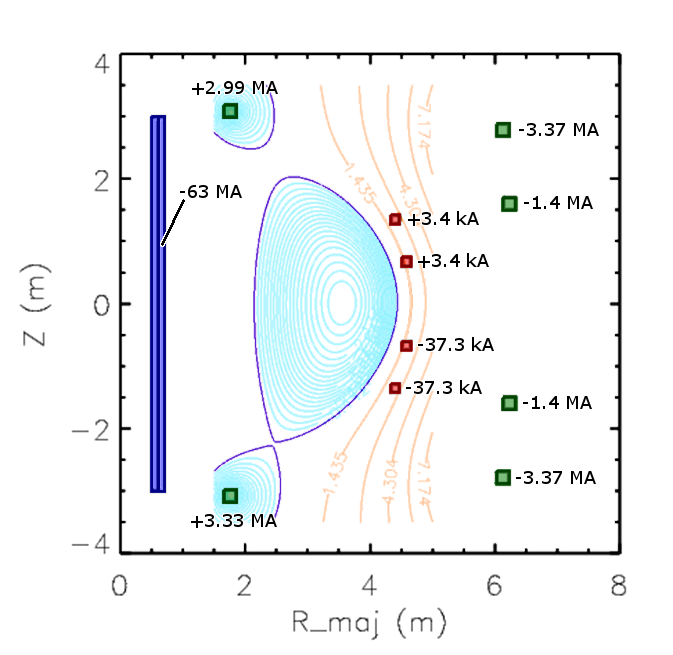}
 \caption{CS, PF, AUX coil requirements from ACCOME, where the CS is in blue, the PF coils are green, and the AUX coils are red. The location of the coils is to scale, but the sizes of the coils are not (except for the CS). The direction of positive current is defined to be the direction of the plasma current.}
 \label{fig:PFCSAUX_ACCOME}
\end{figure}

The PF, CS and AUX coils were designed based on ACCOME requirements, as shown in Fig. \ref{fig:PFCSAUX_ACCOME}. Stress simulations have been performed for the CS as detailed in Section \ref{sec:coilStress}. However, no analysis was performed for the PF and AUX coils because they carry relatively little current and are not constrained by stress limits due to the availability of physical space for structure. 

The Central Solenoid is part of the bucking cylinder, and occupies the space $R$~=~0.5--0.7~m, $Z$~=~-3--3~m, as shown in Fig. \ref{fig:PFCSAUX_ACCOME}. It is layer-wound with REBCO CICC cables, similar to the TF coils conductors. It does not have vertical segmentation since this would weaken its mechanical strength. The solenoid operates from 63~MA/turn to -63~MA/turn. This configuration generates a peak field on the coil of 12.9~T, similar to the ITER solenoid \cite{AymarITERSummary2001}. During reactor start-up, a flux swing of 32~Wb is available if the current in the solenoid is ramped across its full operating range. The maximum stress in the CS structure is 660~MPa, due to the compressive force from the TF coils. However, the CS differs in that it experiences transient stresses. The mission of ARC requires non-inductive scenarios and thus the number of cycles is expected to be limited compared to inductive designs like ITER. This eases fatigue concerns, plus the superconductor already has a large margin in critical current because the REBCO tapes in the CS are wound with the vertical magnetic field of the solenoid lying in the plane of the REBCO tapes.

The REBCO superconducting CS and PF coils provide the principal poloidal field required for plasma shaping. The use of superconducting coils minimizes recirculating power in support of the Pilot plant mission of ARC. The PF coils used for pulling the X-points are situated outside the vacuum vessel, but inside the TF volume at $R \sim 2$~m, $Z = \pm 3$m. These PF coils do not need joints because the TF demountability allows for their modular insertion and extraction. The PF pull coils are shielded by the FLiBe blanket and can be relatively small because they are inside the TF volume. Outboard plasma shaping and equilibrium fields are supplied by PF coils placed outside the TF at $R \sim 6$m.

The copper AUX coils allow for fast control of the plasma, on the resistive time of the vacuum vessel, helping to avoid disruptions. They are effectively unshielded from neutron damage, so they are single-turn to improve radiation survivability. Additionally, they are part of the replaceable vacuum vessel, so they have a short in-vessel lifetime before they are removed/replaced. The degradation in electrical conductivity from the radiation damage is negligible, as at 800--900~K the radiation damage in copper is largely annealed out. These coils require standoffs from the vacuum vessel, which are assumed to be thin stainless steel (with much higher resistance to ground than the loop resistance in the copper coils). The resistivity of the coils is approximately $5.3 \times 10^{-8}$ $\Omega$-m at 800~K \cite{HighTempCopper}. Given a 5~cm circular coil cross section, each lower AUX coil requires approximately 1~MW of electrical power, while the upper AUX coils require about 9~kW. At this size, the AUX coils have very little effect on the neutronics, as discussed in Section \ref{sec:PowerCore}.

\subsection{Magnet cooling}
\label{sec:Mag_MagnetCooling}

The magnet system could be cooled by liquid hydrogen, neon, or gaseous helium. For this study we assume liquid hydrogen, which is relatively inexpensive and abundant, pressurized to 5--10~bar to increase the liquid temperature range. Independent cooling loops will refrigerate each TF coil leg and joint. The coil cooling channels are located in the copper stabilizer, while the joint cooling channels are located in the comb-style steel structure (as shown in Fig. \ref{fig:CombJoint}).

Radiant and conductive heat from the 900~K FLiBe tank (see Section \ref{subsec:Blanket_Thermal_Analysis}) is removed in several stages before it can reach the superconducting coils. The thermal shielding, composed of aluminum silicate surrounding the low-pressure blanket tank, is cooled with water at room temperature. The neutron shield surrounds this and is also cooled with liquid water. A set of three vacuum gaps separate the TF coils and the neutron shield (see Fig. \ref{fig:inboardRadialBuild}). The thermal shield of the first gap (closest to the neutron shield) is cooled with liquid nitrogen and the thermal shield of the outer gap is cooled with liquid hydrogen.

The water cooling loop removes several MW of heat to maintain components at room temperature, although the pumping power required to do this is negligible (see section \ref{sec:Power_NeutronShieldThermalShieldingAndCooling}). In the nitrogen loop, 160~kW of heat must be removed, requiring 1.5~MW of electricity. The hydrogen loop removes 700~W of heat that is radiated through the first vacuum gap, 10~kW from resistive heating in the TF coil joints and 14~kW of distributed nuclear heating throughout the volume of the TF magnets (see section \ref{subsec:Blanket_Thermal_Analysis}). The total electrical power required to remove heat in the hydrogen loop is approximately 1.4~MW. It is noted that these levels of electrical consumption are relatively small (see Table \ref{tab:PowerInventoryLH2}) compared with the 70~MW of power required for the LHCD system (see Table \ref{tab:lowerHybridPowerInventory}) and thus has a small impact on $Q_e$. The practically achievable efficiencies in the cryocooler systems have been calculated with the correlations recommended by Kittel \cite{CryocoolerEfficiency}.

The coil system must also cycle between room temperature and 20~K for maintenance. Additional refrigeration channels and electrical heaters in the magnets structure will be used for this thermal cycling. When warming up, the electrical heaters will be used and gaseous helium at 300~K will be pumped into the refrigeration channels. Once the structure reaches approximately 100~K, nitrogen gas at 300~K will flow into the vacuum chamber, to accelerate the heating process and to prevent humidity from entering the cryostat once it is opened. When cooling down, cold gaseous helium will be pumped into the refrigeration channels until the magnet temperature reaches 20~K, at which point the steady state hydrogen loop will start flowing. Liquid refrigerants are avoided in the 300~K -- 20~K cycle to prevent evaporation from causing bursts in the cooling channels.

\begin{table}
  \centering
  \begin{tabular}{lccc}
    \toprule
    System & Coolant & $P_{th}$ & $P_{e}$ \\
    \midrule
    \midrule
		Nuclear heating & LH\subscript{2} & 14 kW & 0.8 MW \\
		TF coil joints & LH\subscript{2} & 10 kW & 0.6 MW \\
    TF coil joints load & --- & --- & 10 kW \\
    Thermal shielding & Water & 12 kW & $\sim 0$ \\
    Vacuum shielding & LN\subscript{2} & 160 kW & 1.5 MW \\
    Vacuum shielding & LH\subscript{2} & 700 W & 35 kW \\
    AUX coils & FLiBe & 2 MW & $\sim 0$ \\
    AUX coils load & --- & --- & 2 MW \\
    \midrule
    Total Power & --- & 2.1~MW & 5.1 MW \\
    \bottomrule
    \end{tabular}
  \caption{Electrical power, $P\subscript{e}$, required for coil current supply and the removal of waste heat, $P_{th}$, in the magnet systems.}
  \label{tab:PowerInventoryLH2}
\end{table}

\subsection{Alternative Designs}

\begin{figure}[htp]
 \centering
 \includegraphics[width=0.45\textwidth]{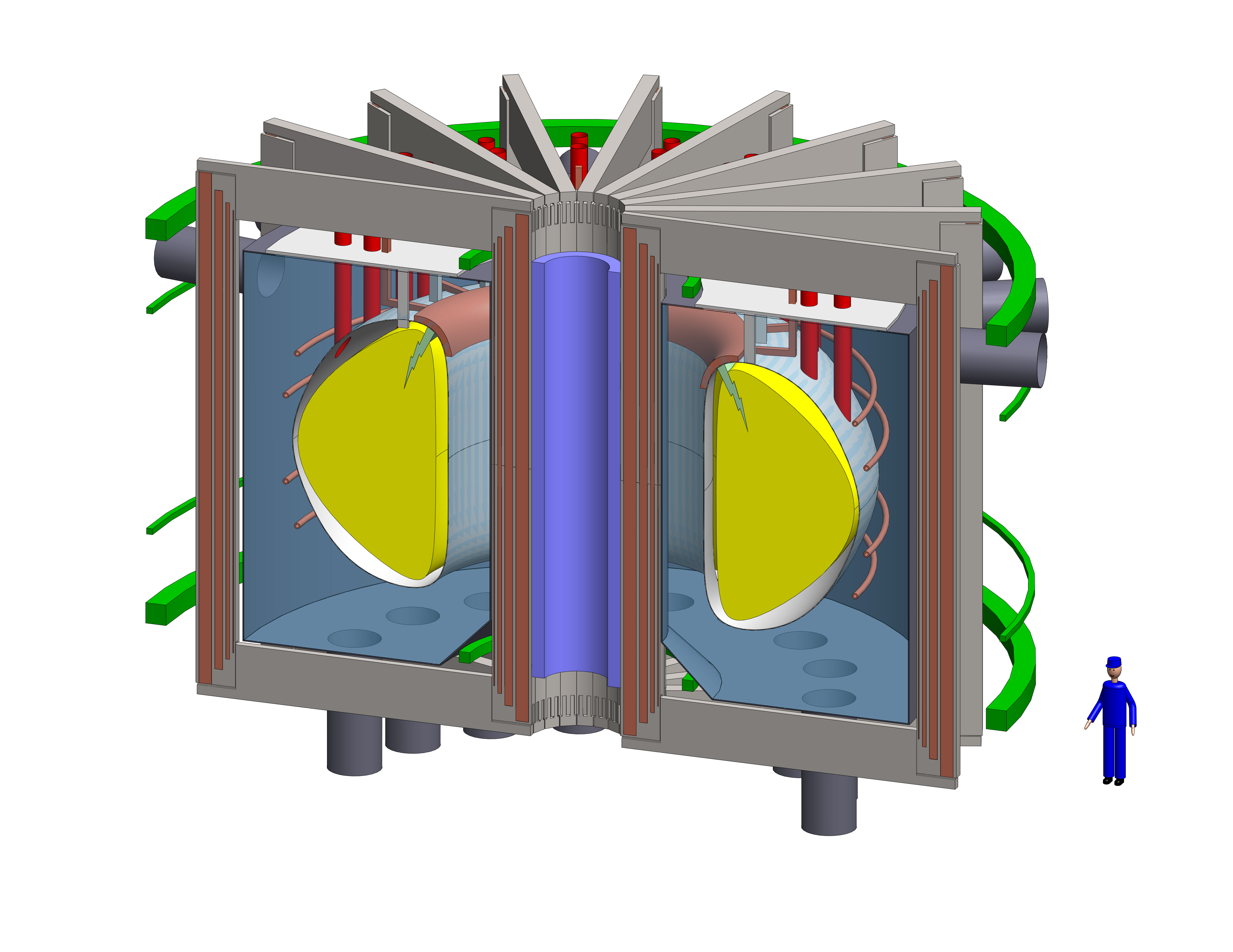}
 \caption{The ``picture frame'' TF coil arrangement represents an alternative possible magnet configuration. The joints on all four corners of the magnets are demountable, allowing for even easier reactor maintenance than the D-shape design. However, this configuration is mechanically weaker than the D-shaped coils, thus limiting the toroidal field and making the ``picture frame'' design more suitable for an FNSF.}
 \label{fig:PictureFrame_Design}
\end{figure}

In addition to the D-shaped coils presented above, a ``picture-frame'' magnet design (see Fig. \ref{fig:PictureFrame_Design}) was also considered. The picture-frame design is based on the TF coils of Alcator C-Mod, modified to use REBCO instead of copper \cite{beck1991alcator}.  We present the picture-frame design to illustrate the wide array of design choices possible when utilizing the idea of a compact, modular, demountable design. While the D-shaped coils are mechanically stronger, allowing for a higher on-axis field, the window-pane coils would be more suitable for a flexible, FNSF design. Unlike the D-shape coils, which only are demountable at two points, the picture-frame coils have sliding joints at every corner.  These joints allow easier access to the blanket tank and vacuum vessel in addition to offering further mechanical stability in the event of a large disruption. However, the additional joints in this design increase the total heat generated by resistive dissipation and the electricity required to remove it.

\section{Fusion power core}
\label{sec:PowerCore}

Traditional tritium breeding and neutron absorbing blankets for fusion reactor designs involve complex components, including significant solid, structural material. Since the blanket is generally contained within the TF coils, these structures must also be separable into toroidal sections so they can be installed through access ports between the TF coils.  This results in challenging engineering constraints, difficult remote handling, and a low tritium breeding ratio (TBR) because the structural part of the blanket does not usually contribute to breeding. ARC utilizes a fully liquid blanket comprised of a molten salt called FLiBe surrounded by demountable superconducting TF coils to facilitate disassembly \cite{sorbom2012liquid}. To maximize the tritium breeding volume, the vacuum vessel is completely immersed in a continuously recycled FLiBe blanket, with the exception of minimal support columns suspending the vacuum vessel in the FLiBe. For modeling simplicity, the support columns are straight, but in a final engineering design, the support columns will be curved to inhibit neutrons from streaming through them.

$F_4Li_2Be$ is a eutectic mixture of lithium fluoride and beryllium fluoride and has been used in fission reactors as a high temperature, high thermal efficiency moderator/coolant. Liquid FLiBe has been considered in fission and fusion nuclear technology applications due to its favorable characteristics: a wide operating temperature range in the liquid state, chemical inertness, and similar thermal-hydraulic characteristics to water (see Table \ref{tab:FLiBe_Properties}) \cite{sze2000flibe,williams2006assessment}. For magnetic fusion, it also has the desirable feature of low electrical conductivity, which will help to minimize any MHD effects caused by the large background magnetic field. The beryllium in FLiBe allows the blanket to multiply and moderate high-energy neutrons via the \superscript{9}Be(n,2n)\superscript{4}He reaction. Through the exothermic \superscript{6}Li(n,t)\superscript{4}He and endothermic \superscript{7}Li(n,t+n)\superscript{4}He reactions, the lithium breeds tritium, multiplies neutrons, and can generate substantial additional heating power.  While breeding sufficient tritium with FLiBe in traditional blanket designs is difficult, we use MCNP neutronics analysis to show that the immersion blanket design, coupled with a non-structural beryllium neutron multiplier, results in TBR $\geq$ 1.1.  This TBR relies on using FLiBe with the lithium enriched to be 90\% \superscript{6}Li by isotopic abundance.  We also show that, when coupled with a TiH\subscript{2} shield, FLiBe is able to protect the TF coils from neutron irradiation, despite the extreme space constraints in the compact ARC design.

\begin{table}
  \centering
  \begin{tabular}{lcc}
    \toprule
    Property & FLiBe & Water \\
    \midrule
    \midrule
    Melting point (K) &  732 &  273 \\
    \midrule
    Boiling point (K) &  1700 &  373 \\
    \midrule
    Density (kg/m\superscript{3}) &  1940 &  1000 \\
    \midrule
    Specific heat (kJ/kg/K) &  2.4 &  4.2 \\
    \midrule
    Thermal conductivity (W/m/K) &  1 &  0.58 \\
    \midrule
    Viscosity (mPa-s) &  6 &  1 \\
    \bottomrule
    \end{tabular}
  \caption{Comparison of the properties of liquid FLiBe and water.  Note that temperature-dependent properties are given for liquid water at 293 K and liquid FLiBe at 950 K.}
  \label{tab:FLiBe_Properties}
\end{table}

Due to its favorable thermodynamic characteristics, FLiBe can be used as an active coolant for the vacuum vessel and divertor, allowing for a simplified single-phase, low-pressure, single-fluid cooling scheme. The immersion blanket permits the entire vacuum vessel to be removed vertically as a single, interchangeable component, allowing for modular replacement. This minimizes the need for lengthy, complex maintenance procedures. In addition, having the vacuum vessel as a single component allows for full off-site testing and quality assurance of all in-vessel components before installation into the reactor. Using an all-liquid blanket eliminates the problem of radiation damage in the blanket and can reduce the amount of solid radioactive waste by a factor of $\sim 20$ (because ARC only has $\sim 5$ cm of vacuum vessel rather than a $\sim 1$ m thick solid blanket). 

\subsection{Open-ended divertor and vacuum vessel design}

The design of commercial-scale fusion reactors requires significant knowledge of plasma material interactions. Presently, the response of Plasma Facing Components (PFCs) to the extreme heat and neutron fluxes of a reactor is unknown \cite{federici2001plasma}. While proposed facilities such as IFMIF \cite{moeslang2006ifmif} and VULCAN \cite{olynyk2012vulcan} could inform this research, ARC would allow several vacuum vessel and divertor designs to be tested in actual reactor conditions.

Demountable coils allow for relatively simple removal and replacement of the vacuum vessel as a single ``plug-and-play'' component.  This allows the ARC reactor to serve as a test bed for several PFC and divertor design configurations. With this in mind, the specific vacuum vessel and divertor designs have been left open-ended. For the purposes of neutronics analysis of the ARC design, the divertor has been modeled as an $8$ cm thick layer of tungsten covering $17$\% of the lower vacuum vessel surface area. An actual divertor will almost certainly be thinner, but, due to the open-ended nature of the design, an overly-thick layer was modeled to give a conservative estimate of tritium breeding. The first wall has been modeled as $1$ cm of tungsten on top of a $1$ cm structural Inconel 718 layer.  Both of these choices have been made based on current experimental designs. However, they represent only one of many possible experimental configurations for the ARC design.

\subsection{TF coil neutron shielding}
\label{sec:TF_Shielding}

\begin{figure}[ht]
 \centering
 \includegraphics[width=0.3\textwidth]{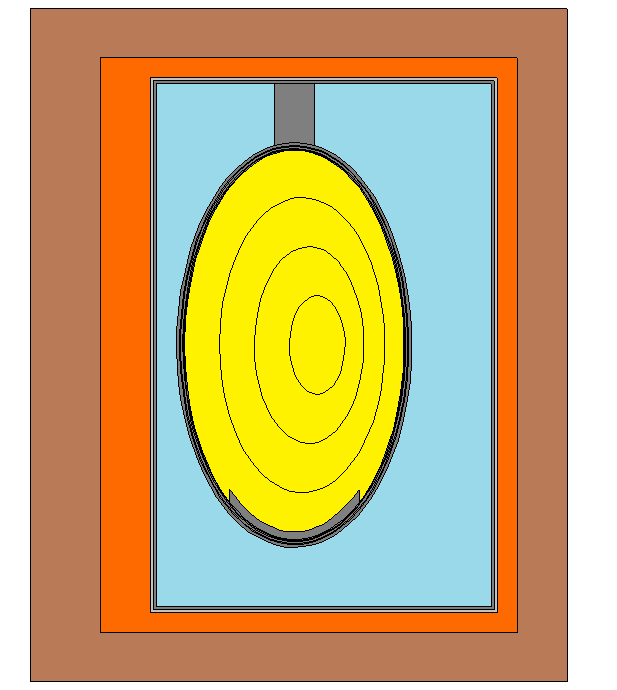}
 \caption{Cross section of the axisymmetric geometry used to model neutron transport in ARC using MCNP.  Yellow indicates the plasma region, orange indicates the TiH\subscript{2} shielding, light blue indicates FLiBe, and brown indicates the superconducting coil structure. Gray indicates either Inconel 718 or tungsten for the vacuum vessel, connection post, and divertor. The ``squaring'' of some corners in this simplified model has a negligible effect on the neutronics analysis.}
 \label{fig:MCNP_Model}
\end{figure}

Superconducting magnets have been shown to be sensitive to high-energy neutron radiation \cite{bromberg2001options}.  In order to assess the radiation survivability of the REBCO tapes, a model axisymmetric reactor geometry was studied using the Monte Carlo neutron transport code MCNP \cite{BrownMCNP2002} (see Fig. \ref{fig:MCNP_Model}).  The most recent irradiation experiments have shown that the critical current of Nb\subscript{3}Sn begins to degrade at a total neutron fluence (considering only ``high energy'' neutrons with energy $> 0.1$ MeV) of $3 \times 10^{18}$ neutrons/cm\superscript{2} \cite{bromberg2001options}.  It is important to note that this value is a lower bound to the neutron fluence on superconductor survivability, rather than a definite point of failure.  The irradiation resistance of REBCO is expected to be at least as good, if not better than Nb\subscript{3}Sn \cite{bromberg2001options}, and ARC can continue to operate with degraded critical current. This is because the operational limits of the tapes are set by magnet stress limits, rather than the critical current of REBCO (see Section \ref{sec:Magnets}).

Given the space constraints of ARC, FLiBe alone is insufficient as a neutron shield for the REBCO superconducting coils, particularly on the inboard side. Therefore, we have added a shield composed of TiH\subscript{2} around the blanket tank (the component that holds the high-temperature molten FLiBe), with additional thickness on the inboard side.  TiH\subscript{2} has an extremely high hydrogen density, appropriate for neutron moderation, and a high cross section for the neutron absorption, making it an ideal shielding material \cite{beck1960research}.  

\begin{figure}[ht]
 \centering
 \includegraphics[width=0.5\textwidth]{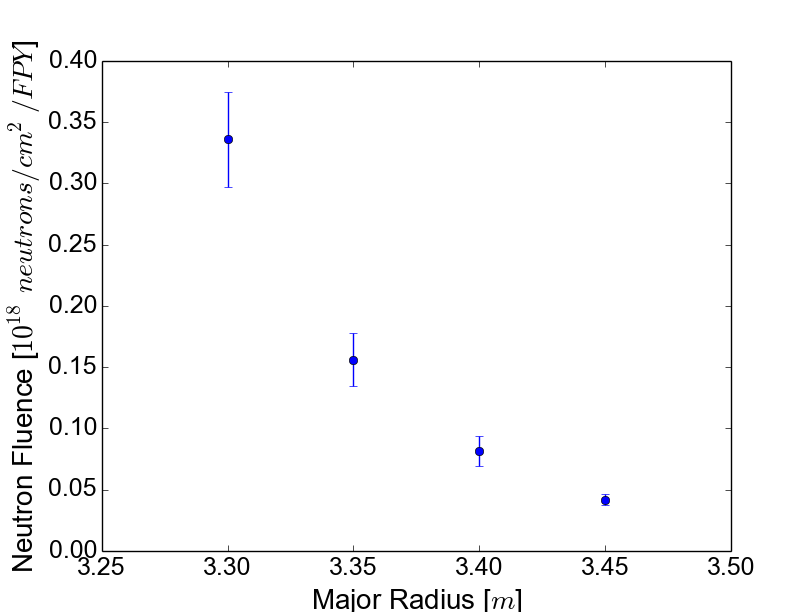}
 \caption{The effect of increasing the reactor major radius (using all of the extra length for TiH\subscript{2} shielding) on the yearly neutron fluence to the TF coils. Note that a small increase in the major radius results in a dramatic reduction of the yearly TF neutron fluence.  While ARC was designed to be as small as possible, a full commercial reactor would likely have a slightly larger major radius to extend the plant lifetime.}
 \label{fig:MajorRPlot}
\end{figure}

We find that, after adding the TiH\subscript{2} shield, it takes $9$ Full-Power Years (FPY) of reactor operation to reach a fluence of $3 \times 10^{18}$ neutrons/cm\superscript{2} in any part of the magnet. While this lifetime is likely insufficient for a dedicated, commercial-scale power plant, $9$ FPY represents a lower bound due to insufficient data on tape irradiation. Furthermore, the lifetime could be extended substantially by scaling the reactor size up by a small amount, increasing the total amount of room for shielding (see Fig. \ref{fig:MajorRPlot}).

\subsection{Tritium breeding}

\begin{figure}[ht]
 \centering
 \includegraphics[width=0.25\textwidth]{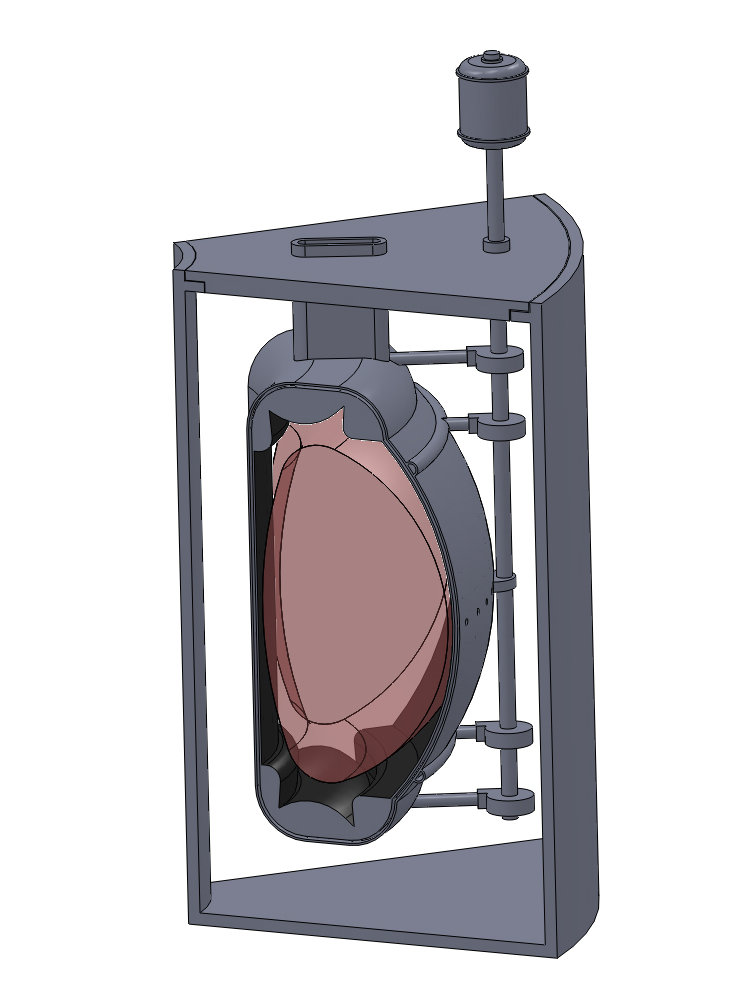}
 \caption{Double-walled vacuum vessel design with FLiBe coolant channel.}
 \label{fig:DoubleVV}
\end{figure}

The same MCNP model described above was used to assess the TBR.  In order to account for uncertainties in cross section libraries as well as the fact that our model is 2-D rather than 3-D, ARC was designed with the goal of obtaining a TBR $\geq 1.1$ \cite{el2009toward}.  In addition, it is critical for the first fusion devices that consume substantial tritium to have good margin for TBR, so they do not jeopardize the world tritium inventory. To maximize TBR, enriched F\subscript{4}Li\subscript{2}Be with a 90\% isotopic abundance of \superscript{6}Li was chosen to enhance the tritium breeding cross section for lower-energy neutrons.  Early designs with a single-walled vacuum vessel proved inadequate to provide both a TBR $\geq 1.1$ and the needed mechanical strength (see Section \ref{subsubsec:disruptionMechAnalysis}). Therefore, a double-walled vacuum vessel design was developed, utilizing a FLiBe channel and a non-structural beryllium neutron multiplier sandwiched between two layers of structural material (see Figs. \ref{fig:DoubleVV} and \ref{fig:ChannelResults}).

\begin{figure}[ht]
 \centering
 \includegraphics[width=0.5\textwidth]{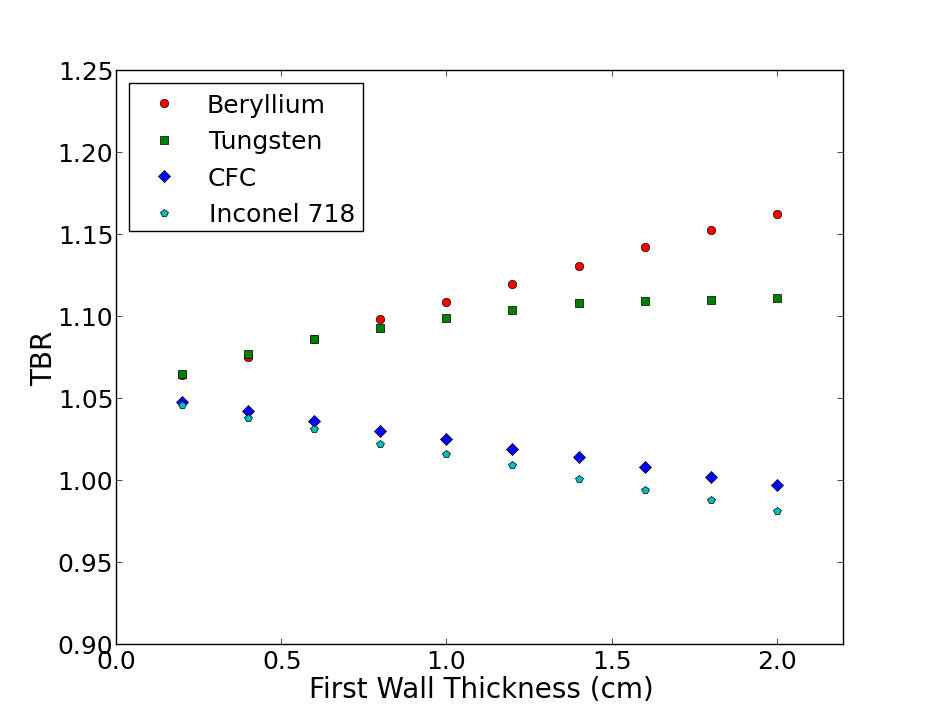}
 \caption{Effect of first wall PFC material and thickness (excluding the divertor) on TBR.}
 \label{fig:FirstWall_Scan}
\end{figure}

TBR is extremely sensitive to the amount and type of material between the neutron source and the breeding blanket. Use of a double-walled, ribbed vacuum vessel addresses this by allowing the FLiBe channel to be very close to the core of the reactor while still maintaining the structural integrity of the vacuum vessel. Even with a FLiBe cooling channel close to the fusion plasma, the choice of first wall material and thickness plays a large role in determining TBR.  As can be seen in Fig. \ref{fig:FirstWall_Scan}, beryllium and tungsten first walls increase TBR, while Carbon Fiber Composite (CFC) and Inconel 718 first walls decrease the TBR. It is surprising that small thicknesses of tungsten increase TBR when using \superscript{6}Li \cite{sato2003impact}, but this is due to the high-energy neutron multiplication cross section of tungsten. The ARC design uses a $1$ cm tungsten first wall to achieve a TBR of $1.1$, but (as shown in Fig. \ref{fig:FirstWall_Scan}) many other configurations are possible if the $TBR \geq 1.1$ limit is relaxed.

\subsection{Steady state thermal analysis}
\label{subsec:Blanket_Thermal_Analysis}

Thermal analysis was performed self-consistently from the plasma facing first wall to the superconducting magnets.

\subsubsection{Nuclear heating}

In order to perform a complete thermal analysis, nuclear heating was evaluated in MCNP for key volumes of the reactor (see Table \ref{tab:Nuclear_Heating}). It is important to note that both neutron and photon heating needed to be taken into account in order to accurately capture the physics of the nuclear reactions induced within the reactor materials. 

\begin{table}
  \centering
  \begin{tabular}{lc}
    \toprule
    Reactor Volume &  Total Heating (MW) \\
    \midrule
    \midrule
    Inner VV &  26 \\
    \midrule
    Coolant Channel &   33 \\
    \midrule
    Neutron Multiplier &   81 \\
    \midrule
    Outer VV &   18 \\
    \midrule
    Neutron Shield &   2.8 \\
    \midrule
    TF Coil &   0.014 \\
    \midrule
    \bottomrule
    \end{tabular}
  \caption{Nuclear (neutron and photon) heating of relevant ARC volumes from MCNP.}
  \label{tab:Nuclear_Heating}
\end{table}

\subsubsection{Bulk blanket thermal analysis}

The input temperature of FLiBe in the bulk blanket was set at $800$ K to give some margin from its freezing point of $732$ K. The temperature rise across the blanket was estimated using simple energy conservation. With the specific heat of FLiBe, an input flow velocity of $0.2$ m/s, and 18 inlets each with an area of $0.3$ m\superscript{2}, the exit temperature is found to be $900$ K. This calculation includes energy generated from breeding reactions within the blanket, which is significant. Turbulent 2-D COMSOL simulations were performed to verify this output temperature and estimate the peak blanket temperature to be less than $1000$ K.

\subsubsection{Vacuum vessel and channel thermal analysis}

In addition to providing favorable TBR and structural characteristics, the double-walled vacuum vessel design allows for the FLiBe in the channel to be used as active liquid cooling for the first wall. Single phase fluid cooling, with low conductivity/MHD issues, is a highly attractive cooling scheme for magnetic fusion. The vacuum vessel has $1$ cm thick tungsten tiles facing the plasma, mounted on a $1$ cm inner vacuum vessel. On the other side of a $2$ cm channel is a non-structural $1$ cm beryllium neutron multiplier attached to the $3$ cm thick outer vacuum vessel. As can be seen in Fig. \ref{fig:DoubleVV}, centrifugal pumps (with an externally located motor and drive shaft) circulate cool FLiBe at 800 K through the channels in the vacuum vessel to remove heat from the first wall. These channels flow in both poloidal directions for half the circumference and exhaust to the bulk blanket. A separate set of channels and pumps exist for divertor cooling.

In order to calculate the vacuum vessel cooling requirements, a 1-D analytic estimate was performed for a slab model of the system using
\begin{align}
   T_{channel,outlet} = T_{channel,inlet} + \frac{\left( \frac{1}{2} P_{tot} \right)}{\dot{m} C_{p}} \label{eq:ChannelTemp}
\end{align}
and
\begin{align}
   T_{VV,outlet} = T_{channel,outlet} + \Delta T_{interface} + \dot{q} \frac{\Delta t}{\kappa_{VV}} . \label{eq:VVTemp}
\end{align}
Here $P_{tot}$ is the total power deposited in the entire vacuum vessel model (the heat flux from the plasma plus the volumetric neutron power from Table \ref{tab:Nuclear_Heating}), $\dot{m}$ is the mass flow rate, $C_{p}$ is the specific heat, $\Delta t$ is the inner vacuum vessel thickness, and $\kappa_{VV}$ is the thermal conductivity. The temperature jump across the channel-wall interface, $\Delta T_{interface}$, is estimated using the Dittus-Boelter correlation,
\begin{align}
   Nu = 0.023 Re^{0.8} Pr^{0.4} , 
\end{align}
and the relation
\begin{align}
   \Delta T_{interface} = \frac{\left(2 \Delta t \right) \dot{q}}{\kappa_{channel} Nu} , \label{eq:convBoundaryLayer}
\end{align}
where $Re$ and $Pr$ are the channel Reynolds and Prandtl numbers. Lastly $\dot{q}$ is the heat flux through the inner vacuum vessel, estimated as $\dot{q} = \dot{q}_{plasma} + P_{VV,inner} / S_{VV}$, where $\dot{q}_{plasma}$ is the heat flux on the inner vacuum vessel surface from the plasma, $P_{VV,inner}$ is the total neutron power deposited in the inner vacuum vessel, and $S_{VV}$ is the vacuum vessel plasma-facing surface area. It was assumed that the inner vacuum vessel surface received a uniform radiative heat flux of 30\% of the plasma heating power (the sum of alpha, ICRF, and LHCD power), which is equal to $0.2$ MW/m\superscript{2}. This yields an average output FLiBe temperature of $823$ K and a maximum Inconel temperature of $1030$ K. 

\begin{figure}[ht]
 \centering
 \includegraphics[width=0.45\textwidth]{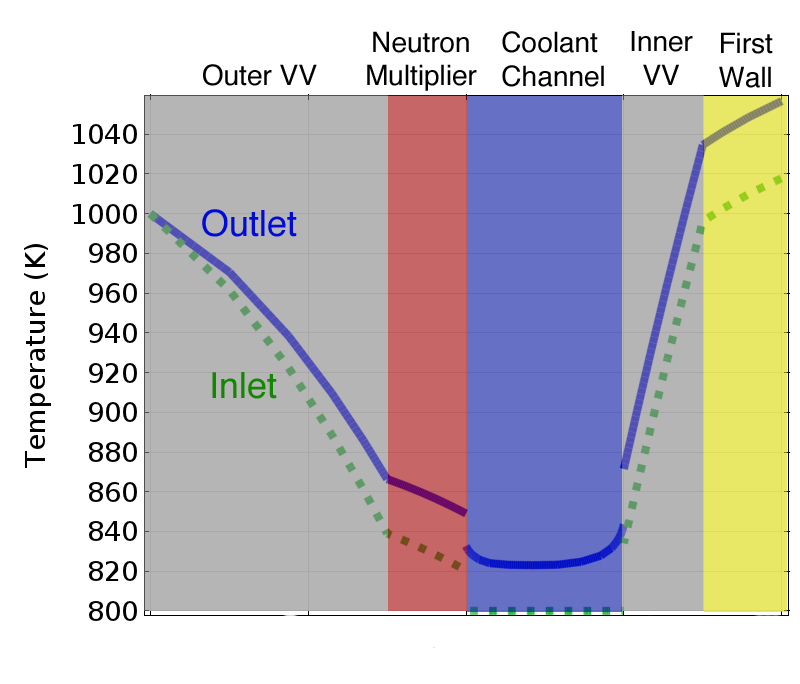}
 \caption{COMSOL model predicted temperature distribution across the vacuum vessel (with the plasma-facing surface on right) at both the channel inlet and outlet.}
 \label{fig:ChannelResults}
\end{figure}

These analytic estimates agree with a higher fidelity 2-D COMSOL calculation (see Fig. \ref{fig:ChannelResults}), with volumetric heat in each layer taken from Table \ref{tab:Nuclear_Heating}. Note the apparent discontinuities across the fluid/solid interface are due to the convective boundary layers, as was treated by Eq. \refEq{eq:convBoundaryLayer}. The outer surface in contact with the main FLiBe blanket was fixed at the maximum blanket temperature of $\sim$ $1000$ K to be conservative, although the average blanket temperature is expected to be $\sim$ 800-900 K. Importantly the maximum temperature on the vacuum vessel, which occurs on the plasma facing surface, is insensitive to the boundary condition on the blanket facing surface.

The fluid flow velocity in the vacuum vessel coolant channel was limited to be $2$ m/s because of concerns about flow-assisted corrosion, although nickel-based alloys have excellent corrosion-resistant properties for molten salts \cite{delpech2010molten} at these temperatures. At this fluid velocity, the COMSOL simulation (see Fig. \ref{fig:ChannelResults}) yields an average FLiBe output temperature of $826$ K and a peak Inconel temperature of $1034$ K at the inner surface near the channel outlet. The tungsten first wall is slightly hotter at $1056$ K, but since it is not structural we only require that it stays well below its melting temperature of $3700$ K.

\subsubsection{Divertor thermal analysis}

Explicit thermal modeling of the heat loading of a FLiBe-cooled divertor is beyond the scope of this study. However, we note that FLiBe would offer significant advantages for heat removal compared with the typical design choice of helium gas: much higher heat capacity, low flow rates, and low pressures. Instead, we will estimate the scale of the divertor heat loading in order to show that the compact size and high magnetic field of ARC do not dramatically intensify the problem. 

The key metric that determines the viability of cooling is the divertor heat flux, given by
\begin{align}
   \frac{P_{div}}{S_{div}} = \frac{P_{heat}}{2 \pi R_{0} \lambda_{SOL}} , \label{eq:divertorHeatFlux}
\end{align}
where $P_{heat} \equiv P_{\alpha} + P_{ext}$. This formula does not attempt to give an accurate prediction for the divertor heat loading. Rather it ignores detailed, albeit important, effects (such as the radiated power fraction) and tries to reveal how ARC compares to other more thoroughly-analyzed machines, when holding these detailed effects constant. Unfortunately, the scrape-off layer width, $\lambda_{SOL}$, cannot currently be predicted with any certainty, so we instead investigate where the ARC design falls between two heat loading figures of merit.  The upper extreme uses the Eich/Goldston scaling which estimates $\lambda_{SOL} \sim 1/B_{p}$ \cite{eich2013scaling}. The lower extreme assumes a pressure limited Scrape-off Layer (SOL) and leads to the estimate $\lambda_{SOL} \sim 2 \pi a \sqrt{\left(1 + \kappa^{2} \right)/2}$ \cite{whyte2012reactor}. Table \ref{tab:divertorSeverity} shows where ARC falls in comparison to a range of current and proposed tokamaks. Based on this analysis, we conclude that the difficulty of the divertor problem in ARC resides between ITER and reactor designs, which seems appropriate for a Pilot power plant. 

As previously noted, a rather large range in steady state fusion power (up to 525 MW) can be used to test divertor solutions. It is important to note that power exhaust issues are improved by operating at high safety factor (which decreases $B_p$ and widens $\lambda_{SOL}$) and by having high $Q_p$ (which reduces the required $P_{heat}$ at fixed $P_f$). Both these design features result from operating at high toroidal fields.

\begin{table*}
  \centering
  \begin{tabular}{l c c c c c}
    \toprule
     & ARC & ARIES-AT & JET & C-Mod & ITER \\
    \midrule
    \midrule
    Major radius, $R_{0}$ (m) & 3.3  & 5.2 & 2.92 & 0.67 & 6.2 \\
    \midrule
    Aspect ratio, 1/$\epsilon$ & 3 & 4 & 3.07 & 3.05 & 3.1\\
    \midrule
    Minor radius, $a$ (m) & 1.13  & 1.3 & 0.95 & 0.22 & 2 \\
    \midrule
    Elongation, $\kappa$ & 1.84  & 2.2 & 1.81 & 1.68 & 1.75 \\
    \midrule
    On-axis magnetic field, $B_{0}$ (T) & 9.2  & 5.8 & 3.6 & 5.4 & 5.3 \\
    \midrule
    Plasma current, $I_{p}$ (MA) & 7.8  & 12.5 & 4 & 1.5 & 15 \\
    \midrule
    $P_{heat}$ (MW) & 143  & 389 & 28.9 & 8 & 150 \\
    \midrule
    $1/B_{p}$ ($T^{-1}$) & 1.07  & 0.89 & 1.74 & 1.01 & 0.95 \\
    \midrule
    \midrule
    $P_{heat} B_{p} / R_{0}$ (MW*T/m) & 41.0  & 84.2 & 5.69 & 11.8 & 25.5 \\
    \midrule
    $P_{heat} / S_{\text{p}}$ (MW/$m^2$) & 0.67  & 0.85 & 0.18 & 1.00 & 0.21 \\
    \bottomrule
  \end{tabular}
  \caption{Inter-machine comparison \cite{AymarITERSummary2001, NajmabadiARIES_AT2006, RebutJETOverview1985, GreenwaldCModHMode1997} of divertor heat loading metrics, where $S_{\text{p}}$ is the surface area of the plasma.}
  \label{tab:divertorSeverity}
\end{table*}

\subsubsection{Neutron shield thermal shielding and cooling}
\label{sec:Power_NeutronShieldThermalShieldingAndCooling}
In order to thermally isolate the 900 K blanket tank from the neutron shielding and magnet systems, the blanket tank is surrounded by 1 cm of aluminum silicate wool. The heat flux through the blanket thermal shield can be analytically estimated as
\begin{align}
   \dot{q} = \frac{\kappa_{TS} \Delta T_{TS}}{\Delta t_{TS}} , \label{eq:BlanketHeatFlux}
\end{align}
where $\kappa_{TS}$ is the thermal conductivity of the aluminum silicate thermal shield, $ \Delta T_{TS}$ is the temperature difference across the shield, and $\Delta t_{TS}$ is the shield thickness. Using the thermal conductivity of 0.1 $W~m^{-1}K^{-1}$ \cite{daryabeigi1999effective} and keeping the outside of the blanket thermal shield at 293 K (room temperature), the heat flux through the blanket thermal shield is calculated to be 6 $kW/m^{2}$. However, this is negligible compared to the 2.8 MW of neutron power (see Table \ref{tab:Nuclear_Heating}) deposited in the neutron shield. This heat is removed by a 1 cm thick, toroidally continuous coolant channel of vertically flowing water. We can use 0-D power balance to find that
\begin{align}
   P_{V} &= \rho_{C} v_{C} \left( 2 \pi R \Delta t_{C} \right) C_{p} \Delta T_{C} ,
\end{align}
where $P_{V} = 2.8$ MW the volumetric neutron heating, $\rho_{C} = 1000$ kg/m\superscript{3} is the coolant density, $v_{C} = 1$ m/s is the coolant flow velocity, $R = 1.5$ m is the major radial position, $\Delta t_{C} = 1$ cm is the coolant channel thickness, $C_{p} = 4200$ J/kg-K is the specific heat, and $\Delta T_{C} = 7$ K is the coolant temperature difference between the inlet and the outlet. Assuming medium-roughness metal pipes, the pressure drop through our cooling channel is calculated to be $\Delta p = 1.1 x 10^4$ Pa. Assuming a conservative pump efficiency of $50\%$, the total power required to cool the neutron shield will be $\sim$ 2.5 kW, which is negligible compared to the other reactor power requirements. Thus, the neutron shield can be cooled using water in the remaining $2$ cm of inboard radial build allocated for thermal insulation of the neutron shield (see Fig. \ref{fig:inboardRadialBuild}).

\subsection{Disruption analysis}
\label{sec:PowerCore_disruption}

Primarily, disruptions pose two distinct threats to a device.  First, during the disruption plasma current can transfer to the vacuum vessel and cause $\vec{j} \times \vec{B}$ forces that exert large electromagnetic stresses on components. Second, the plasma can move and directly contact the wall. This transfers the energy contained in the plasma to the PFCs, which can melt and erode the surface. Here we will show that the small size and high fields of ARC do not dramatically affect disruption survivability compared to other devices.

\subsubsection{Disruption mechanical analysis}
\label{subsubsec:disruptionMechAnalysis}

The most significant steady state force acting on the vacuum vessel is buoyancy, due to the significant volume of FLiBe it displaces. The vacuum vessel and posts must be designed to support themselves. However this requirement is insignificant compared to the transient forces during a plasma disruption. These stresses were modeled in COMSOL and treated analytically using a simplified model. Stresses arise from the $\vec{j} \times \vec{B}$ forces between the background magnetic fields and plasma current that has been transferred to the vessel. Disruptions manifesting themselves as kink modes are particularly troublesome because they can deposit significant {\em poloidal} current, which interacts with the dominant toroidal magnetic field. Toroidal current can only interact with the background vertical field and generally produces forces an order of magnitude below those from the poloidal halo current.

\begin{table}
  \centering
  \begin{tabular}{c | c | c c c c c}
    \toprule
     & ITER & ARC & ARIES-AT & JET & C-Mod \\
    \midrule
    \midrule
    $\Delta_{vf}/117$          & 1  & 0.91 & 0.84 & 0.24 & 0.08 \\
    \midrule
    $\Delta_{vt}^{CQ}/56$  & 1  & 0.89 & 1.72 & 0.26 & 0.53 \\
    \midrule
    $\Delta_{vt}^{TQ}/1.1$ & 1  & 1.50 & 3.14 & 0.13 & 0.18 \\
    \bottomrule
  \end{tabular}
  \caption{Inter-machine comparison \cite{AymarITERSummary2001, NajmabadiARIES_AT2006, RebutJETOverview1985, GreenwaldCModHMode1997} of the severity of the disruption forces and disruption thermal loading (normalized to the value for ITER).}
  \label{tab:machineSeverity}
\end{table}

We assumed a worst case unmitigated, asymmetric disruption with a toroidal mode number of $n=1$, $f_{halo} = 40\%$ (the fraction of the plasma current directed poloidally as halo current), and a toroidal peaking factor of $2$ (the ratio of the maximum halo current to the toroidal average). The most significant stresses occur on the inboard midplane at the toroidal location of peak current. Here the force density distribution becomes
\begin{align}
   \vec{F}_{VV} &= - \frac{ f_{halo}I_{p} B_{0} R_{0}}{\pi \left( R_{0} - a \right)^{2} d_{VV}} \hat{e}_{R} \label{eq:disruptionForce}
\end{align}
where $d_{VV}$ is the vacuum vessel thickness. Modeling the vacuum vessel as a thin cylinder we find the stress to be
\begin{align}
   \sigma_{VV} = \frac{p_{VV} D_{VV}}{2 d_{VV}} = \frac{ f_{halo}I_{p} B_{0}}{\pi \left( 1 - \epsilon \right) d_{VV}} \label{eq:disruptionStress}
\end{align}
where $p_{VV} = F_{VV} d_{VV}$ is the effective pressure acting on the vacuum vessel and $D_{VV} = 2 \left( R_{0} - a \right)$ is the diameter of the cylinder. This demonstrates that the parameter
\begin{align}
   \Delta_{vf} \equiv \frac{\left( I_{p}/\text{MA} \right) \left( B_{0}/\text{T} \right)}{1-\epsilon}
\end{align}
is representative of the severity of disruptions in a given machine. If two machines have the same value of $\Delta_{vf}$ then they both require similarly strong/thick vacuum vessels to withstand a disruption with a given $f_{halo}$. Table \ref{tab:machineSeverity} shows that, despite the small size and high fields of ARC, the difficulty of tolerating the forces of a disruption is comparable to what ITER faces. However, it should be noted that, even though ARC faces a similar problem to ITER, it has half the space (in major radius) available for a vacuum vessel that addresses it.

Calculating the exact magnitude of the stresses caused by a disruption is difficult because the halo current distribution varies with time and may not flow through the full thickness of the vacuum vessel. In the worst case the current will be confined only to the $1$ cm inner vacuum vessel. In the best case it will be distributed over the inner vacuum vessel and the $3$ cm outer vacuum vessel.

\begin{figure}[ht]
 \centering
 \includegraphics[width=0.45\textwidth]{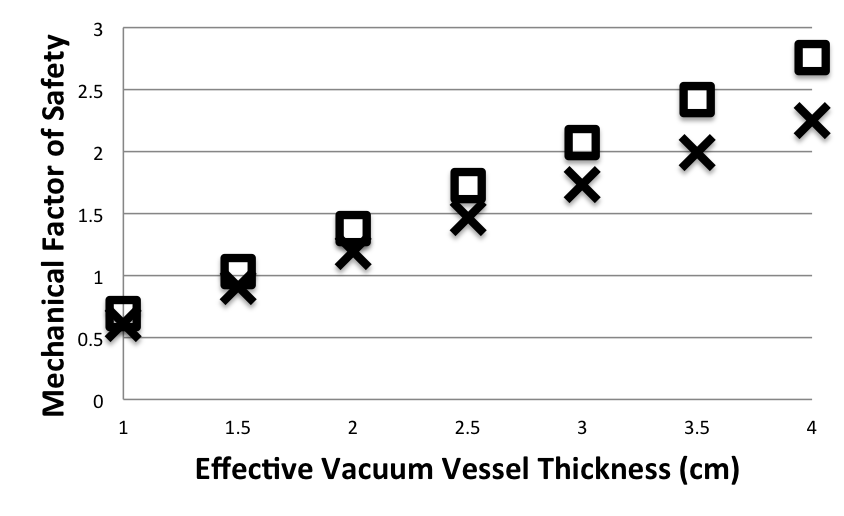}
 \caption{COMSOL (crosses) and analytic (squares) estimates from Eq. \refEq{eq:disruptionStress} of the mechanical factor of safety as a function of the effective vacuum vessel thickness.}
 \label{fig:VVmechanicalSafetyFactor}
\end{figure}

We define the mechanical factor of safety to be the ratio of the yield stress of the material to the peak stress expected during the disruption. The yield stress of Inconel 718 at the maximum vacuum vessel temperature of $1030$ K is roughly $940$ MPa \cite{SpecMetalsInconel718}. Both analytic theory and COMSOL give similar results, shown in Fig. \ref{fig:VVmechanicalSafetyFactor}. We see that, in the most pessimistic case, it seems unlikely that the vacuum vessel will survive. However, ideally a full, worst-case, unmitigated  disruption should never occur and our assumptions have been conservative. Even if it does the connections between the vacuum vessel and the permanent reactor components can be designed to fail gracefully and the vessel can be replaced. This is a much better scenario than similar first wall damage happening to large solid blanket modules.

It is important to note that in the ARC design, the blanket tank, rather than the vacuum vessel, is the primary nuclear safety barrier. Thus, even in the worst-case scenario of a catastrophic vessel breach, both the activated vessel and the tritium-containing FLiBe inside of it would still be contained within the nuclear safety barrier of the blanket tank.

\subsubsection{Disruption thermal analysis}

As the plasma contacts the wall it deposits its energy into the plasma-facing components. Depending on how much energy is deposited and how fast it happens the first wall PFCs can melt. Melting of the first wall is intolerable because it deforms the PFCs, ruining their alignment to the B field. This leads to uneven distribution of heat during subsequent disruptions and further melting. We will study this effect following methods discussed in Ref. \cite{OlynykThesis2013}.

In the limit of an instantaneous heat pulse, the maximum temperature rise on the surface is given by
\begin{align}
   \Delta T_{max} \equiv T_{max} \left( t \right) - T_{0} \propto \frac{H_{e, max}}{\sqrt{t}} , \label{eq:disruptionTempChange}
\end{align}
where $H_{e}$ is the radiant exposure given in units of energy per unit surface area.

There are two types of energy contained by the plasma and each are transferred to the vacuum vessel at different rates. The current quench phase of a disruption transfers the energy contained in the poloidal field, given by
\begin{align}
   W_{pol} = \frac{1}{2} L I_{p}^{2} , \label{eq:disruptionPolEnergy}
\end{align}
where $L = \mu_{0} R_{0} \left( \text{ln} \left( \frac{8 R_{0}}{a} - 2 + \frac{l_{i}}{2} \right) \right)$ is the plasma inductance ignoring shaping effects. The thermal quench phase of a disruption transfers the energy contained in the thermal energy of the plasma given by
\begin{align}
   W_{th} = \frac{3}{2} V \left\langle p \right\rangle_{V} . \label{eq:disruptionThermalEnergy}
\end{align}
The timescales for these two processes are the current quench time, $t_{CQ}$, and the thermal quench time, $t_{TQ}$, respectively. Since the maximum radiant exposure can be approximated as
\begin{align}
   H_{e, max} \propto \frac{W}{S_{VV}} ,
\end{align}
where $S_{VV}$ is the vacuum vessel surface area, the temperature rise from each phase is
\begin{align}
   \Delta T_{max, CQ} \propto \frac{I_{p}^{2}}{a \sqrt{t_{CQ}}}
\end{align}
and
\begin{align}
   \Delta T_{max, TQ} \propto \frac{a B_{0}^{2} \beta_{T}}{\sqrt{t_{TQ}}} .
\end{align}

The current quench timescale is given by the $L/R$ time of the plasma, meaning
\begin{align}
   t_{CQ} = \frac{L}{\mathscr{R}} \propto a^{2} ,
\end{align}
where $\mathscr{R}$ is the plasma resistance. Therefore, we will define
\begin{align}
   \Delta_{vt}^{CQ} \equiv \frac{\left(I_{p} / \text{MA} \right)^{2}}{\left( a / \text{m} \right)^{2}}
\end{align}
to approximate the difficulty of handling the current quench heat loading. The thermal quench time is not precisely known, however Fig. 54 of Ref. \cite{MirnovITERch3PhysicsBasis1999} shows that, to best approximation, $t_{TQ} \propto a$, i.e. the thermal quench follows a convective timescale. Therefore, we will define
\begin{align}
   \Delta_{vt}^{TQ} \equiv \sqrt{a / \text{m}} \left( B_{0} / \text{T} \right)^{2} \beta_{T}
\end{align}
to represent the difficulty of handling the current quench heat loading. We see in Table \ref{tab:machineSeverity} that, because of the low current (high $q_{95}$) in ARC, the current quench should be relatively modest. However, because of the small size and high power density, the thermal quench will be more problematic. One sees that ARC is intermediate with respect to ITER and a large fusion reactor. Safe dissipation of thermal energy, especially during a disruption, remains a research challenge for all burning plasma tokamaks.

\subsection{Material damage and activation}

MCNP was used to compute material damage due to irradiation, keeping track of both DPA and
\superscript{4}He production in several components (see Table \ref{tab:materialdamage}). Little research has been done regarding how Inconel 718 responds to the irradiation environment of a fusion device \cite{iida2004nuclear}. However, studying the response of components to fusion neutron effects is part of the motivation for ARC.

For context, the irradiation lifetime of ferritic steel is estimated at $150$-$200$ DPA \cite{NajmabadiARIES_AT2006}. Additionally, with a helium concentration of 500 appm, martensitic steel survives, but shows some increase in yield strength and some reduction in ductility \cite{JungHeEffectOnSteel2001, MoeslangHeEffectOnSteel1988}. It is unknown if Inconel 718 would behave similarly in a fusion neutron spectrum, but one expects the vacuum vessel would survive for at least 6-12 months.

\begin{table}
  \centering
  \begin{tabular}{l | c | c }
    \toprule
    Component & \superscript{4}He (ppm) & DPA \\
    \midrule
    \midrule
    Inner vacuum vessel  &  280   & 44    \\
    Outer vacuum vessel & 140    & 26    \\
    Blanket tank               & 0.58   & 0.14 \\
    Support column          & 8.3     & 3.0   \\
    \bottomrule
  \end{tabular}
  \caption{Component helium production and DPA in one FPY. The production of \superscript{3}He was found to be negligible.}
  \label{tab:materialdamage}
\end{table}

Initial tests suggest that rewelding steel might become problematic at around 1 He appm \cite{AymarITERSummary2001,el2006nuclear}, which implies that the blanket tank would need replaced after only two years. In order to eliminate this issue, the replaceable components will be bolted, rather than welded to the permanent structure. However, this isn't perfect, as it creates concerns about differential swelling and diffusive bonding (a critical area of active research for both fission and fusion materials). One possible solution would be to increase the blanket tank height to allow for additional FLiBe/shielding between the plasma and the bolted sections.

A primary advantage of the liquid immersion blanket design is that it significantly reduces the amount of solid material near the plasma. The vacuum vessel design for ARC will have only 85 metric tons ($\sim$ 11 $m^3$) of solid material compared to over 2000 metric tons in ITER \cite{SborchiaITER2013}. This drastically lowers the amount of material that could become activated.

\subsection{Material corrosion}
\label{subsec:Corrosion}

While a full flow-assisted corrosion analysis of the structural materials in contact with FLiBe is beyond the scope of this paper, basic corrosion rates were investigated to ensure that the design is reasonable. Recent experimental work as part of the FFHR design study \cite{sagara2006conceptual} investigated the corrosion effects of FLiBe on Inconel 625, an alloy similar to ARC's Inconel 718 (which was chosen for its high-temperature strength). At a temperature of 873 K (similar to that of the ARC blanket), the static corrosion rate of Inconel 625 was found to be 1.1 $\mu$m/yr \cite{kondo2009high}. These results, combined with the extremely slow 0.2 m/s input flow velocity of FLiBe in the ARC blanket give confidence that the Inconel 718 vacuum vessel and blanket tank will survive any corrosive effects of the FLiBe blanket, particularly in the ``FNSF stage''. As discussed in Section \ref{sec:RandDReq}, further research into radiation-assisted corrosion of Inconel is required.

\section{Economics}
\label{sec:Economics}

The main driver for minimizing the size of ARC is to reduce the cost of building the reactor. While a full costing of the ARC reactor is beyond the scope of this paper, a rough costing based on volumes and materials prices has been performed. With a major radius of 3.3 m, ARC is similar in size to experiments that have already been built (JET and TFTR). The following analysis aims to justify that ARC is feasible from a materials cost standpoint. 

In order to assess the bulk materials costs of the ARC reactor, the reactor was broken down into three subsystems: the replaceable vacuum vessel, the blanket, and the magnets/structure. In order to estimate the costs of components requiring extensive machining, a volumetric cost scaling law based on several design studies was used \cite{MeadeScaling}. 

\subsection{Materials Costs}

Material prices were obtained either from estimates of commodity prices or quotes requested from manufacturers (see Table \ref{tab:MaterialsCosts}).  Although the REBCO tape and FLiBe are not technically raw materials they are included in the bulk costing analysis.

\begin{table}
  \centering
  \begin{tabular}{lc}
    \toprule
    Component & Cost\\
    \midrule
    \midrule
    Beryllium \cite{USGSreport} & $\$257/kg$\\
    \midrule
    Inconel 718 \cite{ContinentalSteelQuote} & $\$56/kg$\\
    \midrule
    Tungsten \cite{USGSreport} & $\$29/kg$\\
    \midrule
    Stainless steel 316LN \cite{ContinentalSteelQuote} &  $\$9.6/kg$\\
    \midrule
    Copper \cite{USGSreport} & $\$8.3/kg$\\
    \midrule
    REBCO tape \cite{SuperPowerquote} &  $\$18/m-\$36/m$\\
    \midrule
    FLiBe \cite{sanders1971review} &  $\$154/kg$\\
    \midrule
    TiH\subscript{2} \cite{ADMAquote} &  $\$26.4/kg$\\
    \bottomrule
    \end{tabular}
  \caption{Materials costs in 2014 US dollars. Due to the large amount of REBCO required, the quote was given as a price range. Note the REBCO cost is in \$/m, rather than \$/kg.}
  \label{tab:MaterialsCosts}
\end{table}

\subsection{Fabricated Component Scaling}

In order to provide a better cost estimate than simple materials costs, a rough scaling based on total cost per weight was employed, following Ref. \cite{MeadeScaling}.  In this scaling, the total projected costs of four burning plasma designs (FIRE, BPX, PCAST5, and ARIES-RS) were divided by the weight of the device from the cryostat inwards. As seen in Ref. \cite{MeadeScaling}, these four costs/tonne were similar, which gives confidence that the scaling is not machine specific. As a simple estimate the ARC study averaged the four costs/tonne and adjusted for inflation.  The adjusted scaling for ARC was found to be $\$1.06M/tonne$ in FY2014 US dollars.  This scaling will be referred to as the ``fabricated" component scaling. To calculate the total cost of a component using the fabricated component scaling, we multiply the total weight of the component by $\$1.06M/tonne$. In the case of components that do not require machining (e.g. the FLiBe blanket), the fabricate cost will be the same as the material cost.

\begin{table*}
  \centering
  \begin{tabular}{lccccc}
    \toprule
    \textbf{Component} & \textbf{Volume} & \textbf{Weight} & \textbf{Material} & \textbf{Material Cost} & \textbf{Fabricated Cost} \\
    \midrule
    \midrule
    First wall & $2.01 \  m^3$ & 3.72 tonnes & Tungsten & $\$110k$ & $\$4.0M$ \\
    \midrule
    Inner VV wall & $2.03 \ m^3$ & 16.6 tonnes & Inconel 718 & $\$930k$ & $\$18M$ \\
    \midrule
    Multiplier & $4.09 \ m^3$ & 3.82 tonnes & Beryllium & $\$980k$ & $\$4.1M$ \\
    \midrule
    Outer VV wall & $6.27 \ m^3$ & 51.4 tonnes & Inconel 718 & $\$2.9M$ & $\$55M$ \\
    \midrule
    VV ribbing & $0.83 \ m^3$ & 6.80 tonnes & Inconel 718 & $\$380k$ & $\$7.2M$ \\
    \midrule
    VV posts & $0.51 \ m^3$ & 4.14 tonnes & Inconel 718 & $\$230k$ & $\$4.4M$ \\
    \midrule
    \midrule
    \textbf{Replaceable VV Subtotal} & $\bf{15.7 \ m^3}$ & \textbf{86.5 Tonnes} & \textbf{N/A} & $\bf{\$5.5M}$ & $\bf{\$92M}$ \\
    \midrule
    \midrule
    Blanket tank & $11.8 \ m^3$ & 97.1 tonnes & Inconel 718 & $\$5.4M$ & $\$100M$ \\
    \midrule
    TiH\subscript{2} shield & $101 \ m^3$ & 380 tonnes & TiH\subscript{2} & $\$10M$ & $\$10M$ \\
    \midrule
    Channel FLiBe & $4.09 \ m^3$ & 8.07 tonnes & FLiBe & $\$1.2M$ & $\$1.2M$ \\
    \midrule
    Blanket tank FLiBe & $241 \ m^3$ & 475 tonnes & FLiBe & $\$73M$ & $\$73M$ \\
    \midrule
    Heat exchanger FLiBe & $241 \ m^3$ & 475 tonnes & FLiBe & $\$73M$ & $\$73M$ \\
    \midrule
    \midrule
    \textbf{Blanket Subtotal}  & $\bf{599 \ m^3}$ & \textbf{1440 tonnes} & \textbf{N/A} & $\bf{\$160M}$ & $\bf{\$260M}$ \\
    \midrule
    \midrule
    Magnet structure & $544 \ m^3$ & 4350 tonnes & SS316 LN & $\$42M$ & $\$4.6B$ \\
    \midrule
    Magnet top ring & $120 \ m^3$ & 959 tonnes & SS316 LN & $\$9.2M$ & $\$9.2M$ \\
    \midrule
    REBCO structure & $40 \ m^3$ & 358 tonnes & Copper & $\$3.0M$ & $\$380M$ \\
    \midrule
    REBCO tape & $5730 \ km$ & $\sim 0$ tonnes & REBCO & $\$103M-\$206M$ & $\$100M-\$210M$ \\
    \midrule
    \midrule
    \textbf{Magnet/Structure Subtotal} & $\bf{704 \ m^3}$ & \textbf{5670 tonnes} & \textbf{N/A} & $\bf{\$160M-\$260M}$ & $\bf{\$5.1B-\$5.2B}$ \\
    \midrule
    \midrule
    \textbf{Grand Total} & $\bf{1320 \ m^3}$ & \textbf{7190 tonnes} & \textbf{N/A} & $\bf{\$330M-\$430M}$ & $\bf{\$5.5B-\$5.6B}$ \\
    \bottomrule
    \end{tabular}
  \caption{Cost/weight breakdown table for ARC reactor (excluding the balance of plant equipment).}
  \label{tab:CostTable}
\end{table*}

\subsection{Replaceable Vacuum Vessel}

The replaceable vacuum vessel costs were analyzed using the MCNP neutronics model to estimate the material volumes. The material volumes were multiplied by the material densities and assigned a total cost using Table \ref{tab:MaterialsCosts}.  The Inconel ``ribbing'' structure inside the vacuum vessel cooling channel was estimated to be 10$\%$ of the channel volume and the total vacuum vessel post volume was approximated to be $10\%$ of the value reported by MCNP (because the posts are discrete, not toroidally continuous as in the 2-D model). The divertor cost was left out of the analysis because the design was left as an open question. However, a rough estimate for a 2 cm tungsten divertor covering 20$\%$ of the first wall area is on the order of $\$500k$ for materials, implying a $\$17.5M$ fabricated cost. The replaceable vacuum vessel cost breakdown and subtotal are shown in Table \ref{tab:CostTable}. 

\subsection{Blanket}

The blanket costs were analyzed using the MCNP neutronics model to estimate the material volumes.  The material volumes were multiplied by the material densities and assigned a total cost using Table \ref{tab:MaterialsCosts}. In order to estimate the volume of FLiBe required for the heat exchanger (in the balance of plant), a simple shell and tube model with cooling fins using helium as the secondary fluid was used. With this estimate, the volume of FLiBe in the heat exchanger was calculated to be between 160 m\superscript{3} and 600 m\superscript{3}, depending on the details of the heat exchanger design. It was assumed that the heat exchanger would be designed to minimize the amount of FLiBe required, so a volume of 241 m\superscript{3} (the same volume as in the blanket tank) was chosen as a rough estimate. Since the TiH\subscript{2} is in powder form and the FLiBe is liquid, the fabricated cost for components made from these materials was set equal to the material cost. The blanket cost breakdown and subtotal are shown in Table \ref{tab:CostTable}. 

\subsection{Magnets}

The magnet structure costs were analyzed using the COMSOL magnet stress model to estimate the volume of steel required.  In order to estimate the required length of tape, the area of REBCO needed to produce the given magnetic field was computed (taking into account the geometry of the coils). This was divided by the area of a single tape to find the number of tapes required and then multiplied by the perimeter of the superconducting coil (see Section \ref{sec:Magnets}). Material volumes/lengths were assigned a total cost using Table \ref{tab:MaterialsCosts}. Note that because the magnet tension ring (holding the top coil flanges together) is a large but very simple component, the fabricated cost is expected to be similar to the material cost. The reactor base is treated the same. It is important to note that the reactor base is conservatively modeled as entirely steel for the cost evaluation, but the actual structure would likely be comprised of both concrete and steel.  The magnet/structure cost breakdown and subtotal are shown in Table \ref{tab:CostTable}.

\subsection{Cost Feasibility}

Assuming the higher cost estimate for the REBCO tape, the materials costs for ARC total $\$428M$ and the total fabricated component cost estimates total $\$5.56B$. While these are simple estimates, they provide several critical insights. The material costs of the ``novel'' materials/components in the ARC reactor (REBCO tape, FLiBe, TiH\subscript{2} shielding) are only a small fraction of the total fabricated cost predicted by the fabricated component scaling.  While there is a price to generating higher magnetic fields due to the extra structure in the magnets, this premium is easily overcome by the overall ability to reduce the volume of the plasma, shield, and coils. This can be seen by noting that 9.2 T ARC has a fifth of the $\sim$ $\$24B$ price of the 5.3 T ITER (calculated by applying the fabricated component scaling to the $\sim$ 23,000 tonne ITER). Yet ARC matches ITER's fusion power and produces net electricity. The cost of ARC is approximately one-third the cost of the 8 T ARIES-RS ($\sim \$14B$), but ARIES-RS has approximately four times the electrical output. The smaller ARC is appropriate for an ``entry-level'' fusion Pilot plant, but there likely exists a better economic optimization of magnetic field strength versus mass for a full power plant. Finally, one notes that the ``fabricated'' cost for a commercial version of ARC will be reduced through economies of scale if multiple reactors are built.

\section{Identification of R\&D requirements}
\label{sec:RandDReq}

\subsection{Plasma physics and current drive}

First, the I-mode regime must be further studied, characterized, and demonstrated with non-inductive profiles. As with all small reactor designs the core scenario exploits enhanced confinement from current profile and q control. Therefore, a fully developed and consistent non-inductive scenario with the required physics parameters should be explored more completely. Ideally, we would use a burning plasma experiment in order to also test the self-determining effect of alpha-dominated heating on the plasma profiles. One aspect of I-mode of particularly significant to the ARC design is the maximum power density normalized to density achievable before transition to H-mode \cite{Hubbard_IModeTransition}. The published range spans $P_{heat}/S_{p}/n_{20}$ = $0.2$ to $0.5$ MW/$m^2$ over a range of magnetic fields (up to $6$ T). The operating point of ARC is characterized by $P_{heat}/S_{p}/n_{20}\sim{0.55}$ at 9.2 T. In addition, it is important to understand the mechanisms allowing I-mode to maintain a stationary pedestal without damaging ELMs.

The engineering of the lower hybrid system also requires significant research. Currently, reliable lower hybrid klystron sources exist at $6$ GHz, but the $8$ GHz system incorporated in ARC has yet to be demonstrated. Like all components, there is limited data on the integrated response of possible waveguide materials in the fusion neutron environment. A particular challenge for an RF launcher in a reactor is surviving the high temperatures and radiation damage, while maintaining high electrical conductivity to avoid resistive losses and heating. However, the ARC LHCD design is no more problematic than ARIES-AT design, which uses LHCD launched from the low-field side. Still, since the waveguides will be hot, they will likely be made more resilient because of annealing. Furthermore, we only require that they have a lifetime longer than that of the vacuum vessel, which is only a couple of years in ARC.

\subsection{Magnets}

The most important outstanding magnet research pertaining to ARC is the design and testing of REBCO superconducting joints.  While several designs have been tested in a small-scale ``bench top'' setting \cite{YBCO_JOINTS_GOH}, they must be proven to be robust at reactor-level fields and stresses. The joint insulation must be tested at the high quench voltages (about 2~kV) and has to protect the joint against Paschen and tracking discharges.

In addition, further research into the properties of REBCO at liquid hydrogen temperatures (20~K) is required to assess the feasibility of different temperature regimes of superconductor operation. The performance of REBCO cables at 20~K in a complex and varying (in both space and time) magnetic field needs to be studied. Also, cabling methods must be developed for the industrial production of coil conductors. Lastly, an effective quench detection/protection system must be designed and demonstrated at 20~K.

% Maybe add something about uninsulated coils

\subsection{Fusion power core}

Crucial, both for ARC and any other superconducting reactor, is an accurate understanding of how superconducting tape responds to fast neutron irradiation. The amount of material needed to shield REBCO directly constrains the minimum possible major radius needed to achieve a given TF coil lifetime. Current fluence experiments only establish a conservative irradiation limit, and REBCO has never been tested to failure in a fusion relevant environment \cite{bromberg2001options}.

Additionally, some designs have been proposed for extracting tritium from FLiBe \cite{dolan1992vacuum, fukada2007design}. However, due to the cost of tritium handling, few experiments have been built to assess the turn around time needed for extraction and subsequent refueling. This turn around time directly determines the necessary tritium inventory and the quantity that is needed to initially start a reactor. More experiments are required to demonstrate that this turn around time is sufficiently fast to meet regulatory limits on the total on-site tritium inventory. 

Another source of uncertainty is the effect of a strong background magnetic field on the flow, turbulence, and heat transfer characteristics of FLiBe. Initial computational investigation seems to show that these effects can be neglected in a wide range of fusion relevant parameters \cite{AbdouFLiBeMHD2001}, but a more detailed investigation is required for an engineering design. Specifically, it is unknown if the magnetic field will alter flow assisted corrosion, which could impact the vacuum vessel and divertor coolant channels. In addition, it is unknown how much the resistivity of the FLiBe will be affected by exposure to high neutron and gamma fluxes from the core.

Finally, a better understanding of radiation-assisted corrosion of Inconel in contact with FLiBe is required to reduce uncertainty in vacuum vessel and blanket tank lifetimes. While the experiments in Ref. \cite{kondo2009high} indicate that the corrosion of Inconel in FliBe will be minimal (see Section \ref{subsec:Corrosion}), these experiments do not take into account radiation effects (such as radiation-assisted transport of chromium to the surface). This could significantly speed up corrosion in a fusion reactor. In addition, experiments at higher temperatures (up to 1400 K) are required to assess corrosion for a Pilot configuration of ARC, where the blanket temperature will be increased to allow for more efficient electricity generation. Simple estimates based on a non-ideal Brayton cycle (see Section \ref{sec:DesignOverview}) indicate that increasing the FLiBe blanket operating temperature from 900 K to 1200 K would allow the plant thermal efficiency to improve from 40$\%$ to 50$\%$. This increase in blanket temperature would require more robust structural and first wall materials, further highlighting the need for fusion materials research.

\section{Conclusions}
\label{sec:Conclusions}

With a major radius of $3.3$ m and minor radius of $1.1$ m, ARC is significantly smaller in size and thermal output than most current reactor designs, which typically generate $\sim$ 1 GWe. ARC produces $525$ MW of fusion power ($\sim 200$ MWe), operating in the promising I-mode regime. Steady state plasma current is driven by ICRF fast wave and lower hybrid waves, both launched from the high field side. The reactor has a bootstrap fraction of only $63$\%, which gives operators greater control of the current profile. This, together with the high safety factor of $q_{95} \sim 7$, reduces the likelihood of disruptions. The TF coils use REBCO superconductors, allowing ARC to have an on-axis magnetic field of $9.2$ T and peak field on coil of $\sim$ 23 T. The TF coils are also demountable and the tritium breeding blanket is a tank of liquid FLiBe, which permits all internal components to be installed as a single module. This allows the device to perform as a fusion nuclear science facility, testing many different vacuum vessel and divertor configurations. The initial vacuum vessel is two concentric Inconel 718 shells, separated by structural ribbing and FLiBe coolant channels that enable a tritium breeding ratio of $1.1$. Neutron shielding allows for ARC to operate for $9$ FPY before the reaching the lower bound on the TF coils neutron survivability. This lifetime could be increased dramatically with just a small increase in reactor size.

The ARC reactor design study has shown that high magnetic field, demountable TF coils, and an all-liquid blanket synergistically combine to provide several advantages over traditional tokamak designs. First and foremost, the ARC design allows for much smaller devices. As shown in Section \ref{sec:Economics}, even with the novel materials required for the ARC design, this small size reduces the overall cost of building a reactor. The modular nature of ARC allows for the demonstration reactor to also be used as an FSNF, testing several vacuum vessel/first wall/divertor designs in a reactor-relevant environment. The all-liquid blanket of ARC simplifies cooling and dramatically reduces the amount of activated waste produced. 

While a full engineering design is beyond the scope of the ARC study, the benefits and feasibility of compact, high-field reactor/FNSF designs have been shown. The ARC study has not identified any insurmountable difficulties with the given design, motivating more detailed study into compact, high-field devices.

\section{Acknowledgments}
\label{sec:Acknowledgments}

We thank Leslie Bromberg, Charles Forsberg, Martin Greenwald, Amanda Hubbard, Brian LaBombard, Bruce Lipschultz, Earl Marmar, Joseph Minervini, Geoff Olynyk, Michael Short, Pete Stahle, Makoto Takayasu, and Stephen Wolfe for conversations and comments that improved this paper. We also thank Zach Hartwig for allowing us to use his C++ wrapper for MCNP and for advice regarding neutronics. BNS was supported by U.S. DoE Grant No. DE-FG02-94ER54235. JB was supported by U.S. DoE Grant No. DE-SC008435. FJM was supported by the U. S. Department of Energy, Office of Fusion Energy Science under Grant No. DE-FC02-93ER54186. JMS is supported by the National Science Foundation Graduate Research Fellowship Program, under grant No. 1122374. This work originated from a MIT Nuclear Science and Engineering graduate course. DGW acknowledges the support of the NSE Department and the PSFC.



%% References


\bibliographystyle{elsarticle-num}
\bibliography{ArcManuscript}






% \end{thebibliography}

\end{document}